\PassOptionsToPackage{pdfpagemode=UseOutlines, colorlinks=true, bookmarks=true, bookmarksnumbered=true}{hyperref}
\documentclass{iopjournal}
\usepackage{lmodern}
\usepackage{ragged2e} 

\makeatletter
\long\def\@makecaption#1#2{%
  \vskip\abovecaptionskip
  \justifying\normalfont\small
  \textbf{#1}~#2\par
  \vskip\belowcaptionskip
}
\makeatother

\usepackage{float}

\usepackage{amssymb, amsbsy, amsmath, latexsym, dsfont, array, layout,
graphicx, mathrsfs, color,soul}

\usepackage[toc,page]{appendix}
\usepackage{mathtools}

\usepackage{xcolor}
\usepackage[normalem]{ulem} 
\usepackage{amsfonts}
\usepackage{mathrsfs} 
\usepackage{comment}

\usepackage[most]{tcolorbox}
\usepackage{newfloat}
\usepackage{mdframed}
\usepackage{alltt}
\usepackage{xparse}

\usepackage{booktabs}
\usepackage{tabularx}
\usepackage{tabularray}
\usepackage{colortbl}
\usepackage{cite}

% ...
\usepackage[nameinlink]{cleveref}

\hypersetup{colorlinks, linkcolor=darkblue, filecolor=darkgreen,
urlcolor=darkblue, citecolor=darkblue}

\definecolor{darkred}{rgb}{0.5,0,0}
\definecolor{darkgreen}{rgb}{0,0.5,0}
\definecolor{darkblue}{rgb}{0,0,0.5}

\newcommand{\ket}[1]{\left|{#1}\right\rangle}
\newcommand{\bra}[1]{\left\langle{#1}\right|}

\newcommand{\ketbrad}[1]{\left|{#1}\rangle\!\langle{#1}\right|}
\newcommand{\ketbra}[2]{\left|{#1}\rangle\!\langle{#2}\right|}

\newif\ifinappendix
\inappendixfalse

\let\oldappendix\appendix
\renewcommand{\appendix}{%
  \oldappendix
  \inappendixtrue
}

\newcounter{protocol}
\renewcommand{\theprotocol}{\arabic{protocol}}
\crefname{protocol}{Protocol}{Protocols} 
\tcbuselibrary{skins,breakable}

\newtcolorbox{protocolbox}[2][]{%
  enhanced,
  breakable,
  colback=white,            
  colframe=black,           
  boxrule=0.7pt,
  arc=0mm,
  boxsep=4pt,
  top=2mm,
  bottom=2mm,
  left=2mm,
  right=2mm,
  before skip=10pt,
  after skip=10pt,
  width=\textwidth,         
  halign=center,
  coltitle=black,    
 colbacktitle=white,
  coltitle=black,
  fonttitle=\bfseries,      
  title={\centering \textbf{Protocol \theprotocol: #2}}, 
  overlay unbroken and first={},
  overlay middle={},
  overlay last={},
  #1
}
\usepackage{ragged2e}
\NewDocumentEnvironment{myprotocol}{O{} m +b}{%
  \refstepcounter{protocol}%
  \begin{widetext}
    \begin{protocolbox}[#1]{#2}%
      \begin{flushleft}
      \justifying
      #3
      \end{flushleft}
    \end{protocolbox}
  \end{widetext}
  \addcontentsline{lop}{section}{\protect\numberline{\theprotocol}#2}%
  \ifinappendix
    \onecolumngrid
  \fi
}{}

\newenvironment{widetext}{}{}
\providecommand{\onecolumngrid}{}

\usepackage{tikz}

\definecolor{lime}{HTML}{A6CE39}
\DeclareRobustCommand{\orcidicon}{%
	\begin{tikzpicture}
	\draw[lime, fill=lime] (0,0) 
	circle [radius=0.16] 
	node[white] {{\fontfamily{qag}\selectfont \tiny ID}};
	\draw[white, fill=white] (-0.0625,0.095) 
	circle [radius=0.007];
	\end{tikzpicture}
	\hspace{-2mm}
}

\foreach \x in {A, ..., Z}{%
	\expandafter\xdef\csname orcid\x\endcsname{\noexpand\href{https://orcid.org/\csname orcidauthor\x\endcsname}{\noexpand\orcidicon}}
}

\makeatletter
\def\ps@headings{%
  \def\@oddhead{}%
  \def\@evenhead{}%
  \def\@oddfoot{\thepage}%
  \def\@evenfoot{\thepage}%
}
\makeatother

\date{}
\begin{document}
  \pagestyle{plain}
  \title{QuSquare: Scalable Quality-Oriented Benchmark Suite for Pre-Fault-Tolerant Quantum Devices}

    \author{ David Aguirre$^{1,2,\dagger}$\orcidB{}, Rub\'en Pe\~na$^{2,\dagger,*}$\orcidC{} and Mikel Sanz$^{1,2,3,4}$\orcidD{}}
    
    \affil{$^1$Department of Physical Chemistry, University of the Basque Country UPV/EHU, Apartado 644, 48080 Bilbao, Spain}
    
    \affil{$^2$BCAM - Basque Center for Applied Mathematics, Mazarredo, 14 E48009 Bilbao, Basque Country – Spain}
    
    \affil{$^3$EHU Quantum Center, University of the Basque Country UPV/EHU, Apartado 644, 48080 Bilbao, Spain}
    
    \affil{$^4$IKERBASQUE, Basque Foundation for Science, Plaza Euskadi 5, 48009, Bilbao, Spain}
    
    \affil{$^*$Author to whom any correspondence should be addressed.}
    
     \affil{$^\dagger$These authors contributed equally to this work.}

    \email{rpena@bcamath.org}


  \begin{abstract}
    As quantum technologies continue to advance, the proliferation of hardware architectures with diverse capabilities and limitations has underscored the importance of benchmarking as a tool to compare performance across platforms. Achieving fair, scalable and consistent evaluations is a key open problem in quantum computing, particularly in the pre-fault-tolerant era. To address this challenge, we introduce QuSquare, a quality-oriented benchmark suite whose protocols are designed to satisfy key quality attributes, including relevance, reproducibility, fairness, verifiability, and scalability, for assessing the performance of quantum devices across hardware architectures. QuSquare consists of four benchmark tests that evaluate quantum hardware performance at both the system and application levels: Partial Clifford Randomized, Multipartite Entanglement, Transverse Field Ising Model (TFIM) Hamiltonian Simulation, and Data Re-uploading Quantum Neural Network (QNN). Together, these benchmarks provide a comprehensive assessment of quantum hardware by evaluating both system- and application-level performance through well-defined protocols that enable fair comparisons across different quantum computing technologies and contribute to the development of future benchmarking standards.

  \end{abstract}
    
  \section{Introduction}
The rapid development of quantum computing has led to a proliferation of
hardware architectures, each with different capabilities and limitations. This
raises a key challenge: how the performance of different quantum
computers should be fairly compared. As in classical computing, quantum benchmarks are
essential not only to track technological progress but also to identify the
strengths, limitations, and areas for improvement of hardware platforms. However, designing fair and consistent benchmarks is challenging, since
current quantum devices differ in scale, underlying technology,
connectivity, supported gate sets, and other features. This diversity makes it challenging to establish common standards for evaluation. In the absence of rigorous quantum benchmarks, there is a risk of relying on misleading performance metrics that may distort research priorities, create a false perception of technological progress, complicate the development of technology roadmaps for industry, and lead to biased comparisons between architectures \cite{David}.

Several efforts have been made to evaluate the performance of real quantum
processors across different technologies, including superconducting circuits, trapped-ion, cold atoms, and other emerging platforms \cite{
McCaskey2019QuantumChemistry, mesman2022qpackquantumapproximateoptimization,
Mills_2021, Hamerly_2019, mckay2023benchmarkingquantumprocessorperformance,
Chen_2024, Wright_2019, Jurcevic_2021,  Linke_2017, abdurakhimov2024,
Wagner_2024, galetsky2025photonicprocessorbenchmarkingvariational,
selvam2025quantumcircuitbenchmarkingibm, Mooney_2021, Moses_2023,
Chatterjee_2025, donkers2022qpackscoresquantitativeperformance,
rozanov2025benchmarkingsinglequbitgatesneutral,mesman2024quasquantumapplicationscore}. While these studies provide valuable insights into hardware performance for specific workloads, direct cross-platform comparisons remain challenging due to the diversity of evaluation methodologies and metrics. This limitation has motivated the development of benchmark suites
\cite{supermarq_, Granet:2025vpn,Lubinski_2023,
lubinski2024quantumalgorithmexplorationusing,
Quetschlich2023mqtbench,barbaresco2024bacqapplicationorientedbenchmarks,
ExperimentalEvaluation, Murali2019FullStackRQ,10.1007/978-3-031-90200-0_3,Sawaya_2024,kurowski2023applicationperformancebenchmarksquantum},
which include several workloads designed to assess performance at multiple
levels --- from component level tests that characterize individual hardware elements
to system- and application-level tests that evaluate the whole quantum processor.
Despite these advances, the lack of standardized methodologies for designing
benchmark suites and defining evaluation criteria remains a key challenge for
fair comparison of quantum hardware performance. To address this challenge, recent works have emphasized the need for benchmark suites to satisfy well-defined quality attributes \cite{Proctor2025Benchmarking,amico2023definingstandardstrategiesquantum,David,hashim2024practicalintroductionbenchmarkingcharacterization} and for greater standardization of performance evaluation methodologies \cite{David,lorenz2025systematicbenchmarkingquantumcomputers,9779849,StandrdForQuantumComputingPerformanceIEEE}. Among the most important attributes are scalability, fairness, and reproducibility, requiring benchmarks that can evaluate increasingly large devices, avoid favoring particular technologies, and rely on clear protocols and compilation rules so that quantum computers are assessed on a common basis.

Considering the quality attributes that a quantum benchmark suite should
satisfy, we propose QuSquare, a suite composed of both system-
and application-level tests to assess the performance of pre-fault-tolerant
quantum devices. Taken together, these tests are
designed to ensure that the
suite satisfies the following quality criteria:
scalability, fairness, relevance, verifiability and
reproducibility.

To this end, the QuSquare suite consists of a set of tests designed to evaluate different aspects of quantum hardware performance. The test suite includes four benchmarks: Partial Clifford Randomized, Multipartite Entanglement, Transverse Field Ising Model (TFIM) Hamiltonian Simulation, and Data Re-uploading Quantum Neural Network (QNN). The Partial
Clifford benchmark assesses the accuracy with which a quantum processor
implements a subset of Clifford gates, which is particularly relevant in
contexts such as quantum error correction. The Multipartite Entanglement benchmark evaluates the ability of quantum processors to generate genuine multipartite entanglement, a key resource in quantum computing. The TFIM Hamiltonian Simulation benchmark evaluates the capability of quantum hardware to simulate a quantum many-body dynamics by means of the transverse field Ising model. Finally, the Data Re-uploading QNN benchmark assesses the performance of the hardware in classification problems. To facilitate adoption and reproducibility of benchmark results, we additionally provide an open-source implementation of the QuSquare benchmark suite~\cite{qusquare_repo}.

This article is organized as follows. In Section~\ref{Benchmark_Suite_design}, we describe the design criteria adopted for the QuSquare suite. We discuss the scope and abstraction levels considered, the quality attributes that guided its
design, and the strategies adopted to ensure hardware-agnostic and fair
performance evaluation across different quantum computing platforms. In
Section~\ref{Benchmark_Suite}, we introduce the QuSquare suite and provide detailed descriptions of each benchmark. For each benchmark, we present the protocol for its implementation on a quantum computer, describe the parameters that can be selected by the user and the strategies used to choose them, describe the performance metrics used for quantitative evaluation, specify the reporting requirements, and analyze both the properties of the benchmark and those of the associated metrics. Finally, in Section~\ref{Conclusions}, we present our conclusions and outline directions for future work.

  \section{Benchmark Suite design principles}
\label{Benchmark_Suite_design}
The QuSquare suite has been designed to evaluate the quality of quantum computers by assessing the noise in implementing general circuits and the ability to execute specific applications. Other aspects have not been included because noise level is the most critical characteristic for pre-fault-tolerant devices. Furthermore, evaluating additional characteristics is meaningful only when comparing devices with similar error rates, making noise assessment the first and most important step. The suite was developed following the guidelines proposed by Acuaviva et al.~\cite{David}, which outline principles for creating reliable quantum benchmarks. QuSquare adopts the perspective of a standard cloud user, and its design prioritizes reproducibility and fair comparison by preventing advantages arising from hardware-level access or optimizations unavailable through standard cloud interfaces.

\begin{itemize}
    \item \textbf{Quality attributes}: The development of the QuSquare suite has been guided by key quality attributes that ensure the reliability of the benchmark suite. These include \emph{relevance}, which guarantees that the benchmark suite targets meaningful workloads and captures performance aspects that are significant for quantum computation; \emph{reproducibility}, which is achieved by ensuring that all benchmark definitions, configurations, and execution procedures are clearly and unambiguously specified, ensuring that results can be independently reproduced by any user or provider; \emph{fairness}, which ensures impartial comparison across different technologies and mitigates bias toward specific architectures; and \emph{scalability}, which enables the performance of quantum computers with a large number of qubits to be measured.

\item \textbf{Compilation rules}: Defining compilation rules is crucial to ensure a fair comparison between quantum computers. The QuSquare suite focuses on evaluating the peak performance of each quantum processor under best available compilation settings. The following compilation rules are allowed:
\begin{itemize}
    \item \emph{Classical subroutine simulation} is not allowed, unless explicitly specified by the benchmark. All circuits must be executed on quantum hardware without replacing quantum subroutines with classically simulated equivalents. Examples of prohibited substitutions include: replacing a Clifford subcircuit with a precomputed lookup table of measurement outcomes, or using classical simulation to skip the execution of part of the circuit on hardware.
    \item \emph{Circuit equivalence transformations} are allowed, provided that the equivalent circuit produces the same final quantum state. For instance, approximate circuit transformations are not allowed. Circuit cancellation is allowed, enabling the elimination of subcircuits that have no effect on the final quantum state, such as a unitary operation immediately followed by its inverse. Removing gates that do not affect the final measurements is not allowed. The reason is that each benchmark circuit is designed to certify the implementation of a specific unitary operation: a gate that appears irrelevant to the appended measurement may nonetheless be an integral part of the intended unitary if the circuit is used as a subroutine within a larger algorithm.
    \item \emph{Choosing physical qubits}: Any set of physical qubits can be selected for the execution of the benchmark, provided that the same qubits are used for every measurement or estimation associated with a given performance metric. The user may evaluate multiple different qubit subsets and report the best result obtained, provided that all tested subsets and their corresponding results are also reported, ensuring transparency while capturing peak performance.

    \item \emph{Low-level optimizations}: Low-level optimizations specifically tailored to the benchmark — such as pulse-level tuning — are not allowed. Standard low-level operations applied uniformly by the platform (e.g., dynamical decoupling sequences enabled by default) are permitted, as they are accessible to any cloud user and cannot be disabled.

\end{itemize}
These compilation rules ensure that all benchmarks are compiled under the same conditions, avoiding artificial performance improvements. Both the provider and the user must follow these compilation rules, unless explicitly specified otherwise by the benchmark. This ensures fair and directly comparable benchmark results by preventing advantages arising from hardware-control capabilities unavailable through cloud access.

     \item \textbf{Scope and abstraction levels}: The QuSquare suite is
       designed to operate at different levels of abstraction, covering both
       system-level and application-level benchmarks. At the system-level, the suite evaluates hardware capabilities such as entanglement
       generation and Clifford gate implementation accuracy. At the application level, it assesses the performance of application
       algorithms, such as Hamiltonian simulation and quantum machine learning
       tasks, reflecting the practical computational power of the quantum
       hardware.

     \item \textbf{Reporting requirements}: The final report of each benchmark
       is essential to ensure reproducibility and transparency of the
       benchmark suite. To this end, every benchmark execution must include:
      \begin{itemize}
        \item Information about the hardware device used.
        \item The transpiled quantum circuit executed on the device.
        \item The set of physical qubits employed for each circuit execution.
       \item The calibration data corresponding to the time of each circuit execution, including single-qubit gate error rates, two-qubit gate error rates, readout error rates, qubit coherence times ($T_1$ and $T_2$), and the calibration timestamp. Additional device-specific calibration data may also be reported when available and relevant to the benchmark.
        \item The compilation configuration and settings applied to each circuit.
        \item All selected benchmark parameters and their corresponding values.
      \end{itemize}
      Additional reporting requirements specific to each test will be detailed
      in subsequent sections.

     \item \textbf{Error correction and mitigation rules}: Error mitigation and error correction are not permitted, unless explicitly stated otherwise by a specific benchmark. This reduces ambiguity and enables direct and unambiguous cross-platform comparison by avoiding performance improvements introduced by external error-mitigation or error-correction procedures.

\end{itemize}






  
  \section{Description of the benchmark suite}
\label{Benchmark_Suite}

The QuSquare suite consists of four benchmarks: Partial Clifford Randomized,
Multipartite Entanglement, TFIM Hamiltonian Simulation, and Data Re-uploading QNN. The first two
benchmarks evaluate quantum hardware at the system level, while the latter two evaluate it at the application level.

Each benchmark is organized into three subsections: \textit{Background and
motivation}, which outlines the theoretical foundations and relevance of the
benchmark in the quantum computing field; \textit{Benchmark description}, which
provides stepwise instructions for executing the benchmark on a quantum
processor, describes how benchmark parameters should be selected, explains the strategies employed to execute it, defines the metric, and specifies the reporting requirements to ensure reproducibility and transparency of the benchmark;
and \textit{Benchmark property analysis}, which discusses the properties that
the benchmark satisfies.

\subsection{Partial Clifford Randomized Benchmark}
\label{sec:Binary_Randomized_Benchmark}

\subsubsection{Background and motivation}

The $N$-qubit Clifford group \cite{PhysRevA.57.127}, denoted as $\mathbb{C}_N$,
consists of unitary operations that normalize the $N$-qubit Pauli group
$\mathcal{P}_N$. Formally, this means that any Clifford operation $C \in
\mathbb{C}_N$ maps a Pauli operator $P \in \mathcal{P}_N$ to another Pauli
operator under conjugation:
\begin{equation}
    \forall C \in \mathbb{C}_N, \quad CPC^\dagger \in \mathcal{P}_N, \quad
    \forall P \in \mathcal{P}_N.
\end{equation}

Typical examples of Clifford gates that are not in the Pauli group include the
Hadamard ($H$), the phase gate ($S = \sqrt{Z}$), the controlled-NOT (CNOT), the
SWAP, and the $i$SWAP gates. Notably, the subgroup $\{H, S, \text{CNOT}\}$ is
sufficient to generate the full Clifford group on any number of qubits. The
number of elements in the $N$-qubit Clifford group is given by
\cite{cliffordsize}:
\begin{equation}
    |\mathbb{C}_N| = 2^{N^2+2N} \prod_{j=1}^N (4^j - 1),
\end{equation}
which, despite being finite, grows exponentially with the number of qubits.

The Clifford group is particularly significant due to the Gottesman-Knill
theorem \cite{gottesman1998heisenbergrepresentationquantumcomputers}, which
states that quantum circuits composed solely of Clifford gates and measurements
in the Pauli basis can be efficiently simulated on a classical computer in
polynomial time. Consequently, Clifford circuits alone do not provide a quantum
computational advantage over classical computation. To achieve universal
quantum computation, a gate set must include at least one non-Clifford gate,
such as the $T$ gate ($T = \sqrt{S}$).

Despite their limited computational power, Clifford gates play a fundamental
role in quantum computing. They are essential for:
\begin{itemize}
    \item \textbf{Quantum error correction}: Stabilizer codes rely on Clifford
      operations for encoding, syndrome extraction, and decoding
      \cite{Rengaswamy_2020}.
      
    \item \textbf{Classical shadows}: Classical shadows is a technique for
      estimating many observables from few copies of a quantum state using
      randomized measurements \cite{Huang_2020}. These measurements are
      typically based on unitary transformations drawn from ensembles such as
      local or global Clifford unitaries \cite{Huang_2020,
      west2025realclassicalshadows}, which are widely used due to their
      efficient implementation and favorable theoretical properties.
    
    \item \textbf{Quantum cryptography and key distribution}: Many
      cryptographic protocols, such as entanglement distillation
      \cite{Bennett_1996}, Six-State Protocol \cite{Bru__1998}, and Quantum
      Secret Sharing Scheme \cite{Hillery_1999}, leverage Clifford operations.

    \item \textbf{Quantum benchmarking}: An important property of Clifford
      gates is that they form a unitary 3-design. This property makes them
      particularly useful for benchmarking quantum devices. Techniques such as
      randomized benchmarking \cite{Emerson_2005} use Clifford gates to
      characterize error rates in quantum gate implementations.

    \item \textbf{Quantum circuits}: Although Clifford operations alone are not
      universal, they provide the essential framework upon which arbitrary
      unitaries are constructed when combined with a non-Clifford gate such as
      $T$. In practice, this decomposition is particularly relevant for real
      quantum devices, where Clifford gates are typically implemented with high
      fidelity and used as the fundamental calibrated primitives, while
      non-Clifford gates are synthesized or injected as higher-level resources
      to achieve full universality.
\end{itemize}

The restriction of this benchmark to Clifford gates, rather
  than to a more universal gate set, is a deliberate design choice motivated
  by the need for unbiased and scalable error estimation. Scalable techniques
  for benchmarking non-Clifford operations, such as mirror circuit protocols,
  require appending the inverse of the tested circuit, which itself
  introduces additional error sources that cannot be cleanly separated from
  those of the forward circuit and result in a systematic bias in the
  estimated error rate. Restricting to Clifford gates instead allows us to
  exploit the algebraic structure of the Clifford group to obtain unbiased
  error estimates while preserving scalability. To this end, we focus on the
  family of Randomized Benchmarking (RB) protocols, which are specifically tailored
  to the efficient characterization of Clifford-gate error rates and provide
  the methodological foundation of this benchmark.

Randomized Benchmarking encompasses a broad class of methods that use random circuits of varying depths to quantify error rates in a given gate set. Initially developed in the mid-to-late 2000s
\cite{Emerson_2005, PhysRevA.80.012304,PhysRevA.77.012307}, RB was designed to
overcome key limitations of quantum process tomography \cite{Chuang_1997,
PhysRevLett.78.390}. Over time, numerous distinct RB protocols have emerged,
each with specific applications, strengths, and limitations. Some major
families of RB protocols include: \textit{Standard RB}
\cite{PhysRevLett.106.180504}, \textit{Native Gate RB protocols}
\cite{hashim2024practicalintroductionbenchmarkingcharacterization,
polloreno2023theorydirectrandomizedbenchmarking, PRXQuantum.5.030334,
PhysRevLett.129.150502, Arute_2019}, \textit{RB for General Groups}
\cite{hashim2024practicalintroductionbenchmarkingcharacterization,
PhysRevA.92.060302, helsen2019newclassefficientrandomized},
\textit{Simultaneous RB} \cite{PhysRevLett.109.240504}, \textit{Interleaved RB}
\cite{PhysRevLett.109.080505}, \textit{Cycle Benchmarking} \cite{Erhard_2019},
\textit{Purity Benchmarking protocols}
\cite{Wallman_2015,PhysRevLett.117.260501} and \textit{RB protocols for
Non-Markovian Errors} \cite{Wallman_2016,
hashim2024practicalintroductionbenchmarkingcharacterization}. Despite their
structural similarities, these protocols measure different error
characteristics.\\

In this benchmark, we utilize a slight modification of a recently proposed RB
protocol known as \textit{Binary Randomized Benchmarking}
(BiRB)\cite{PRXQuantum.5.030334}, which is based on Direct Fidelity Estimation
(DFE)~\cite{Flammia_2011} method, and serves for measuring error rates in
Clifford gates. Notably, BiRB falls within the category of \textit{Native Gate
RB} protocols, which are specifically designed to benchmark layers of gates
sampled from a distribution.

The BiRB protocol presents several advantages over other RB protocols, which
motivate its selection for this benchmark:
\begin{itemize}
    \item \textbf{Scalability}: BiRB does not require mirror circuits,
      stabilizer state preparation, or complex stabilizer state measurements,
      unlike many other RB protocols that measure similar error quantities,
      such as Direct RB \cite{polloreno2023theorydirectrandomizedbenchmarking}
      or Clifford Group RB \cite{PhysRevLett.106.180504}. The only additional
      overhead, beyond the implementation of the Clifford gates whose errors
      are to be characterized, consists of the measurement in the Pauli basis,
      which can be performed with a single layer of single-qubit gates. This
      scalability allows us to evaluate Clifford gate error rates on larger
      quantum processors.
    
    \item \textbf{Robustness to State Preparation and Measurement (SPAM)
      Errors}: BiRB is designed to mitigate the influence of SPAM errors,
      thereby allowing an accurate characterization of the errors in the
      implementation of Clifford gates, while minimizing the impact of other
      error sources that could otherwise lead to misleading results.    
\end{itemize}

\subsubsection{Benchmarking description}
In this benchmark, our objective is to evaluate the largest portion of a
Clifford unitary that the quantum computer can implement with high fidelity. To
this end, we estimate the mean error rate of circuit layers corresponding to a
fraction of circuits that realize global random Clifford unitaries, which we
refer to as \emph{$\mu$-fraction random Clifford circuits}, or simply
\emph{$\mu$-fraction circuits}. The central idea is to report the largest
fraction of the Clifford circuit --ideally the entire circuit-- that the device
can implement with fidelity exceeding a given threshold. We now provide a
detailed definition of the $\mu$-fraction random Clifford circuits and describe
the quantity used to characterize their noise on a quantum device.

\textit{$\mu$-fraction random Clifford circuits}.--- A $\mu$-fraction random
Clifford circuit on $N$ qubits is defined by the following construction. First,
a random Clifford unitary is sampled uniformly from the $N$-qubit Clifford
group and transpiled into a circuit implementation of depth $d$, composed of
Clifford gates. Since such implementations are not unique, we typically select
one that is expected to minimize errors on the target device for example, one
with minimum depth or fewer two-qubit gates. We then compute $d_\mu = \lceil
\mu d \rceil$, where $\mu \in (0,1]$ specifies the fraction of the circuit to
be retained. An integer $s$ is chosen uniformly at random from the interval
$[0, d - d_\mu]$, and the $\mu$-fraction random Clifford circuit is defined as
the contiguous subsequence of gates between depths $s$ and $s + d_\mu$ of the
original circuit.

\textit{Estimating the average entanglement infidelity of $\mu$-fraction
circuits}.--- The central quantity that this benchmark targets is $\epsilon_\mu$, which
characterizes the average noise level experienced during the execution of
$\mu$-fraction random Clifford circuits on the device. Under the assumption that
stochastic Pauli errors are the dominant noise source, $\epsilon_\mu$ approximates
the average entanglement infidelity of these circuits:
\begin{equation}
  \varepsilon_\mu = 1- \mathbb{E}_{C}F_e(\mathcal{U}(C), \Phi(C)),
\end{equation}
where $C$ denotes a $\mu$-fraction random Clifford circuit, $\mathcal{U}(C)$
is the superoperator of its ideal (noiseless) implementation, and $\Phi(C)$ is
its actual noisy implementation on the quantum hardware. To extract $\epsilon_\mu$
from experimental data, we consider the average entanglement fidelity over $d$
independently drawn layers of $\mu$-fraction circuits:
\begin{equation}
    \bar{F}_d = \mathbb{E}_{C_d}F_e(\mathcal{U}(C_d), \Phi(C_d)).
\end{equation}
Building on the theoretical framework of the BiRB protocol~\cite{PRXQuantum.5.030334},
this average fidelity decays exponentially with the number of circuit layers $d$
under Pauli noise, as proven in Appendix~\ref{app:Apendix Clifford RB}:
$\bar{F}_d \approx A p_{rc}^d$ for real constants $A$ and $p_{rc}$. The
average error rate of $\mu$-fraction circuit layers is then defined as:
\begin{equation}
    \epsilon_\mu = \frac{4^N-1}{4^N}(1-p_{rc}).
\end{equation}
This particular rescaling of $p_{rc}$ is motivated by the following physical
  interpretation. The \emph{process polarization} $f$ of a quantum channel $\mathcal{E}$ \cite{hashim2024practicalintroductionbenchmarkingcharacterization}, is
related to the entanglement infidelity by the formula
$f(\mathcal{E}) = \frac{4^N}{4^N-1}F_e(\mathcal{E}) - \frac{1}{4^N-1}$,
which equals $1$ for the identity channel and $0$ for a fully depolarizing channel.
The decay parameter $p_{rc}$ equals the process polarization of the effective error map of a
$\mu$-fraction circuit layer, and $\epsilon_\mu$ is then the infidelity of the
corresponding effective depolarizing channel — that is, the average infidelity
per $\mu$-fraction circuit layer. Further details and derivations can be found in
Refs.~\cite{PhysRevX.13.041030, PRXQuantum.5.030334, polloreno2023theorydirectrandomizedbenchmarking}.

It is important to note that when $\mu=1$, we are considering the complete
Clifford group and $\varepsilon_\mu$ measures the \textit{average
gate set infidelity} (AGSI), analogous to Standard RB.

\begin{figure}[ht]
    \centering 
    \includegraphics[width=0.7\columnwidth]{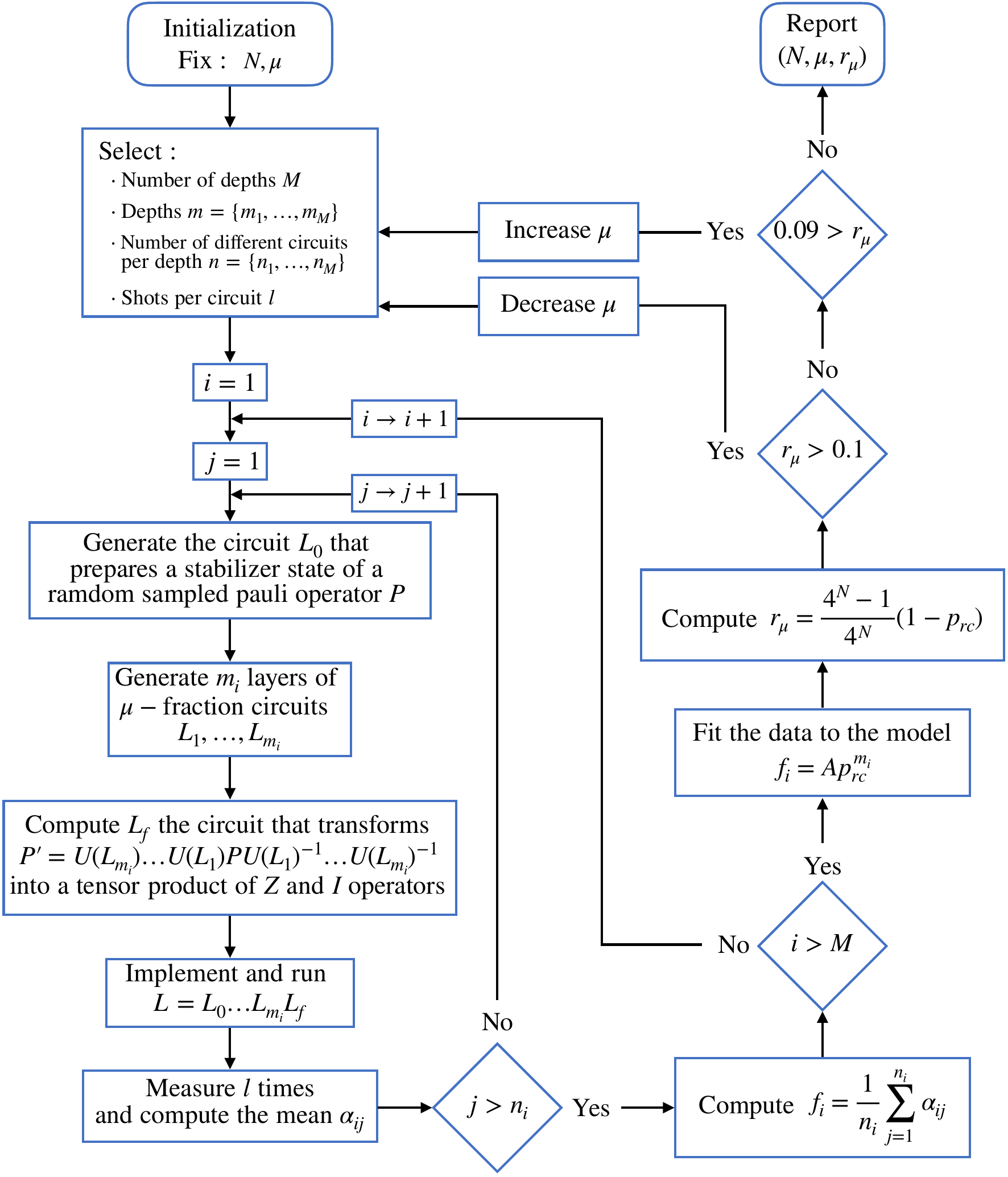}
    \caption{Flowchart of the Partial Clifford Randomized Benchmark. The diagram summarizes the steps described in Protocol \ref{protocol:2}, including parameter initialization, circuit generation and execution, measurement estimation, and the iterative update of $\mu$ based on the estimated infidelity $r_\mu$.}
    \label{fig:clifford}
\end{figure}

\paragraph{Stepwise instructions.}
The detailed description of the benchmark is presented
in the following box, and a flowchart of the procedure is shown in
Figure~\ref{fig:clifford}.\\

\begin{myprotocol}{Partial Clifford Randomized Benchmark protocol}
   \label{protocol:2}
  
    \textbf{Initialization:} Choose the number of qubits to be tested, $N$, and
    the value of $\mu$ based on prior knowledge.

   \noindent\textbf{Return:} The largest value of $\mu$ for which the estimation $r_\mu$
   of the infidelity $\epsilon_\mu$ remains less than or equal to $0.1$.

  \noindent\textbf{Procedure:} 
  \begin{enumerate}
    \item \textbf{Choosing Parameters:} Select \( M \), the maximum number of
      $\mu$-fraction circuits layers (depths), and define a sequence of positive
      integers \( m = \{m_1, \dots, m_M\} \) such that \( m_i < m_j \) for \( i
      < j \), representing the different layer depths of the $\mu$-fraction
      circuits. Additionally, define another sequence of positive integers \( n
      = \{n_1, \dots, n_M\} \), where each \( n_i \) specifies, for the depth
      $m_i$, the number of times different $m_i$ layers of $\mu$-fraction
      circuits are selected. Finally, choose \( l \), the number of times each
      layered circuit is executed.

    \item \textbf{Construct the circuits and get the data:}\\
      For each $i$ from $1$ to $M$:
      \begin{enumerate}
        \item For each $j$ from $1$ to $n_i$:
          \begin{enumerate}

            \item Sample a uniform random $N$-qubit Pauli string $P \in \{\pm
              (I, X, Y, Z)^{\otimes N}\} \setminus \{\pm I^{\otimes N}\}$,
              i.e., a Pauli operator with a global sign, different from the
              identity. Then, generate a uniformly random state $\ket{\psi(P)}$
              from the set of tensor-product stabilizer states stabilized by
              $P$, such that $P \ket{\psi(P)} = \ket{\psi(P)}$. Compute $L_0$,
              a circuit of single qubit gates that prepares $\ket{\psi(P)}$,
              such that $L_0|0\rangle^{\otimes N}=\ket{\psi(P)}$ .
            
            \item\label{step:2aiiClifford} Select $m_i$ Clifford unitaries $U_1,\dots, U_{m_i}$ randomly
              from the $N$-qubits Clifford group $\mathbb{C}_N$. For each
              Clifford unitary, generate the most efficient quantum circuit,
              composed of Clifford gates and optimized for the specific quantum
              processor, obtaining the circuits $C_1, \dots C_{m_i}$, with
              depths $d_1,\dots,d_{m_i}$. For each circuit $C_k$, select uniformly
              randomly a number $s_k$ from the interval $[0, \lfloor d_k(1 -
              \mu)\rfloor]$. The $\mu$-fraction random Clifford circuit $L_k$
              is defined by the contiguous sequence of gates from depth $s_k$ to
              $s_k + \lfloor d_k \mu\rfloor$ within the original circuit $C_k$.\\

            \item\label{step:2aiiiClifford} Compute classically the final layer $L_{f}$ of
              single-qubit gates that transform 
            \begin{equation}
                 P' = U(L_{m_i})\cdots U(L_1)PU(L_1)^{-1}\cdots U(L_{m_i})^{-1}
            \end{equation}
            into a tensor product of $Z$ and $I$ operators, where $U(L)$
            denotes the unitary implemented by circuit $L$.

            \item Prepare the $\ket{0}^{\otimes N}$ state on the processor and construct the circuit by appending the sequence $L = L_0 \cdots L_{m_i} L_f$ to it.

            \item Run the circuit $L$ and measure in the computational basis
              obtaining $l$ eigenvalues of $P'$, that will lie in $\{-1,1\}$.
              We call $\alpha_{ij}$ the mean of these $l$ values obtained.
          \end{enumerate}
        \item\label{step:2bClifford} Compute the mean $f_i = \frac{1}{n_i} \sum_{j=1}^{n_i}\alpha_{ij}$.
      \end{enumerate}
      
    \item \textbf{Fit the data to the model:} Fit the data obtained in previous
      step to an exponential model of the form
        \begin{equation}
          f_i = A p_{rc}^{m_i},
        \end{equation}
      where the parameters $A$ and $p_{rc}$ are determined by performing a
      linear regression on the logarithm of the model.

    \item \textbf{Compute $r_\mu$:} Compute an estimation of $\epsilon_\mu$:
      \begin{equation*}
        r_\mu = \frac{4^N-1}{4^N} (1-p_{rc}).
      \end{equation*}
   \item\label{step:5Clifford} \textbf{Check $r_\mu$ conditions:}  If $r_\mu > 0.1$, select a smaller
     value of $\mu$ and repeat the procedure from Step~2. If $r_\mu \leq 0.1$
     and $\mu = 1$, return $\mu = 1$ together with $r_\mu$. If $r_\mu < 0.1$
     and $0.1 - r_\mu > 0.01$, increase $\mu$ and repeat the procedure from
     Step~2, else, return the current values of $\mu$ and $r_\mu$.  
    \end{enumerate}
  \end{myprotocol}
  
Further details regarding the validity of Steps~2 and~3 can be found in
Ref.~\cite{PRXQuantum.5.030334}.

\paragraph{Selection of parameters and strategies.} The strategies and parameters restrictions are as follows:

\begin{itemize}
    \item  The first step, namely the choice of parameters, is left to the
      user; however, in order to ensure precise and accurate results, we
      provide several recommendations. Since our protocol employs the BiRB
      structure to estimate the infidelity of the circuits, one can refer to
      the original BiRB work for guidance on parameter selection. As stated
      there, it is statistically optimal to maximize the number of random
      circuits while setting the number of shots per circuit to one. In
      practice, however, this approach may be impractical due to the time
      required to generate and load a large number of distinct circuits onto
      the processor. For this reason, it is acceptable to instead use fewer
      circuits with more shots each. Although we do not prescribe a fixed
      number of circuits or shots, further guidance on their selection can be
      found in Appendix~B of \cite{PRXQuantum.5.030334}. In practice, we
      recommend that the product of the number of circuits and the number of
      shots exceed $10^5$ in order to achieve low variance and thus obtain
      precise estimates. Another parameter of interest is the number of layer
      depths to be tested. As established in BiRB work, two depths are
      sufficient, provided that they are separated by $O(\log(1/p))$, where $p$
      is the decay parameter estimated in Step~3 and is related to the
      processor fidelity. Since we are interested in executing circuits with
      relatively high fidelity, this separation is small, i.e., of order
      $O(1)$. We therefore recommend using at least two depths, with their
      values separated by at least three layers.
 
    \item We do not prescribe a specific method for transpiling each Clifford
      unitary before extracting the corresponding portion of the circuit.
      Instead, given a device with a particular topology and native gate set,
      we leave it to the user to determine the most suitable transpilation
      strategy. This includes selecting the qubits on which the algorithm will
      run and optimizing the resulting circuit implementation. The only
      required conditions are as follows:
      
      \begin{enumerate}
        \item For a given $N$ and $\mu$, all circuits of different depths must
          be transpiled and executed on the same set of qubits. 

        \item The transpilation process must be performed independently for
          each layer. This allows for optimizations within the transpilation of
          individual layers, but not across the joint implementation of all
          layers.

        \item Prior to truncation, the transpilation of a unitary into a
          quantum circuit must yield a circuit that exactly realizes the target
          unitary transformation. Moreover, the native gate set into which the
          circuit is transpiled must consist solely of Clifford gates, ensuring
          that the truncated circuit continues to implement a Clifford unitary.
          If this condition cannot be readily fulfilled, an alternative
          approach is to construct an intermediate representation of the
          circuit composed exclusively of Clifford gates, perform the
          truncation within this representation, and subsequently transpile the
          resulting circuit into native gates, while aiming to preserve a
          circuit depth comparable to that obtained by directly truncating the
          native circuit.

        \end{enumerate}
    
    \item The fifth step of the algorithm does not prescribe a specific
      strategy for increasing or decreasing $\mu$ when the conditions for
      returning to Step~2 are met. This choice is left to the user, who should
      explicitly report the strategy employed. We recommend adopting a binary
      search over the variable $\mu$, as it requires only a logarithmic number
      of steps in the desired precision.

    \item To reduce the number of different values of $\mu$ that must be tested
      when running the benchmark with $N$ qubits to achieve an infidelity close
      to $0.1$, it is advantageous to heuristically estimate an appropriate
      initial value of $\mu$. Several strategies can be employed for this
      purpose. For example, one may select a value of $\mu$ close to that
      obtained for $N-1$ qubits, or, if $N$ is sufficiently small, perform
      classical simulations of the device to estimate a suitable value. After
      testing the initial estimate on the target device, the user may then
      refine $\mu$ using any of the adjustment strategies discussed previously. 
\end{itemize}

Table \ref{tab:clifford-parameters} presents a summary of the
parameters, strategy restrictions, and recommendations associated with this
benchmark.


\definecolor{Alto}{rgb}{0.85,0.85,0.85}
\begin{table}
\centering
\begin{tblr}{
  width = \linewidth,
  colspec = {Q[4.3cm,valign=m]Q[500,valign=m]},
  cell{1}{1} = {c},
  cell{1}{2} = {c},
  cell{2}{2} = {c},
  cell{3}{2} = {c},
  cell{4}{2} = {c},
  cell{5}{2} = {c},
  hline{2} = {-}{0.05em},
  hline{1,11} = {-}{0.08em},
  }
 \textbf{Parameters and strategies} & \textbf{Rules}  \\ 
 $M$ & $\ge 2$ \\
$m = \{m_1,\dots,m_M\}$ & $m_{M} - m_{1} \ge 3$\\
$n = \{n_1,\dots,n_M\}$ & $n_i \cdot l \ge 10^{5},\ \forall\, n_i \in n$\\
$l$ & $n_i \cdot l \ge 10^{5},\ \forall\, n_i \in n$\\
Unitary transpilation & Each unitary is independently transpiled into Clifford gates prior to truncation. \\
Total circuit transpilation & The circuit resulting from the concatenation of layers cannot be optimized. \\
Set of qubits & All circuits must be transpiled to target the same set of qubits. \\
Select initial $\mu$ & While the choice is left to the user, leveraging information from the benchmark execution with one fewer qubit or from classical simulations is recommended. \\
Strategy to update $\mu$ & The choice is left to the user; however, a binary search is recommended. \\
\end{tblr}
\caption{Summary of the parameters and strategies for the Partial Clifford Randomized Benchmark. 
The table includes the rules for selecting the parameters $M$, $m=\{m_i\}_{i=1}^M$, 
$n=\{n_i\}_{i=1}^M$, and $l$. Furthermore, it includes the rules for unitary and total 
circuit transpilation, the qubit-selection requirements, and the recommended strategies 
for choosing the initial value of $\mu$ and updating it during the benchmark.}
\label{tab:clifford-parameters}
\end{table}

\paragraph{Metric definition.}\label{sec:metric_definition_clifford}
The metric reported by this benchmark, for a given number of qubits $N$, is the
triple $(N, \mu_{\max}, r_\mu)$, where $0 \leq \mu_{\max} \leq 1$ denotes the
maximum value of $\mu$ for which a $\mu$-circuit can be executed on a processor
with $N$ qubits such that the resulting infidelity $r_\mu$ is less than or
equal to $0.1$, within a tolerance of $0.01$.

\paragraph{Reporting requirements.}
For the Partial Clifford Benchmark, experimental teams or hardware
providers are required to submit a comprehensive report that includes the
elements specified in Section~\ref{Benchmark_Suite_design}, as well as the
following additional items:

\begin{itemize}
  \item The random unitaries selected in Step~\ref{step:2aiiClifford} of
    Protocol~\ref{protocol:2}, prior to transpilation and truncation.

  \item The strategy employed to generate the $\mu$-fraction circuits in Step~\ref{step:2aiiClifford}
    of Protocol~\ref{protocol:2}; that is, the method by which the Clifford unitary
    is translated into a circuit and the procedure used to extract the corresponding
    portion of the circuit.

  \item The values of $f_i$ computed in Step~\ref{step:2bClifford} of Protocol~\ref{protocol:2}
    for each circuit depth $m_i$.

  \item The tuples $(\mu, r_\mu)$ computed each time Protocol~\ref{protocol:2}
    reaches Step~\ref{step:5Clifford}.
\end{itemize}

\subsubsection{Benchmark and metric property analysis}
In this section, we examine which of the benchmark and metric properties defined in \cite{David} are satisfied. We begin by analyzing the benchmark properties, considering the procedure from circuit specification through to the acquisition of measurement results from the quantum computer.

\begin{enumerate}
    \item {\bf Relevance:} As established, Clifford gates play a central role
      in quantum computation, and their reliable implementation constitutes a
      critical step toward the realization of fault-tolerant quantum computing.
      Consequently, since this benchmark overcomes the limitations of existing
      randomized benchmarking protocols in assessing error rates on large-scale
      quantum processors in a fair and architecture-independent manner, we
      consider it to be highly relevant.

    \item {\bf Reproducibility:} If the benchmark is executed on the same
      quantum computer with the same configuration, and a sufficiently large
      number of Clifford circuits and shots per circuit are sampled, the
      reported results will be statistically consistent. Any variations between
      runs will be limited to statistical fluctuations that depend only on the
      number of sampled circuits, the number of implemented layers, and the
      number of shots performed.

    \item {\bf Fairness:} The benchmark is parameterized by several
      user-selectable options, such as the transpilation strategy or the choice
      of gate set used to implement the circuit. This flexibility enables the
      user to obtain the best possible results while avoiding fixed rules that
      apply only to specific devices. In contrast to other benchmarks based on
      RB protocols that focus on circuits built from specific gates, this
      benchmark evaluates the performance of quantum processors on a general
      task --implementing a Clifford unitary, or a portion thereof-- without
      imposing assumptions about the circuit topology or native gate set. These
      choices are left to the user, enabling a fair comparison of devices based
      on their ability to perform this task, independent of their architecture.

    \item {\bf Verifiability:} The benchmark is well defined, with strong
      support from proven theorems established in other benchmarks such as
      BiRB, as well as from the analytical results derived for our class of
      circuits in this section and in the appendices. In addition, we provide
      practical recommendations for selecting the parameters of the benchmark,
      ensuring that the reported results are accurate and reliable.

    \item {\bf Usability:} By relying on Clifford circuits, we ensure that all
      steps requiring classical computations can be performed efficiently in
      polynomial time, with no prohibitively large constants or exponents. For
      the quantum part of the benchmark, the circuits are implemented on a
      user-chosen number of qubits, with their depth determined by one of the
      benchmark parameters. This design makes the protocol straightforward to
      implement on quantum processors of arbitrary size and connectivity.

\end{enumerate}

Using the data obtained from the quantum device, the metric defined in Sec.~\ref{sec:metric_definition_clifford} is computed classically. This metric is analyzed with respect to the following properties, as shown in Table~\ref{T1}:

\begin{enumerate}
    \item \textbf{Practical:} The metric is practical, as computing sample means and fitting the model via linear regression is computationally scalable. Moreover, the number of times we recompute $r_\mu$ scales logarithmically with the desired precision, further ensuring overall efficiency and scalability.
    
    \item \textbf{Repeatable:} The metric is repeatable because the computations performed on the measurement data obtained from the quantum processor are fully deterministic.
    
    \item \textbf{Reliable:} The metric is reliable because a quantum processor with lower noise should be capable of implementing deeper circuits, which in turn implies that it can realize circuits with a larger value of $\mu_{\max}$ compared to a noisier device.
    
    \item \textbf{Consistent:} No hardware-specific bias is introduced, and the metric is computed identically across all architectures. Therefore, the metric is consistent.
\end{enumerate}

\subsection{Multipartite Entanglement Benchmark}

\subsubsection{Background and motivation}
\label{sec:ghz_background}
Entanglement is one of the fundamental resources in quantum mechanics and plays
a central role in quantum computation. Indeed, entanglement has been
demonstrated to be a fundamental resource enabling quantum algorithms to
overcome classical computation~\cite{Entanglement}. The
Greenberger--Horne--Zeilinger (GHZ) state~\cite{Greenberger1989} is an entangled
quantum state that is useful as a resource for a variety of applications, including
quantum computing \cite{Knill2005}, quantum
metrology~\cite{RevModPhys.90.035005}, quantum
teleportation~\cite{QuantumTeleportation}, and other tasks. Due to
technological advances, GHZ states have been generated across various
platforms, including trapped ions~\cite{Moses2023,Monz2011}, superconducting
circuits~\cite{Song2019,PhysRevLett.119.180511, Barends2014SuperconductingQC,
kam2023generationpreservationlargeentangled}, Rydberg atoms~\cite{Omran2019,Graham2022}, photonic
systems~\cite{Zhong2018,PhysRevLett.117.210502}, and silicon-based
architectures~\cite{Takeda2022}.

The $N$-qubit GHZ state is defined as
\begin{equation*}
  |{\rm GHZ_N}\rangle=\frac{1}{\sqrt{2}}\left(\ket{0}^{\otimes  N} +
  \ket{1}^{\otimes N}\right).
\end{equation*}
Several quantum algorithms have been proposed to generate GHZ
states~\cite{Ozaeta2019,Cruz2019, Blatt2008, Moses2023, Mooney2021}; however,
the most efficient circuit for preparing a GHZ state strongly depends on the
characteristics of the quantum processor, its native gate set, qubit
connectivity topology, and other hardware-specific aspects. Once a circuit
designed to prepare the GHZ state is implemented on a given processor, it is
essential to evaluate how accurately the state is prepared. This accuracy can
be quantitatively assessed through the quantum state fidelity.

The quantum state fidelity provides a quantitative measure of how close two
quantum states are. It can therefore be used to assess how accurately a target
quantum state has been prepared. The fidelity of a quantum state, given by its
density matrix $\rho$, with respect to the known GHZ state is defined as
\begin{equation*}
\mathcal{F}=\langle {\rm GHZ_N} |\rho|{\rm GHZ_N}\rangle.
\end{equation*}
The fidelity reaches a value of $1$ when $\rho$ is identical to the pure state
$|{\rm GHZ}_N\rangle \langle {\rm GHZ}_N|$. There are various methods to
estimate the fidelity such as parity
oscillations~\cite{Leibfried2005,Monz2011}, multiple quantum coherences
(MQC)~\cite{PhysRevA.101.032343}, quantum state
tomography~\cite{PhysRevLett.119.180511}, classical shadow
tomography~\cite{Huang2020}, shadow
overlap~\cite{huang2024certifyingquantumstatessinglequbit} or Direct Fidelity
Estimation (DFE)~\cite{Flammia_2011}. Among these, one of the most scalable
approaches to measuring the fidelity of a quantum state is Direct Fidelity
Estimation (DFE), whose complexity --particularly for states such as the GHZ
state-- is independent of the number of qubits. This method involves performing
measurements with respect to Pauli observables that can be efficiently computed
classically. The required Pauli-basis measurements can be efficiently
implemented on a quantum processor by applying a single-qubit Pauli gate to
each qubit, followed by a measurement in the computational basis, ensuring that
the procedure does not increase the circuit depth. Furthermore, the classical
computation of the employed Pauli operators is efficient, as the simulations
involve only Clifford circuits, and the Gottesman–Knill
theorem~\cite{Gottesman1999} guarantees classical efficiency. Given its
advantages for fidelity estimation, Direct Fidelity Estimation is employed in
this benchmark.

\textit{Direct Fidelity Estimation}.--- Direct fidelity estimation is a
technique that allows us to compute the fidelity of an arbitrary quantum state
by performing at most $O(2^N)$ measurements in the Pauli basis. However, when
the states under consideration are stabilizer states, such as the GHZ state,
the number of required measurements depends only on the desired precision of
the fidelity estimation.  The procedure to estimate the fidelity between a
given pure state $\ket{\psi}$ and a classical unknown density matrix state
$\rho$ is described below.

Let $\mathbb{P}_N = \{P_i \mid i = 1, \dots, 2^{2N}\}$ denote the set of all
possible Pauli strings on $N$ qubits, and define $ p_i = \bra{\psi} P_i
\ket{\psi} $, which is computed classically. The goal is to construct an
estimator for the fidelity, denoted by $Y$, that satisfies
\begin{equation}
  \mathrm{Pr}(|Y - \mathcal{F}(\rho, \ketbrad{\psi})| \leq \varepsilon) \geq
  1 - \delta,
  \label{eq:prob_ghz}
\end{equation}
where $\varepsilon$ and $\delta$ denote the maximum estimation error allowed and the confidence level of the estimator, respectively. The number of
measurements required depend on these two parameters. The estimation of the fidelity
is computed using the following procedure:

\begin{enumerate}
  \item Choose $\ell = \lceil 8/(\varepsilon^2 \delta) \rceil$ Pauli strings from
    $\mathbb{P}_N$ randomly, with the probability of selecting equal to
    $p_i^2/2^N$.

  \item For each of the Pauli strings $P_i$ chosen, prepare the state
    $\rho$ on the quantum computer $m_i$ times, where
    \begin{equation*}
    m_i = \left\lceil \frac{8}{p_i^2 \ell \varepsilon^2}
    \log\left(\frac{4}{\delta}\right)\right\rceil.
    \end{equation*}
    Each copy of the state $\rho$ is measured in the basis of the Pauli strings $P_i$,
    obtaining measurement outcomes $A_{ij} \in \{1,-1\}$ for $j = 1, \dots,
    m_i$.

  \item Compute $Y$ with these results
    \begin{equation*}
      Y = \frac{1}{\ell} \sum_{i=1}^{\ell} \left(\frac{1}{m_i
      p_i}\sum_{j=1}^{m_i} A_{ij} \right).
    \end{equation*}
\end{enumerate}
The total number of times the state $\rho$ must be prepared and measured in a
Pauli basis on the quantum computer is $m = \sum_{i=1}^{\ell}m_i$, while the
number of $p_i$ values that need to be computed classically is $\ell$. 

The estimator $Y$ is an unbiased estimator of the fidelity, a
  result that follows from the Pauli decomposition of the target state
  projector~\cite{Flammia_2011}. Any pure state $|\psi\rangle$ satisfies
  $|\psi\rangle\langle\psi| = \frac{1}{2^N}\sum_i p_i P_i$, so
  $\mathcal{F}(\rho, |\psi\rangle\langle\psi|) = \frac{1}{2^N}\sum_i
  p_i\,\mathrm{Tr}(\rho P_i)$. When the state $\ket{\psi}$ is a stabilizer
  state, such as the case of the GHZ state, the $p_i$ are either $0$ or $1$ for
  all $i$. Therefore, there are $2^N$ Pauli strings with selection probability
  $2^{-N}$, while all others have zero probability. Since each measurement
  outcome $A_i$ satisfies $\mathbb{E}[A_i] = \mathrm{Tr}(\rho P_i)$, the
  average $Y$ is an unbiased estimator of the fidelity $\mathcal{F}$. 

  Furthermore, the value of $m$
  becomes independent of the number of qubits, and both $\ell$ and $m$ remain
  significantly smaller \cite{Flammia_2011}. To preserve the bound given in
  Eq.~\eqref{eq:prob_ghz}, the number of Pauli strings to be selected is given
  by $\ell = \left\lceil \frac{8 \log(4 / \delta)}{\varepsilon^2}\right\rceil$,
  and $m_i = 1$. Therefore, the total number of measurements is $m =
  \left\lceil \frac{8 \log(4 / \delta)}{\varepsilon^2}\right\rceil$, which does
  not depend on $N$. See the Ref.~\cite{Flammia_2011} for more details.

\subsubsection{Benchmark description}
The Multipartite Entanglement Benchmark aims to report the largest GHZ state
with genuine multipartite entanglement that can be generated on the quantum
processor. It provides standardized information about the size of the qubit
cluster that can be entangled in a quantum state relevant for various
applications, and which is, in principle, both straightforward to generate and
measure.

\begin{figure}[ht]
    \centering 
     \includegraphics[width=0.6 \columnwidth]{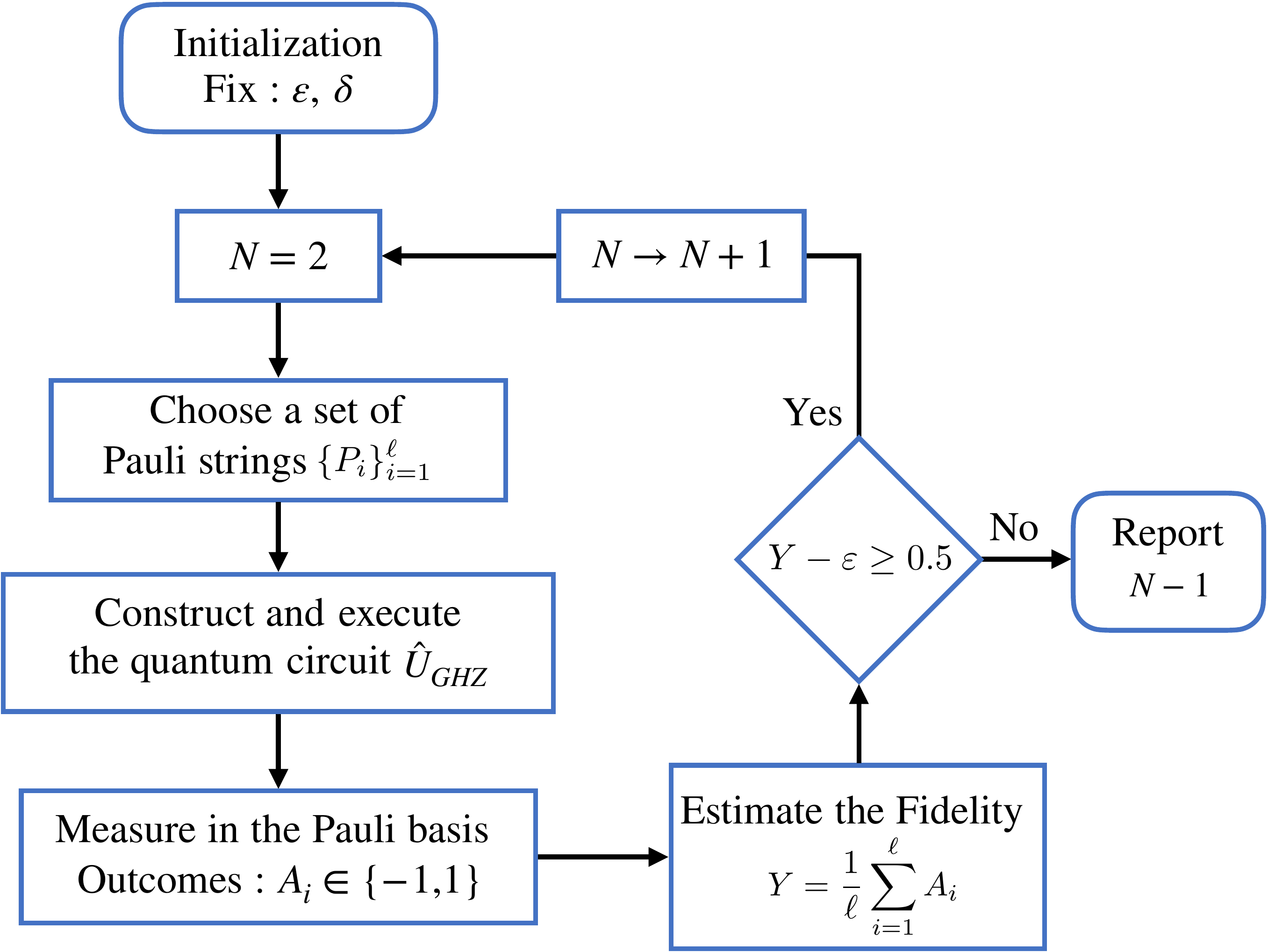}
    \caption{Flowchart of the Multipartite Entanglement Benchmark. The diagram
    summarizes the steps described in Protocol \ref{GHZ_protocol}, including
    the selection of Pauli strings, the construction and execution of the quantum
    circuit, the computation of the fidelity estimator, and the iterative increase
    of the number of qubits $N$.}
    \label{fig:ghz}
\end{figure}

\paragraph{Stepwise instructions}
A detailed description of the benchmark is provided in the following box, and a
flow chart of the procedure is shown in Figure~\ref{fig:ghz}.\\
\begin{myprotocol}
{Multipartite Entanglement Benchmark protocol}
\label{GHZ_protocol}
   \textbf{Initialization:} Set $N=2$, and fix $\varepsilon$ and $\delta$.

   \noindent\textbf{Return:} Maximum number of qubits that can generate the GHZ
   state with fidelity greater than $1/2$.  

  \noindent\textbf{Procedure:} 
  \begin{enumerate}
    \item \textbf{Selection of Pauli strings:} 
    \label{selectionPauli}Since the quantum state \(\ket{\rm GHZ_N}\) is a
    stabilizer state, and it can be expressed as $\ket{\rm GHZ_N} = G_N\ket{0}$,
    where \(G_N\) is a Clifford unitary. Therefore, there exist exactly $2^N$
    Pauli strings $P$ for which the expectation value
      \[
        \bra{\rm GHZ_N} P \ket{\rm GHZ_N} = 1,       
      \] 
      while the expectation value vanishes for all other Pauli strings. The set
      of Pauli strings with nonzero expectation values, from which we sample to
      perform direct fidelity estimation, is given by
       \begin{equation*}
         \mathcal{T}= \left\{G_N P G_N^\dag : P \in \bigotimes_{i=1}^N S_i, \quad S_i \in
          \{I,Z\}\setminus \{ I^{\otimes N}\} \right\} .
      \end{equation*}
      Notice that the identity operator \( I^{\otimes N} \) is excluded to avoid bias in the fidelity estimation. The selection procedure consists of choosing
        \begin{equation*}
        \ell = \left\lceil \frac{8\log(4/\delta)}{\varepsilon^2} \right\rceil
      \end{equation*}
      Pauli strings $\{P_i\}_{i=1}^{\ell}$ uniformly at random from the set
      $\mathcal{T}$. This sampling can be performed efficiently on a classical
      computer, as all the gates involved are Clifford operators.

    \item \textbf{Construct the circuit:} Design the most efficient quantum
      circuit $\mathcal{C}$ that the provider can implement on the target
      processor to prepare the $N$-qubit GHZ state.

    \item \textbf{Quantum data acquisition:} For each $i$ from $1$ to $\ell$,
      execute the quantum circuit $\mathcal{C}$ on the quantum computer and
      perform a single measurement in the basis corresponding to each Pauli
      string $P_i$ selected in Step \ref{selectionPauli}. Each measurement
      yields outcomes $A_{i} \in \{-1,1\}$. If the quantum processor only
      allows measurements in the computational basis, apply an additional layer
      of single-qubit gates to transform the measurement basis accordingly
      before performing the measurement.

    \item\label{step:4GHZ} \textbf{Compute the estimator of the fidelity:} Using the data
      obtained in the previous step, compute the estimator $Y$ of the fidelity
      $\mathcal{F}(\ketbrad{\rm GHZ_N}, \rho)$ between the GHZ state $\ket{\rm
      GHZ_N}$ and the state $\rho$ prepared by the quantum computer during the
      execution of the circuit $\mathcal{C}$, as follows:
      \begin{equation*}
        Y = \frac{1}{\ell} \sum_{i=1}^{\ell} A_{i}.
      \end{equation*}

    \item \textbf{Evaluate fidelity:} If the fidelity estimator minus the
      allowed error, $Y - \varepsilon$, exceeds $1/2$, this implies that, with
      probability $1 - \delta$, the fidelity surpasses $1/2$, and the test for
      $N$ qubits was successful. Repeat the procedure from Step
      \ref{selectionPauli} for successive values of $N$ and report the largest
      $N$ such that the test is successful.
       
    \end{enumerate}
  \end{myprotocol}

\paragraph{Selection of parameters and strategies.} The benchmark requires the
selection of two parameters $\varepsilon$ and $\delta$ and the strategy for updating $N$. Upper bounds for $\varepsilon$ and
$\delta$ are provided to ensure accurate results. Specifically, it is
recommended to set $\varepsilon \leq 0.05$ to maintain an estimation error
below $5\%$, with a confidence level of at least $90\%$ (i.e., $\delta \leq
0.1$). With these settings, fewer than $2 \cdot 10^4$ measurement shots are
required, an amount manageable by most current quantum processors. The strategy
for updating the parameter $N$ is left to the user, although a binary search
procedure is recommended. A summary of the parameters and execution strategies
for this benchmark is provided in Table \ref{tab:ghz-parameters}.


\paragraph{Metric definition.}\label{sec:metric_definition_ghz}
The purpose of this metric is to quantify the maximum number of qubits for
which a quantum computer can generate entanglement in a GHZ state. This
provides an indication of the device’s capability to produce genuine
multipartite entanglement. The metric is defined as the size of the largest GHZ
state that the provider can prepare with high confidence and with fidelity
greater than $1/2$~\cite{Guhne2010,Leibfried2005}.

\paragraph{Reporting requirements.}
For the Multipartite Entanglement Benchmark, experimental teams or hardware
providers are required to submit a comprehensive report that includes the
elements specified in Section~\ref{Benchmark_Suite_design}, as well as the
following additional items:
\begin{itemize}
    \item A detailed description of the quantum circuit used to generate the
      GHZ state.
    \item The resulting fidelity for each qubit number, corresponding to the
      $Y$ value computed in Step~\ref{step:4GHZ} of Protocol~\ref{GHZ_protocol}.
  \end{itemize}


\subsubsection{Benchmark and metric property analysis}
An analysis of the properties satisfied by this benchmark is presented below.

\definecolor{Alto}{rgb}{0.85,0.85,0.85}
\begin{table}
\centering
\begin{tblr}{
  width = \linewidth,
  colspec = {Q[4.3cm,valign=m]Q[500,valign=m]},
  cell{1}{1} = {c},
  cell{1}{2} = {c},
  cell{2}{2} = {c},
  cell{3}{2} = {c},
  hline{2} = {-}{0.05em},
  hline{1,5} = {-}{0.08em},
  }
 \textbf{Parameters and strategies} & \textbf{Rules}  \\ 
  $\delta$ &  $\leq 0.1$ \\ 
 $\varepsilon$ & $\leq 0.05$\\ 
Strategy for updating $N$ & The choice is left to the user; however, a binary
search is recommended.\\    
\end{tblr}
\caption{Summary of the parameters and strategies for the Multipartite
Entanglement Benchmark. The table includes the rules for the confidence
level $\delta$ and the accuracy $\varepsilon$, and a recommended strategy for updating the
number of qubits $N$.}
\label{tab:ghz-parameters}
\end{table}
\begin{enumerate}
    \item {\bf Relevance:} This benchmark is particularly relevant because the
      prepared state corresponds to a widely used quantum state in various
      applications, as discussed in Section~\ref{sec:ghz_background}. Moreover,
      this state is especially relevant in the current era of quantum
      computing, as it is both feasible to generate and practically useful. In
      addition, the benchmark provides valuable insight into the overall
      quality of a quantum processor, offering an estimate of the number of
      functional qubits that can be reliably employed for computation.

    \item {\bf Reproducibility:} This benchmark is reproducible provided that
      the same qubits are selected for the test and the same circuit is used
      across different runs. Minor variations in the reported results are
      expected due to hardware calibration and statistical fluctuations;
      however, the reported metric $N$ across independent executions should not
      differ by more than one.
    
    \item {\bf Fairness:} The benchmark is parameterized by several
      user-defined options, including the specific circuit used to prepare the
      target state, the transpilation strategy, the choice of native gate set,
      and the selection of qubits on which the circuit is executed. This
      flexibility allows users to obtain the best possible performance from
      each device while avoiding rigid rules tailored to particular
      architectures. By leaving these choices to the user, the benchmark
      enables a fair comparison of quantum processors based solely on their
      ability to perform the prescribed task, irrespective of their underlying
      hardware design.

    \item {\bf Verifiability:} The benchmark procedure is well-defined, as it
      is grounded in rigorously established mathematical tools. The lower
      bounds provided for the benchmark parameters ensure confidence in the
      measured results through statistical analysis. Furthermore, the
      methodology is transparent and precisely specified, with each step
      clearly stated to avoid ambiguity or subjective interpretation.

    \item {\bf Usability:} This benchmark is practically feasible because the
      target state to be prepared on the quantum processor is a well-studied
      stabilizer state, for which numerous efficient preparation circuits are
      known and can be readily implemented on most current quantum
      architectures. Moreover, the generation of random Pauli matrices on a
      classical computer is computationally efficient under the specified
      procedure. The number of Pauli operators and measurement shots required
      to estimate the state fidelity scales quadratically with the desired
      precision, allowing the benchmark to achieve low error rates while
      remaining computationally tractable.
\end{enumerate}

Using the data obtained from the quantum device, the metric defined in Sec.~\ref{sec:metric_definition_ghz} is computed classically. This metric is analyzed with respect to the following properties, as shown in Table~\ref{T1}:

\begin{enumerate}
    \item \textbf{Practical:} The metric is practical because computing sample means is computationally scalable. Moreover, the number of times we need to run the benchmark and evaluate the sample mean to calculate the metric grows only linearly with the number of qubits.
    
    \item \textbf{Repeatable:} The metric is repeatable because both the computation of the sample mean and the procedure for determining whether to extend the benchmark to additional qubits are fully deterministic.
    
    \item \textbf{Reliable:} When assessing the largest number of qubits for which a quantum computer can generate entanglement in a GHZ state, this metric is reliable, as it directly measures this quantity. However, this does not imply that the metric is reliable for comparing which device is better at generating entanglement more generally.
    
    \item \textbf{Consistent:} No hardware-specific bias is introduced, and the metric is computed identically across all architectures. Therefore, the metric is consistent.
\end{enumerate}

\subsection{TFIM Hamiltonian Simulation Benchmark}
\label{sec:Hamiltonian_Simulation}
\subsubsection{Background and motivation}
\label{sec:BG}
Hamiltonian simulation is a relevant application of quantum computing \cite{feynman1982simulating}, which aims to reproduce the time evolution of quantum systems governed by a Hamiltonian $\hat{H}$. Specifically, given an initial state $|\psi(0)\rangle$, the objective is to efficiently approximate the evolved state $|\psi(t)\rangle = \hat{U}(t)|\psi(0)\rangle$ using a quantum circuit, where
the time-evolution operator $\hat{U}(t)$ for a time-independent Hamiltonian is described by
$e^{-i\hat{H}t}$. 
This problem has a broad range of scientific and technological domains, including
quantum chemistry
\cite{RevModPhys.92.015003,Cao_2019,Bauer_2020,Ollitrault2021,PhysRevResearch.3.023165},
materials science \cite{Bassman_Oftelie_2021,Bassman_Oftelie_2020,Maskara2025}, condensed matter physics \cite{Fauseweh2024,Tilly2019_many_body_dynamics},
among others. 

Several quantum algorithms have been proposed to simulate
Hamiltonian dynamics  with different accuracy and resource requirements,
including Trotter-Suzuki decomposition \cite{Lloyd1996UniversalQS}, quantum
walk \cite{Qwalk}, linear combination of unitaries (LCU) \cite{2012,Long}, quantum
signal processing (QSP) \cite{Low_2017,Low_2019,Martyn_2021,PhysRevX.6.041067}.
Among these, Quantum Signal Processing (QSP) stands out as a powerful technique, which we employ in this work.

Quantum Signal Processing (QSP) is a powerful quantum algorithm that enables
the implementation of polynomial approximations of matrices using a sequence of
single-qubit rotations controlled by an ancilla qubit \cite{Low_2017,Low_2019,Martyn_2021}. In the
context of Hamiltonian simulation, QSP can be used to approximate the
time-evolution operator $e^{-i\hat{H}t}$ through a polynomial
transformation of the Hamiltonian and has already been implemented on noisy quantum hardware~\cite{Kikuchi2023QSP}. This is achieved by employing the block
encoding technique to embed the Hamiltonian $\hat{H}$ into a larger unitary
matrix
\begin{equation}
\hat{U}_H = \begin{pmatrix}
\hat{H}/\alpha & * \\
* & *
\end{pmatrix},
\end{equation}
where $*$ denotes matrix elements and $\alpha$ is a normalization factor. The block-encoding $\hat{U}_H$ can be constructed using the Linear Combination of Unitaries (LCU) technique. This technique provides a method to implement block-encodings of non-unitary operators, such as Hamiltonians, by expressing them as a linear combination of $K$ unitary operators:
\begin{equation}
\hat{H}= \sum_{i=1}^{K} \alpha_i \hat{U}_i,
\end{equation}
where each $\hat{U}_j$ is a unitary operator acting on the main system qubits and the coefficients $\alpha_j$ are real numbers.  The LCU technique requires $m$ ancillary qubits, where $m=\lceil\log_2{K}\rceil$. To embed the Hamiltonian into a larger unitary matrix, we employ two operators known as $\hat{U}_{\rm Prep}$ and $\hat{U}_{\rm Select}$. 

The operator $\hat{U}_{\rm Prep}$ prepares $m$ ancillary qubits in a superposition state that encodes the coefficients $\alpha_i$ of the Hamiltonian decomposition, while the operator $\hat{U}_{\rm Select}$ applies the unitary operators $\hat{U}_i$ on $s$ qubits conditioned on the state of the ancillary qubits. Here, $s$ denotes the number of qubits needed to represent the Hamiltonian of interest. These operators are defined as follows
\begin{equation}
\hat{U}_{\rm Prep}|0\rangle^{\otimes m}= \sum_i \sqrt{\frac{\alpha_i}{\alpha}}|i\rangle,
\label{Prep}
\end{equation}
where $\alpha=\sum_i|\alpha_i|$. The operator $\hat{U}_{\rm Select}$ is described as follows 
\begin{equation}
\hat{U}_{\rm Select}= \sum_i^K |i\rangle \langle i | \otimes \hat{U}_i.
\end{equation}
Using these two operators, the Hamiltonian can be embedded into a larger unitary matrix as
\begin{equation}
\hat{U}_H = \begin{pmatrix}
\hat{H}/\alpha & * \\
* & * 
\end{pmatrix}
=(\hat{U}_{\rm Prep}^\dagger \otimes \mathbb{I}_{s\times s})\,\hat{U}_{\rm Select}\,(\hat{U}_{\rm Prep} \otimes  \mathbb{I}_{s\times s}).
\label{BlockEncoding}
\end{equation}

Then, an extra ancillary qubit $q_A$ is introduced to implement a sequence of single-qubit rotations that approximate the exponential function. The phase angles of these rotations are determined through classical processing.

\begin{figure}[ht]
    \centering  \includegraphics[width=0.7\columnwidth]{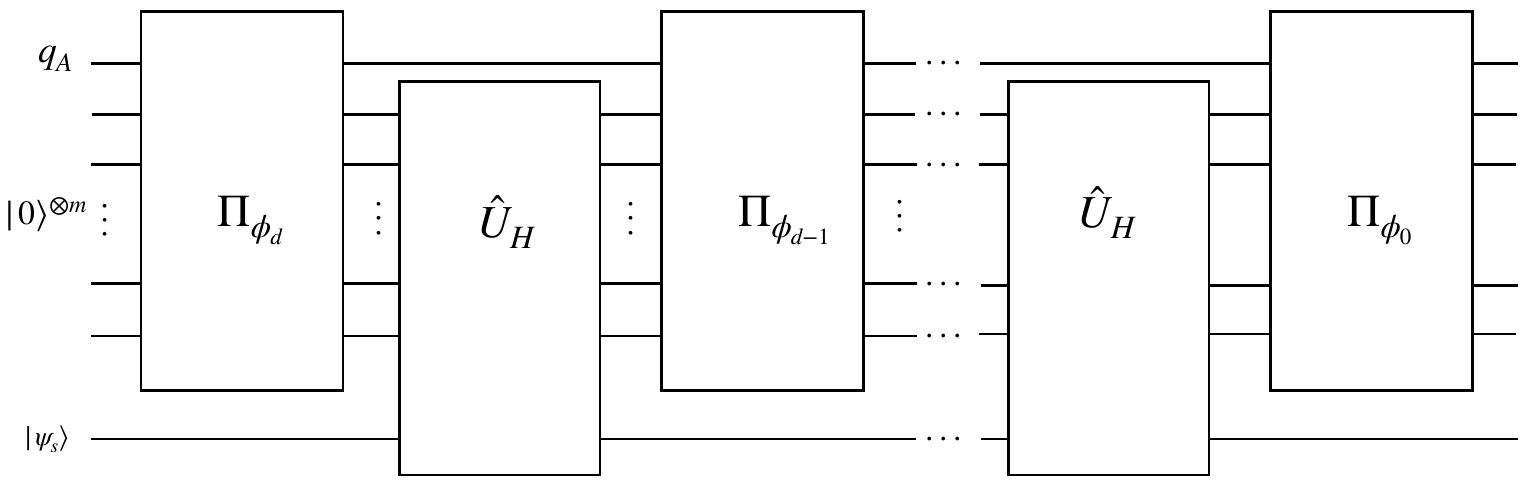}
    \caption{Circuit representation of the QSP sequence described in Eq.~\eqref{QSP}, consisting of alternating  phase-controlled projectors $\Pi_{\phi_k}$ with $k=0,\dots,d$ and the block-encoding unitary $\hat{U}_H$. The circuit uses a single ancilla qubit $q_A$ for controlled operations, $m$ ancillary qubits to implement the block encoding of the Hamiltonian, and the qubits encoding the system state $\ket{\psi_s}$.}
    \label{fig:QSP_general}
\end{figure}

To implement a polynomial function $P$ of degree $d$ on the block encoding defined in Eq.~\ref{BlockEncoding} using QSP, the polynomial must satisfy the following conditions
\begin{enumerate}
    \item \label{cond:parity}~$P$ has parity $d\mod 2$.
    \item \label{cond:norm}~$|P(x)|\leq 1$ for all $x\in [-1,1]$.
\end{enumerate}
Following the QSP formalism of Refs.~\cite{Low_2017,Low_2019,Martyn_2021,PhysRevX.6.041067}, the corresponding QSP operator acting on the block encoding of Eq.~\ref{BlockEncoding} is defined as 
\begin{equation}
  \hat{U}_{\text{QSP}} =\begin{pmatrix}
  P(\hat{H}/\alpha) & * \\
  *& * \\
  \end{pmatrix}= \Pi_{\phi_0}\prod_{j=1}^{d}\hat{U}_H\; \Pi_{\phi_j} ,
  \label{QSP}
\end{equation}
where $\Pi_{\phi_k}=e^{i\phi_k\Pi}$ with $k=0,\dots,d$ are phase-controlled projectors acting on ancillary qubits $q_A$ and $q_m$, $\Pi=2|0\rangle\langle 0| -
\mathbb{I}$ is a reflection, and $\boldsymbol{\phi}=\{\phi_0,\phi_1,... , \phi_d$\} $\in$ $\mathbb{R}$ are phases classically computed such that the upper left block of the unitary
$\hat{U}_{\rm QSP}$ embed the target polynomial $P(\hat{H}/\alpha)$. The circuit implementing the unitary $\hat{U}_{\text{QSP}}$, as defined in Eq.~\ref{QSP}, is shown in Fig.~\ref{fig:QSP_general}.

\begin{figure}[ht]
    \centering  \includegraphics[width=0.5 \columnwidth]{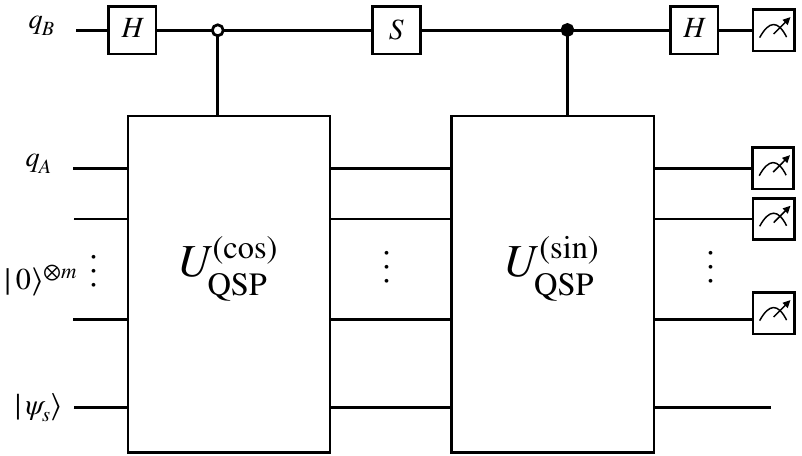}
\caption{Quantum circuit representation of the QSP implementation used to approximate the time-evolution operator $e^{-i\hat{H}t}$. The block $\hat{U}_{\rm QSP}^{(\cos)}$ and $\hat{U}_{\rm QSP}^{(\sin)}$ denotes the quantum circuit used to implement $P_{\cos}^{\epsilon_0}((\alpha t)\hat{H}/\alpha)$ and $P_{\sin}^{\epsilon_0}((\alpha t)\hat{H}/\alpha)$ respectively using quantum signal processing. The evolved quantum state $|\psi(t)\rangle$ is obtained upon post-selection of the ancillary qubit $q_B$ in the state $|1\rangle$, while $q_A$ and all $m$ ancilla qubits in the state $|0\rangle$. }
\label{fig:UTOTAL}
\end{figure}

In the context of Hamiltonian simulation, the evolution operator $e^{i\hat{H}t}$ for a given $t$ is approximated by a degree-$d$ polynomial $P(t\hat{H})$ that satisfies
\begin{equation}
    \|P(t\hat{H}) - e^{-i\hat{H}t} \|_2\leq \epsilon_0,
    \label{eq:errorEpsilon0}
\end{equation}
where $\epsilon_0$ denotes the tolerated error. However, since the exponential function does not possess a definite parity, it does not satisfy Condition~\ref{cond:parity}. To overcome this issue, the exponential can be expressed as
\begin{equation}
     e^{-ixt} = \cos{(xt)} - i\sin{(xt)}
\end{equation}
and two separate polynomial approximations can be used: one even polynomial to approximate $\cos(xt)$ and one odd polynomial to approximate $\sin(xt)$, both of which have definite parity. The approximation of this polynomials can be done through the Jacobi–Anger expansion in terms of Chebyshev polynomials of the first kind \cite{Abramowitz1966}. For $x$ $\in [-1,1]$, this expansion reads as
\begin{equation}
\begin{aligned}
    \cos(xt) &= J_0(t) + 2\sum_{k=1}^{\infty}(-1)^k J_{2k}(t)\,T_{2k}(x), \\
    \sin(xt) &= 2\sum_{k=0}^{\infty}(-1)^k J_{2k+1}(t)\,T_{2k+1}(x),
\end{aligned}
\end{equation}
where $J_k(t)$ denotes the Bessel function of the first kind of order $k$ and $T_k(x)$ denotes the Chebyshev polynomial of the first kind of order $k$. Truncating each expansion at $k=d/2$ yields polynomial approximations $P_{\cos}(xt)$ and $P_{\sin}(xt)$, both with an error $\epsilon$ that depends on $d$. 
To implement QSP with these two polynomials, they must satisfy Condition \ref{cond:norm}. Since $\cos(xt)$ and $\sin(xt)$ are both bounded in magnitude by~1, and $P_{\cos}(xt)$ and $P_{\sin}(xt)$ are $\epsilon$-approximations of these functions, the condition can be ensured by rescaling the polynomials as $P_{\cos}(xt)/(1+\epsilon)$ and $P_{\sin}(xt)/(1+\epsilon)$. This rescaling increases the approximation error of both polynomials by at most a factor of~2.
Taking this into account, in order to achieve a total error $\epsilon_0$ in Eq.~\ref{eq:errorEpsilon0}, we implement the polynomials
\begin{equation}
    P_{\cos}^{\epsilon_0} (xt) := \frac{P_{\cos}(xt)}{1+\epsilon_0/4} \text{ and }  P_{\sin}^{\epsilon_0} (xt) := \frac{P_{\sin}(xt)}{1+\epsilon_0/4},
\end{equation}
of degrees $d$ and $d+1$, respectively, where $d$ is the truncation order given by
\begin{equation}
d = 2\left\lfloor \frac{\log\left(\frac{16}{5\epsilon_0}\right)}{2W\left(\frac{2}{e|t|}\log\left(\frac{16}{5\epsilon_0}\right)\right)} \right\rfloor,
\end{equation}
with $W$ denoting the Lambert function~\cite{10.1145/3313276.3316366, Martyn_2021}.
Notice that the parameter $d$ directly affects the depth of the quantum circuit, as shown in Eq.~\ref{QSP}. To implement the approximation polynomial $P(t\hat{H})$ on a quantum computer, given a block encoding of $\hat{H}/\alpha$, one can realize $P((\alpha t)\hat{H}/\alpha)$ by encoding the real and imaginary components of the evolution operator separately. These correspond to $P_{\cos}^{\epsilon_0}((\alpha t)\hat{H}/\alpha)$ and $-iP_{\sin}^{\epsilon_0}((\alpha t)\hat{H}/\alpha)$, respectively. The two components are then combined using an ancillary qubit that controls the application of the corresponding operators, as illustrated in Fig.~\ref{fig:UTOTAL}.

Having established the general formalism of Quantum Signal Processing for Hamiltonian simulation, we now apply it to a specific and physically relevant model. For the purposes of this benchmark, we focus on a paradigmatic Hamiltonian widely used in condensed matter physics and quantum simulation, the transverse-field Ising model (TFIM).

\textit{Ising Model}.--- The transverse-field Ising model (TFIM) is one of the most fundamental Hamiltonians in condensed matter physics \cite{Pfeuty1970}, which describes $L$ interacting spin-$\tfrac{1}{2}$ particles and is widely used to study magnetism, phase transitions, and critical phenomena. Considering periodic boundary conditions, the Hamiltonian reads as

\begin{equation}
\hat{H}_{\rm TFIM}=\hat{H}_{0} + J\sum_{j=1}^{L}\hat{\sigma}^{x}_j\hat{\sigma}^{x}_{j+1},
\label{TFIM}
\end{equation}
where $\hat{H}_{0}= g\sum^L_{j=1}\hat{\sigma}_j^z$ represents the local energy term and periodic boundary conditions are given by $\hat{\sigma}^{x}_{L+1}=\hat{\sigma}^{x}_{1}$. The transverse magnetic field  and the bare exchange coupling are denoted by $g$, $J$, respectively. The operators $\hat{\sigma}_j^{\alpha}$ with $\alpha=x,z$, denote the Pauli matrices at lattice site $j$. An analytical
solution of the TFIM can be obtained by mapping the model to free fermions via
the Jordan-Wigner and Bogoliubov transformations, as detailed in Ref.
\cite{Farreras_2025}. This solution enables the derivation of an analytical
expression for the expectation value of the total magnetization
$\hat{M}_z(t)=\frac{1}{L}\sum_{i=1}^L\sigma_i^z$, providing an exact reference for benchmarking and
ensuring the scalability of this benchmark. Moreover, due to the periodic
boundary condition, the expectation value of the local magnetization
$\hat{\sigma}^z$ is identical at every site. 

From this analytical reference, the aim of the benchmark is to quantify how accurately the quantum processor reproduces the dynamics of the TFIM Hamiltonian. To this end, we consider all sources of error that contribute to deviations of the total magnetization estimate from its analytical value, excluding hardware-errors. These include statistical sampling error, QSP approximation error, and observable estimation error. The expectation value of the magnetization
is estimated using Hoeffding’s inequality for a given accuracy $\varepsilon$ and confidence level $1-\delta$, which quantify the statistical precision and reliability of the measurement results. The remaining two error sources, the QSP approximation error and the observable estimation error, are discussed in detail in Appendix \ref{sec:ErrorBound}.

\subsubsection{Benchmark description}
The Hamiltonian simulation benchmark aims to assess the capability of the quantum processor to simulate the time evolution by measuring the expectation value of magnetization $M_z(t)$ in the one-dimensional Transverse-Field Ising Model (TFIM) with periodic boundary conditions (PBC).

\begin{figure}[ht]
    \centering 
    \includegraphics[width=0.6 \columnwidth]{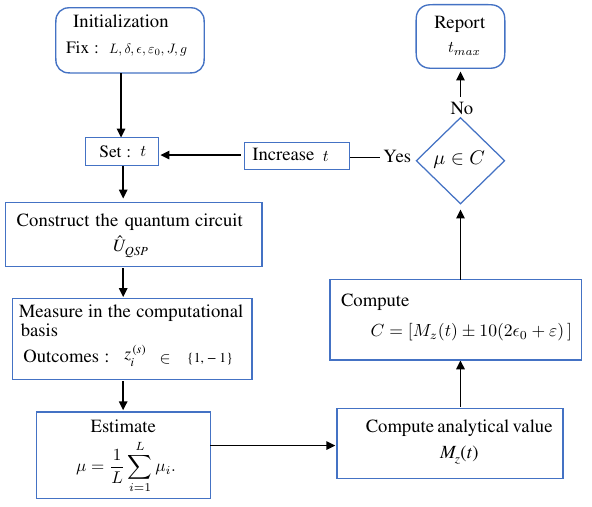}
    \caption{Flowchart of the TFIM Hamiltonian simulation benchmark following 
the steps described in Protocol~\ref{protocol:TFIM_protocol}. 
The diagram illustrates the iterative procedure used to determine the maximum evolution time \(t_{\max}\) for which the estimated total magnetization \(\mu\) remains within the interval 
$C = [\, M_z(t) - 10(2\epsilon_0 + \varepsilon),\; M_z(t) + 10(2\epsilon_0 + \varepsilon)\,]$ }
\label{fig:Hamiltonian}
\end{figure}

\paragraph{Stepwise instructions.}
A detailed description of the benchmark is provided in the following box, and a
flow chart of the procedure is shown in Fig \ref{fig:Hamiltonian}.\\

  \begin{myprotocol}{TFIM Hamiltonian simulation benchmark protocol}
   \label{protocol:TFIM_protocol}
   \textbf{Initialization:} Set the time evolution $t> 0$, choose the number of spins $L\geq 3$,  the tolerance error $\epsilon_0 \leq 7\cdot 10^{-5}$ and fix the parameters $\varepsilon \leq 0.01$ and $ \delta \leq 0.1$.

   \noindent\textbf{Return:} The maximum evolution time $t_{max}$ for which the
   quantum hardware reproduces the global magnetization $M_z(t)$ with a bounded error.

  \textbf{Procedure:} 
  \begin{enumerate}
      \item\label{step:1Hamiltonian} \textbf{Initial condition of the system:} The initial condition of the system is given by
    \[
    |\psi(0)\rangle = \bigotimes_{j=1}^{L} \,|0\rangle_j .
    \]
     \item \textbf{Construct the quantum circuit.} The Fig.
       \ref{fig:QSP_general} shows the quantum circuit that approximates the
       time-evolution operator $e^{-i\hat{H}t}$ using the QSP technique.

    \begin{enumerate}

      \item \textbf{Quantum circuit to implement the operator $\hat{U}_H$.}
        Considering the Transverse Field Ising Hamiltonian
      \[
        \hat{H}_{\rm TFIM} = \sum_{j=1}^L g \hat{Z}_j + \sum_{j=1}^{L} J \hat{X}_j\hat{X}_{j+1},
        = \sum_{k=1}^{2L}\alpha_k \hat{U}_k,\quad \hat{U}_k\in\{\hat{Z}_j,\,\hat{X}_j\hat{X}_{j+1}\}.
      \]
      with $J=g=\frac{1}{L\cdot e}$, so every $\alpha_k = \frac{1}{L\cdot e}$ and $\alpha=\sum_{k=1}^{2L}|\alpha_k| = \frac{2}{e}$.
      Define the vectors
      \begin{equation*}
      \mathbf{P}=
      (\sqrt{\alpha_1/\alpha},\dots, \sqrt{\alpha_{2L}/\alpha})
        =\left(\sqrt{\frac{1}{2L}},\,\ldots,\,\sqrt{\frac{1}{2L}}\right) \quad
        \text{and} \quad \mathbf{U} = (\hat{U}_1,\dots, \hat{U}_{2L}).
    \end{equation*}
    Using the vectors $\mathbf{P}$ and $\mathbf{U}$ construct the quantum
    circuit implementing $\hat{U}_{\rm Prep}$ and $\hat{U}_{\rm Select}$
    following the protocols described in the Appendices \ref{UPREP} and
    \ref{USELECT}, respectively. Then, use these quantum circuits to construct
    the quantum circuit implementing the operator $\hat{U}_H$ as in Eq. \ref{BlockEncoding}.
    
    \item \textbf{Quantum circuit to implement the phase-controlled projector $\Pi_{\phi_{i}}$.}
      Expand the functions $\cos(x t)$ and $\sin(x t)$ in the Chebyshev basis up to order $d/2$:
      \begin{align}
        \cos(xt) \approx  J_0(t) + 2 \sum_{k=1}^{d/2} (-1)^k J_{2k}(t) T_{2k}(x), \quad
        \sin(xt) \approx 2 \sum_{k=0}^{d/2} (-1)^kJ_{2k+1}(t) T_{2k+1}(x),
      \end{align}
      where 
      \begin{equation}
          d = 2\left\lfloor \frac{\log(\frac{16}{5\epsilon_0})}{2W\left(\frac{1}{|t|}\log(\frac{16}{5\epsilon_0})\right)}\right\rfloor.
      \end{equation}
      Store the
      coefficients of the Chebyshev expansion divided by $(1+\epsilon_0/4)$ in the vectors
      $\mathbf{C}^{\cos{}}$ and $\mathbf{C}^{\sin{}}$, ordered from lowest to
      highest degree. Using the above vectors, compute the phase angles
      $\{\phi_k^{\cos}\}_{k=0,\dots, d}$ and $\{\phi_k^{\sin}\}_{k=0,\dots, d+1}$ using the procedure described
      in Appendix \ref{Angles}. Then, using procedure described in Appendix
      \ref{Projector}, construct the quantum circuits to implement the
      phase-controlled projector $\Pi_{\phi}$.
      
    \item \textbf{Quantum circuit to implement QSP.} Construct the quantum
      circuit using Quantum Signal Processing to approximate the functions
      $\cos(\hat{H}t)$ and $\sin(\hat{H}t)$ with polynomials of degree $d$ and $d+1$ respectively
      \[
      U_{\mathrm{QSP}}^{(\cos)}
      = \Pi_{\phi_0} 
      \prod_{k=1}^{d}
      \hat{U}_H \,
      \Pi_{\phi_{k}^{\cos}},
      \quad
      U_{\mathrm{QSP}}^{(\sin)}
      = \Pi_{\phi_0} 
      \prod_{k=1}^{d+1}
      \hat{U}_H \,
      \Pi_{\phi_{k}^{\sin}}.
      \]
      With these two circuits and using an ancillary qubit $q_b$, construct the quantum circuit shown in Fig.~\ref{fig:UTOTAL}.
      
    \end{enumerate}
    \item \textbf{Measurement of total magnetization $\langle M_z\rangle$.}

\begin{enumerate}
  \item \textit{Measurement basis.} All qubits are measured in the computational basis, and only shots with $q_b$ in $\ket{1}$ state, and all other ancilla qubits in $\ket{0}$ state are selected. It can be seen that the probability of finding the ancillary qubits in these states is lower bounded by $\frac{(1 - \epsilon_0)^2}{2}$. The resulting subset of measurements is denoted by $S_{\mathrm{eff}}$.
    \item \textit{Shot setting.} 
    The total number of effective shots is computed using Hoeffding’s inequality for a given $\varepsilon$ and $\delta$:
    \[
       |S_{\mathrm{eff}}| = \frac{2}{\varepsilon^2}\,\ln\frac{2}{\delta}.
    \]

  \item \textit{Estimator.}\label{step:3c} For each shot $s \in S_{\rm eff}$, obtain a bitstring $b^{(s)}=(b^{(s)}_1,\ldots,b^{(s)}_L)\in\{0,1\}^L$ and map each bit to 
    \[
       z_i^{(s)} = (-1)^{\,b^{(s)}_i}\in\{1,-1\},
    \]
    where $z_i^{(s)}$ denotes the sample value for the qubit $i$ in shot $s$. The
    expectation value of the magnetization of each qubit is estimated as
    follows
    \[
       \mu_i=\frac{1}{|S_{\rm{eff}}|}\sum_{s\in S_{\text{eff}}} z_i^{(s)},
    \] where $\mu_i$ is the mean value of the shots. We finally compute the total magnetization of the system, i.e, the mean over all the qubits as

    \begin{equation}
        \mu = \frac{1}{L}\sum_{i=1}^L \mu_i.
    \end{equation}
    
\end{enumerate}
  \item \textbf{Compute analytical value.}
    Compute the analytical expectation value of the total magnetization at time $t$ as follows  
    \begin{equation}
       M_z(t) = 1 - \frac{2}{L} \sum_{m=0}^{L-1} \frac{\Delta_m^{2}}{\epsilon_m^{2} +
       \Delta_m^{2}}\, \sin^{2}(E_m t),
    \end{equation}
    where
    \begin{equation}
      \epsilon_m = g - \cos\left(\frac{\pi (2m+1)}{L}\right),\quad  \Delta_m = \sin\left(\frac{\pi (2m+1)}{L}\right), \quad E_m = 2J
      \sqrt{\epsilon_m^2 + \Delta_m^2}.
    \end{equation}

\item\label{step:4Hamiltonian} \textbf{Evaluate the estimator.}
If the calculated total magnetization $\mu$ lies inside the interval 
$$C=[ M_z(t) -10(2\epsilon_0 + \varepsilon) \, ,\, M_z(t) + 10(2\epsilon_0 + \varepsilon)],$$
where $2\epsilon_0 + \varepsilon$ accounts for the method and measurement errors, respectively, and is independent of noise, see Appendix~\ref{sec:ErrorBound}.

\item \textbf{Increment time.}\label{step:6Hamiltonian}
 If $\epsilon_{t,\min} < 7 \cdot 10^{-5}$, increment $t$; otherwise, decrement $t$, and repeat from Step~\ref{step:1Hamiltonian}. Continue this procedure until a value of $t$ is found such that $\epsilon_{t,\min} < 7 \cdot 10^{-5}$ and $\epsilon_{t+s,\min} \geq 7 \cdot 10^{-5}$, where $s$ is a user-selected small increment satisfying $s < 0.01$.

\end{enumerate}
\end{myprotocol}

\paragraph{Selection of parameters and strategies.}
The strategies and parameters restrictions are as follows:

\begin{itemize}
    \item The choice $L \geq 3$ guarantees a non-trivial TFIM instance under periodic boundary conditions. The total number of qubits required to execute the TFIM Hamiltonian simulation benchmark protocol is
    \[
    N_T = L + \lceil \log_2(2L) \rceil + 2.
    \]

    \item The transverse magnetic field and the bare exchange coupling are fixed $g = J = (Le)^{-1}$. With this choice, the normalization factor is given by $\alpha = 2/e$. As a consequence, the degree of the polynomial approximations $d$ used in QSP depend on integer evolution times, simplifying the analysis and allowing the benchmark to simulate the dynamics by incrementing $t \in \mathbb{N}$. 
    It also simplifies the determination of $t_{\max}$, since we avoid searching over real-valued times that may satisfy the error condition but produce negligible changes in circuit depth.
    
    \item The strategy for selecting and updating $\epsilon_0$ in order to determine $\epsilon_{t,\min}$ for a given evolution time $t$ is left to the user. However, we recommend employing a binary-search procedure, which requires only a logarithmic number of iterations to achieve the desired precision. To further reduce the number of updates of this parameter, one may initialize $\epsilon_0$ using information obtained from executions at other values of $t$, or even from classical simulations when the parameter $L$ is sufficiently small. 
    
    \item The strategy for updating $t$ in Step~\ref{step:6Hamiltonian} is left to the user; however, a binary search is recommended, as it scales logarithmically with the desired precision $s$. The precision $s$, which determines the accuracy in identifying the largest admissible value of $t$, is user-defined and must satisfy $s \leq 0.01$.

    \item By setting $\varepsilon \leq 0.01$ and $\delta \leq 0.1$, we ensure high-precision results and mitigate large fluctuations in the metric arising from statistical variability. With these two parameters, the number of effective shots remains below $6 \cdot 10^4$, which is a reasonable scale for current quantum hardware.

    \item Although we provide complete circuits and a detailed methodology for constructing the final quantum circuit, these should be regarded as recommendations rather than strict requirements. The only mandatory aspect is that the circuit must employ QSP using the constructions shown in Fig.~\ref{fig:UTOTAL} and Fig.~\ref{fig:QSP_general}. The implementations of the block encoding and the phase-controlled projectors may be freely adapted by the user, provided that they perform the same operations as the reference constructions.   
\end{itemize}
A summary of the parameters and execution strategies for this benchmark is provided in Table \ref{tab:hamiltonian-parameters}.

\paragraph{Metric definition}
\label{Metri_Hamiltonian}
The metric of the TFIM Hamiltonian simulation benchmark reports for a selected number of spins $L$, the maximum evolution time, denoted $t_{\mathrm{max}}$. It is defined as the largest evolution time for which the quantum hardware can successfully reproduce the global magnetization $M_z(t)$ of the Ising model with an error not exceeding certain quantity.
Although $t_{\mathrm{max}}$ is the primary metric for comparison, the vector of approximation errors $\epsilon_{t,\min}$ for each evolution time $t \in [1, t_{\mathrm{max}}]$ provides valuable information and can serve as a secondary metric to further characterize the quantum computer’s ability to reproduce the global magnetization $M_z(t)$ of the Ising model.  This metric provides insight into how well the quantum computer can perform Hamiltonian simulation using the QSP technique.

\paragraph{Reporting requirements.}
For the TFIM Hamiltonian simulation benchmark, users or hardware providers are required to submit a comprehensive report that includes the elements specified in Section \ref{Benchmark_Suite_design}, as well as the following additional items
\begin{itemize}
    \item The estimated observables $\hat{\mu}_i$ and its mean $\mu$, of Step \ref{step:3c} for each qubit $i$, time $t$ and $\epsilon_0$.
    \item The strategy used to update $t$ in Step~\ref{step:6Hamiltonian}.
    \item The strategy used to update $\epsilon_0$ in Step~\ref{step:4Hamiltonian}.
    \item Methods and circuits for implementing the block encoding and the phase-controlled projector in QSP.


\end{itemize}

\definecolor{Alto}{rgb}{0.85,0.85,0.85}
\begin{table}
\SetTblrInner{rowsep=1pt} 
\centering
\begin{tblr}{
  width = \linewidth,
  colspec = {Q[4.3cm,valign=m]Q[300,valign=h]},
  cell{1}{1} = {c},
  cell{1}{2} = {c},
  cell{2}{2} = {c},
  cell{3}{2} = {c},
  cell{4}{2} = {c},
  cell{5}{2} = {c},
  cell{6}{2} = {c},
  cell{7}{2} = {c},
  cell{8}{2} = {c},
  cell{9}{2} = {c},
  cell{10}{2} = {c},
  cell{11}{2} = {c},
  cell{12}{2} = {c},
  hline{2} = {-}{0.05em},
  hline{1,16} = {-}{0.08em},
}
\textbf{Parameters and strategies} & \textbf{Rules}  \\
$ L $                              & $\geq 3$ \\
$\delta$                           & $\leq 0.1$ \\
$\varepsilon$                      & $\leq 0.01$ \\
$g$                                & $1$ \\
$J$                                & $1$ \\
$\epsilon_0$                       & $\leq 7\cdot 10^{-5}$ \\
$q_A$                              & $1$  \\
$q_B$                              & $1$ \\
$q_m$                              & $\lceil \log_2{2L} \rceil$ \\
$t$                                & $> 0$ \\
$s$                                & $\leq 0.01$ \\
Strategy to update $t$    & The choice is left to the user; however, a binary search is recommended.\\
Strategy to update $\epsilon_0$    & The choice is left to the user; however, a binary search is recommended.\\
Select initial $\epsilon_0$        & While the choice is left to the user, leveraging information from the benchmark execution with fewer qubits or from classical simulations is recommended. 
\end{tblr}
\caption{Summary of the parameters and strategies for the TFIM Hamiltonian simulation benchmark. The table includes the rules for the system size $L$, the confidence level $\delta$, the accuracy $\varepsilon$, the transverse field $g$ and the exchange coupling $J$ in the TFIM Hamiltonian, the tolerance $\epsilon_0$, the ancillary qubits $q_A$, $q_B$, $q_m$, the evolution time $t$, and the precision parameter $s$. Furthermore, the table includes strategies for selecting and updating $t$ and $\epsilon_0$.}
\label{tab:hamiltonian-parameters}
\end{table}
\subsubsection{Benchmark and metric property analysis}
\label{sec:Hamiltonian_property_analysis}
\begin{enumerate}
 \item {\bf Relevance:}  The Hamiltonian   simulation Benchmark is relevant
   because simulating Hamiltonian dynamics represents one of the most
   fundamental and meaningful workloads in quantum computing. It is directly
   applicable to both scientific and industrial domains, including condensed
   matter physics, quantum chemistry, and materials science. Comparing the measured values with an exact analytical solution of the total magnetization validates the simulated dynamics along this observable, but does not certify the full quantum state. However, this validation provides confidence that the same QSP framework can be extended to probe additional observables of the TFIM. Furthermore, the use of Quantum Signal Processing as the underlying simulation technique makes this benchmark advanced, as QSP is one of the most powerful and
   algorithmically optimal methods for implementing Hamiltonian dynamics on
   quantum devices. 
   
 \item {\bf Reproducibility:} The benchmark is reproducible because the entire procedure can be repeated under identical conditions using the same qubits, circuit construction and parameter settings. Together, these well-defined steps yield consistent estimates of $t_{\max}$. Small deviations may arise from calibration drift or quantum-hardware noise. Consequently, independent repetitions of the benchmark are expected to produce closely aligned outcomes, demonstrating that the procedure is stable and repeatable across multiple runs.
    
 \item {\bf Fairness:} This benchmark is fair because the quantum circuit used to simulate the dynamics of the transverse-field Ising model can be directly transpiled into the native gate set of any quantum computing platform. In addition, users are free to choose their own strategy for updating $\epsilon_0$, thereby reducing potential bias toward specific hardware platforms. The use of Quantum Signal Processing (QSP) ensures that all architectures execute the same workload, independent of their technological details. Moreover, the compilation rules and reporting avoid architecture-specific optimizations, guaranteeing that performance differences arise solely from hardware quality rather than implementation choices.

 \item {\bf Verifiability:} The benchmark is verifiable because every component of the procedure is well defined, from the quantum circuit to implement the QSP technique and observable estimation. The experimental measurement of magnetization is accompanied by explicit statistical guarantees via Hoeffding’s inequality, ensuring controlled and quantifiable uncertainty. Moreover, the reference value for the TFIM magnetization is obtained from an exact analytical solution, enabling a direct comparison between hardware outputs and the theoretical prediction. Together, these features ensure that independent implementations can reliably reproduce and validate the benchmark outcomes

 \item {\bf Usability:} The Hamiltonian simulation benchmark is easy to use because its procedure is fully specified and can be implemented directly on any quantum computing platform. The construction of the block encoding, the generation of QSP phases, and the measurement protocol are classically efficient. In addition, the availability of an exact analytical solution for the magnetization makes the benchmark scalable: the reference value can be computed efficiently for any system size, enabling straightforward comparison between theoretical and experimental results. The design ensures that both experimental teams and hardware providers can apply the benchmark across devices of different sizes and connectivity.
\end{enumerate}

 Using the data obtained from the quantum device, the metric defined in Sec.~\ref{Metri_Hamiltonian} is computed classically. This metric is analyzed with respect to the following properties, as shown in Table~\ref{T1}:

\begin{enumerate}
    \item \textbf{Practical:} The metric is practical,
    as computing sample means from the bitstrings and
    evaluating the analytical solution of the TFIM
    magnetization are both computationally scalable.
    Moreover, the determination of the parameter
    $\epsilon_{t,\min}$ can be performed using an
    efficient strategy, such as a binary search,
    ensuring that the required accuracy is found
    without incurring computational
    overhead.

    \item \textbf{Repeatable:} The metric is
    repeatable, since given the same set of bitstrings
    obtained from the quantum device, the procedure
    used to compute the metric produces the same value.

    \item \textbf{Reliable:} The metric is reliable in the sense that a higher value for a given quantum device indicates its ability to simulate the global magnetization of the Ising model for longer times. However, it is not a reliable measure of the device’s overall capability to perform Hamiltonian simulation.

    \item \textbf{Consistent:} The metric is consistent, as it preserves the same meaning across different hardware platforms.
\end{enumerate}

\subsection{Data Re-uploading QNN Benchmark}
\label{sec:Quantum_machine_learning}
\subsubsection{Background and motivation}
\label{sec:BGQML}
Quantum Machine Learning (QML) is a research field at the intersection of
quantum computing and machine learning that explores algorithms and models
leveraging quantum mechanical principles to perform learning tasks, with the
potential to outperform classical methods \cite{Biamonte_2017, Abbas_2021, PhysRevLett.131.140601}. QML has been applied across a wide range of domains, including image recognition \cite{Senokosov_2024}, computer vision \cite{Henderson2019QuanvolutionalNN}, natural language processing \cite{Li2022QuantumSN}, quantum chemistry \cite{couzinié2025improvedquantummachinelearning},  materials science \cite{vitz2024hybridquantumgraphneural}, finance \cite{mourya2025contextualquantumneuralnetworks} and medical applications \cite{Landman2022quantummethods}. 

Within QML, several model classes have been proposed to address different learning
paradigms, including Quantum Kernel Methods \cite{kubler2021,naguleswaran2024},
Quantum Reinforcement Learning \cite{meyer2024g}, Quantum Generative Models
\cite{islam2025surveyquantumgenerativeadversarial}, and Quantum Neural Networks
(QNNs) \cite{P_rez_Salinas_2020,Zhou_2023}, among others. 
QNNs, in particular, have emerged as one of the most prominent models due to their structural similarity to classical neural networks and their adaptability to the variational hybrid framework suitable for Noisy Intermediate-Scale Quantum (NISQ) devices. QNNs are based on parameterized quantum circuits (PQCs), which are trained through classical or quantum optimization routines to minimize a cost function. However, this training process is often challenging due to the barren plateau phenomenon \cite{McClean_2018, Ragone_2024}. One approach to mitigate this issue is the data re-uploading QNN architecture \cite{P_rez_Salinas_2020}, in which data-encoding unitaries are interleaved with trainable unitaries. This technique is particularly effective because it introduces nonlinearities into the model, enabling it to capture complex patterns in data. Moreover, it has been shown to possess universal approximation capabilities for a single-qubit quantum classifier \cite{P_rez_Salinas_2021}. Owing to these properties, data re-uploading QNNs are strong candidates for near-term quantum applications in classification tasks. For this reason, we adopt a binary classifier as the basis for our benchmark.

The simplest encoding strategy for this architecture, in the single-qubit case, is given by
\begin{equation}
    \mathrm{QNN}_{\boldsymbol{\theta}} (\boldsymbol{x}):= \prod_{l=1}^L U(\boldsymbol{\theta}_l)U(\boldsymbol{x})
    \label{eq:one-qubit-QNN}
\end{equation}
where $L$ denotes the number of layers, $U$ represent generic $\mathrm{SU}(2)$ unitary, $\boldsymbol{x}$ encodes classical input data, and the vector $\boldsymbol{\theta} = \{\boldsymbol{\theta}_1 \dots \boldsymbol{\theta}_L\}$ encompasses the trainable parameters. This model can be used to construct a binary classifier by selecting two label states that are maximally separated in Hilbert space. The training objective is to adjust the parameters so that the model rotates data points from each class toward their corresponding label state, thereby enabling effective classification.

This class of QNNs can be extended beyond the single-qubit setting by employing multi-qubit parameterized circuits with entangling layers, which increase the number of trainable parameters per layer and enhance the expressivity of the model through entanglement \cite{panadero2023}. The multi-qubit extension of the data re-uploading architecture can be written as
\begin{equation}
\text{QNN}_{\boldsymbol{\theta},\boldsymbol{\varphi}}(\boldsymbol{x}) :=
\prod_{l=1}^{L}
\left(
\prod_{s=1}^{N-1} \text{CU}_{s+1}\!(\boldsymbol{\varphi}_l^{(s)})
\left(\bigotimes_{r=1}^N U\!(\boldsymbol{\theta}_l^{(r)})\right)
U(\boldsymbol{x})^{\otimes N}
\right),
\label{eq:multi_qubit_qnn}
\end{equation}
where $L$ denotes the number of layers and $N$ the number of qubits,
$U(\boldsymbol{\theta}_l^{(r)})$ represents \textcolor{blue}{a }single-qubit trainable unitary
applied to qubit $r$ in layer $l$. The unitary
$\text{CU}_{s+1}(\boldsymbol{\varphi}_l^{(s)})$ denotes the controlled version of the general $\mathrm{SU}(2)$ unitary where the
$(s+1)$-th qubit acts as the control and the $s$-th qubit as the target. The parameter sets $\boldsymbol{\theta}$ and $\boldsymbol{\varphi}$ denote the trainable parameters associated with the single-qubit and two-qubit gates, respectively. The total number of trainable parameters in this architecture is $3(2N-1)L$ \cite{xphb-x2g4}. Figure \ref{fig:QNN} shows the quantum circuit representation of a multi-qubit quantum neural network with data re-uploading.

This multi-qubit QNN, when applied to classification tasks and binary classification in particular, can improve upon the performance of its single-qubit counterpart \cite{xphb-x2g4}. The construction and training procedure follows a progressive strategy: starting from the one-qubit QNN, the architecture is expanded by adding qubits incrementally. When a new qubit is introduced, the corresponding entangling gates are initialized as identity operations, and training begins from the optimal parameters obtained in the previous stage. This structured, qubit-by-qubit scaling enables a systematic and hardware-efficient extension of the QNN, while mitigating the onset of barren plateaus.

The training process optimizes the parameters $\boldsymbol{\theta}$ and $\boldsymbol{\varphi}$ through an iterative procedure that minimizes a cost function quantifying the discrepancy between the QNN output and the target labels. Parameter updates are performed by computing the gradient of the cost function with respect to $\boldsymbol{\theta}$ and $\boldsymbol{\varphi}$. These partial derivatives can be efficiently estimated using the parameter-shift rule \cite{Schuld_2019, Wierichs_2022}, a technique that evaluates each derivative by executing the same quantum circuit twice, with the parameter of interest shifted by fixed amounts. This enables gradient-based training without requiring access to the full analytical form of the cost function.

One limitation of these QNN architectures arises from the fact that a general single-qubit unitary can be parametrized by at most three rotation angles. Consequently, each data-encoding unitary can encode at most three features of the input data. The same restriction applies to controlled unitaries, which also allow encoding only up to three features. Therefore, when working with data sets that have more than three features, a classical preprocessing step (e.g., feature selection or dimensionality reduction) becomes necessary to reduce the dimensionality of the input. This ensures that the data can be efficiently mapped into a feature space compatible with the quantum encoding capabilities of the available qubits.

\subsubsection{Benchmark description}

The Data Re-uploading QNN Benchmark aims to evaluate how effectively a quantum computer can perform binary classification tasks, i.e., classify input data into two categories. To this end, the benchmark uses several binary datasets with assigned labels. For each dataset, we construct a multi-qubit data-reuploading QNN as defined in Eq.~\ref{eq:multi_qubit_qnn}, train the network using a subset of the dataset (the training set), and then evaluate its performance on a separate subset (the test set). The test accuracy is computed as the number of correct predictions divided by the total number of test samples. Finally, the benchmark metric is obtained by averaging the test accuracies across all datasets, providing an overall indicator of how well the quantum computer can perform binary classification tasks.
\begin{figure}[t]
    \centering     
    \includegraphics[width= 0.75 \columnwidth]{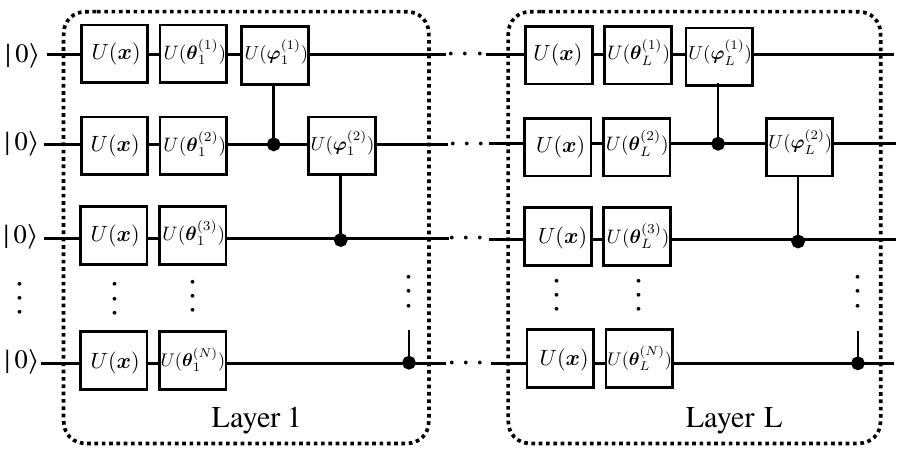}
    \caption{Circuit representation of the multi-qubit quantum neural network with data re-uploading described in Eq.~\eqref{eq:multi_qubit_qnn}. Here, $U(\boldsymbol{x})$ denotes the data-encoding unitary, while $U(\boldsymbol{\theta}_l^{(r)})$ and $U(\boldsymbol{\varphi}_l^{(s)})$ represent the single-qubit trainable unitaries and the controlled trainable unitaries, respectively, with $r = 1,\dots,N$, $s = 1,\dots,N-1$, and $l = 1,\dots,L$.}
    \label{fig:QNN}
\end{figure}

The circuit we employ corresponds to a particular instance of the QNN architecture in Eq.~\ref{eq:multi_qubit_qnn}, where each data-encoding unitary is decomposed as
\begin{equation}
         U(\boldsymbol{x}) = R_z(\pi x_1)\, R_y(\pi x_2)\, R_z(\pi x_3),
         \label{eq:encode_data}
\end{equation}
with $\boldsymbol{x} = \{x_1,x_2,x_3\}$ denoting the input classical data consisting of three features that range from $-1$ to $1$. Similarly, each trainable single-qubit unitary is decomposed as
\begin{equation}
              U(\boldsymbol{\theta}_l^{(r)}) = R_z(\theta_{l,1}^{(r)})\, R_y(\theta_{l,2}^{(r)})\, R_z(\theta_{l,3}^{(r)}),
              \label{eq:encode_parameters}
\end{equation}
where the parameters $\boldsymbol{\theta}_l^{(r)} = \{\theta_{l,1}^{(r)},\theta_{l,2}^{(r)},\theta_{l,3}^{(r)}\}$, which take values in $[-\pi, \pi]$, are optimized during the training process. Finally, the controlled unitaries, where qubit $s+1$ acts as the control and qubit $s$ as the target, are defined as
\begin{multline}
    \text{CU}_{s+1}(\boldsymbol{\varphi}_l^{(s)}) =\\
    |0\rangle\langle
    0|_{s+1} \otimes \mathbb{I}_{s} + |1\rangle\langle 1|_{s+1} \otimes
    U(\varphi_{l,1}^{(s)},\varphi_{l,2}^{(s)},\varphi_{l,3}^{(s)}),
    \label{eq:encode_parameters_multi}
\end{multline}
where the single-qubit rotation applied to the target qubit is given by
\begin{equation}
 \begin{aligned}
 &U(\varphi_{l,1}^{(s)},\varphi_{l,2}^{(s)},\varphi_{l,3}^{(s)})=\\
   &
   \begin{bmatrix}
     e^{-i(\varphi_{l,1}^{(s)}+\varphi_{l,3}^{(s)})/2}\cos(\tfrac{\varphi_{l,2}^{(s)}}{2})
     &
     -e^{i(\varphi_{l,1}^{(s)}-\varphi_{l,3}^{(s)})/2}\sin(\tfrac{\varphi_{l,2}^{(s)}}{2})
     \\[6pt]
   e^{-i(\varphi_{l,1}^{(s)}-\varphi_{l,3}^{(s)})/2}\sin(\tfrac{\varphi_{l,2}^{(s)}}{2})
     &
   \;\;e^{i(\varphi_{l,1}^{(s)}+\varphi_{l,3}^{(s)})/2}\cos(\tfrac{\varphi_{l,2}^{(s)}}{2})
   \end{bmatrix}.
   \end{aligned}
 \end{equation}
The parameters $\boldsymbol{\varphi}_l^{(s)} = \{\varphi_{l,1}^{(s)},\varphi_{l,2}^{(s)},\varphi_{l,3}^{(s)}\}$, which take values in $[-\pi, \pi]$, are optimized during the training process.

To maintain a fair and hardware-focused benchmark, we use synthetic datasets with exactly three features so that no classical dimensionality-reduction techniques are necessary. Additionally, to obtain reliable performance estimates, each dataset is partitioned multiple times into training and test sets using a $70/30$ split, where $70\%$ of the samples are used for training and $30\%$ for testing.

\begin{figure}[ht]
    \centering 
    \includegraphics[width=0.8 \columnwidth]{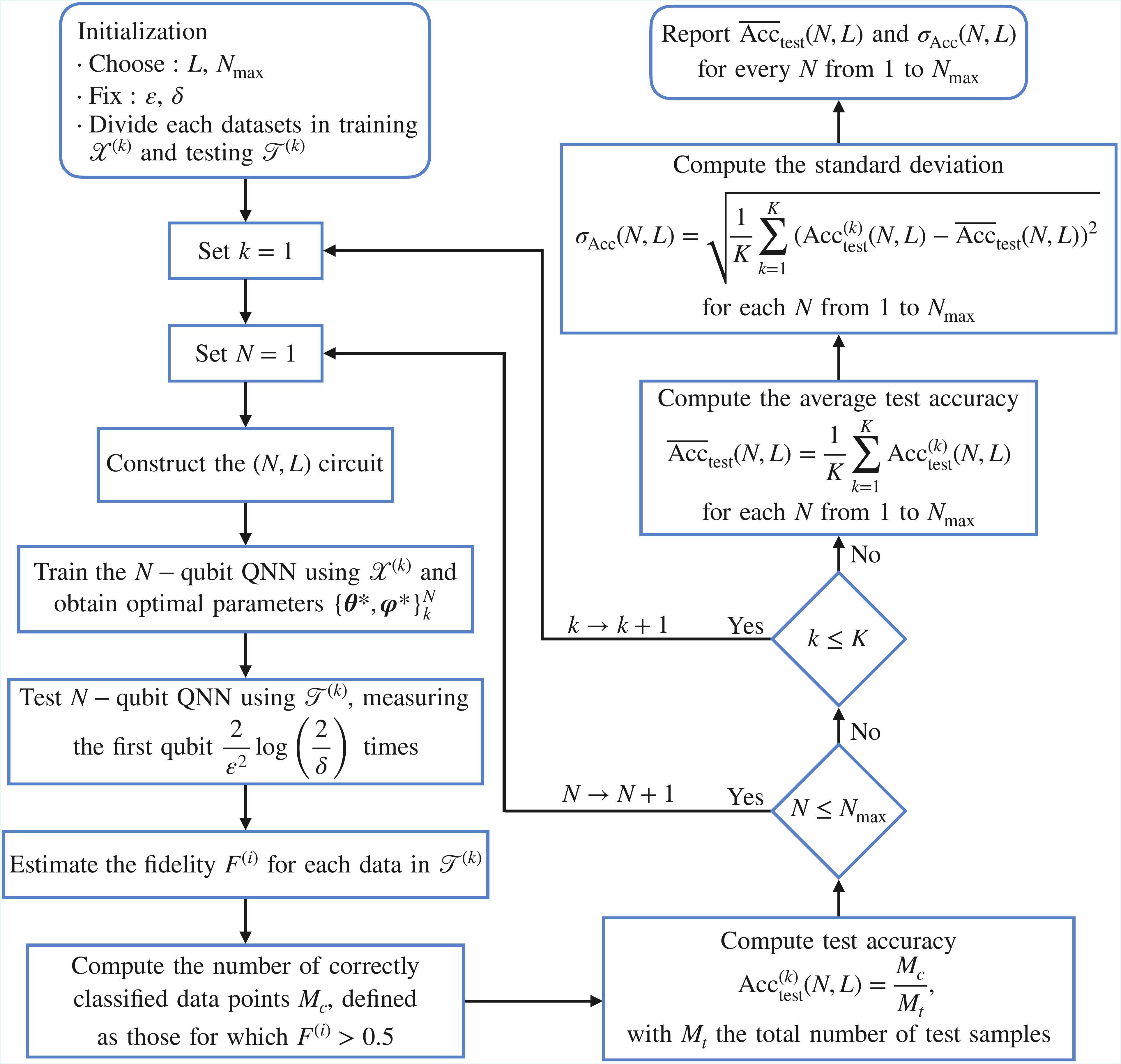}
    \caption{Flowchart of the Data Re-Uploading QNN Benchmark. The diagram summarizes the steps described in Protocol~\ref{protocol:4}, including the construction, training, and testing of the $N$-qubit QNN architecture, the computation of the test accuracy for each dataset, and the evaluation of the average test accuracy $\overline{\mathrm{Acc}}_{\mathrm{test}}(N,L)$ and its sample standard deviation $\sigma_{\mathrm{Acc}}(N,L)$ across all datasets.}
    \label{fig:QNNDiagram}
\end{figure}

\paragraph{Stepwise instructions.}
A detailed description of the benchmark is provided in the following box, and a
flow chart of the procedure is shown in Fig \ref{fig:QNNDiagram}.\\

\begin{myprotocol}{Data Re-Uploading QNN Benchmark Protocol}
    \label{protocol:4}
    \textbf{Initialization:} Choose the maximum number of qubits $N_{\max}$, the layer depth $L$, and the parameters $\varepsilon \leq 0.05$ and $\delta\leq 0.1$, which denote the allowed classification error and the confidence level of the estimation, respectively.
    For each dataset indexed by $k$, the training dataset $\mathcal{X}^{(k)}$ and test
    dataset $\mathcal{T}^{(k)}$ contain features rescaled to the interval $[-1,1]$, which are defined as follows
    \begin{equation}
      \mathcal{X}^{(k)} = \{(\boldsymbol{x}^{(i)},y_x^{(i)})\}_{i=1}^{M_x}, 
      \hspace{1cm}
      \mathcal{T}^{(k)} = \{(\boldsymbol{t}^{(i)},y_t^{(i)})\}_{i=1}^{M_t},
    \end{equation}
    where $M_x$ and $M_t$ represent the number of training and test samples in the $k$-th dataset split, respectively. The vectors $\boldsymbol{x}^{(i)}$ and $\boldsymbol{t}^{(i)}$ denote the $i$-th input samples used for training and testing, and $y_x^{(i)}$ and $y_t^{(i)}$ are their corresponding labels that belong to the set $\{-1,1\}$. \\
    
    \noindent\textbf{Return:} For each $N$-qubit QNN architecture with $N = 1,\dots,N_{\max}$, return the average test accuracy $\overline{\mathrm{Acc}}_{\mathrm{test}}(N,L)$ 
    computed over all dataset partitions, together with its sampled standard deviation $\sigma_{\mathrm{Acc}}(N,L)$.\\

  \noindent\textbf{Procedure:} 
  \begin{enumerate}
    \item \textbf{Training and testing.} For each dataset indexed by $k$:
    
    \begin{enumerate}
      \item For each number of qubits $N$ from $1$ to $N_{\text{max}}$:
      \begin{enumerate}
     
      \item \textbf{Circuit construction.} Construct the quantum circuit to implement the $N$-qubit QNN architecture composed of $L$ layers, see Fig \ref{fig:QNN}. Notice that for $N=1$, the same circuit is implemented, but without the controlled gates.
        
      \item\textbf{Train $N$-qubit QNN architecture.} Obtain the optimal parameters
      $\{\boldsymbol{\theta}^{*},\boldsymbol{\varphi}^{*}\}_k^N$ training the $N$-qubit QNN architecture following the Protocol~\ref{protocol:QML_training_1Q} if $N=1$, or Protocol~\ref{protocol:QML_training_Nq} if $N>1$.
    
      \item \textbf{Test $N$-qubit QNN architecture.}  Evaluate the performance of the trained $N$-qubit QNN architecture on the test dataset $\mathcal{T}^{(k)} = \{(\boldsymbol{t}^{(i)}, y_t^{(i)})\}_{i=1}^{M_t}$ using its corresponding optimal parameters $\{\boldsymbol{\theta}^{*}, \boldsymbol{\varphi}^{*}\}_k^N$. For each input
      $\boldsymbol{t}^{(i)}$, measure the first qubit in the computational basis 
      \begin{equation}
          \frac{2}{\varepsilon^2}\log\left(\frac{2}{\delta}\right),
      \end{equation}
      times and compute the sample mean $\hat{\mu}^{(i)}$ that estimates the expectation value of the Pauli $Z$ operator, as we described in Step \ref{step:Measure} of Protocol \ref{protocol:QML_training_Nq}. Estimate the fidelity with respect to the corresponding label $y_t^{(i)}$ as follows     
      \[
      F^{(i)}(\boldsymbol{t}^{(i)};\theta) 
      = \frac{1+y_t^{(i)}\,\hat{\mu}^{(i)}}{2}, 
      \quad 
      y_t^{(i)} \in \{-1,1\}.
      \]
      A test sample is considered correctly classified if its fidelity satisfies $F^{(i)} \geq 0.5$; otherwise, it is considered incorrectly classified.

    \item \textbf{Compute test accuracy.} The test accuracy is given by
      \[
         \mathrm{Acc}^{(k)}_{\mathrm{test}}(N,L) = \frac{M_c}{M_t},
      \]
      where $M_c$ denotes the number of correctly classified test samples and $M_t$ the total number of test samples.
    
    \end{enumerate}
\end{enumerate}

\item \textbf{Compute the average test accuracy.} For each $N$-qubit QNN architecture for $N$ from 1 to $N_{\max}$, compute the average accuracy by averaging over all
dataset partitions
\begin{equation}
    \overline{\mathrm{Acc}}_{\mathrm{test}}(N,L)
    = \frac{1}{K}\sum_{k=1}^{K} 
      \mathrm{Acc}^{(k)}_{\mathrm{test}}(N,L),
\end{equation}
and its sample standard deviation
\begin{equation}
    \sigma_{\mathrm{Acc}}(N,L)
    = \sqrt{\frac{1}{K}\sum_{k=1}^{K}
      \left(
         \mathrm{Acc}^{(k)}_{\mathrm{test}}(N,L)
         - \overline{\mathrm{Acc}}_{\mathrm{test}}(N,L)
      \right)^2 }.
\end{equation}
where $K$ denotes the total number of datasets, and return both.
\end{enumerate}

\end{myprotocol}

\begin{table*}[t]
\centering
\begin{tabular}{@{}lcccc@{}}
\toprule[1.5pt]
\multicolumn{5}{c}{Benchmarking Metrics Properties Analysis} \\
\midrule[1.25pt]
 Metric & Practical & Repeatable & Reliable & Consistent \\
\midrule[1.25pt]
Partial Clifford  & \checkmark &  \checkmark & \checkmark & \checkmark  \\
Multipartite Entanglement  & \checkmark  &  \checkmark & \checkmark & \checkmark  \\
TFIM Hamiltonian Simulation \hspace{0.5cm} & \checkmark & \checkmark & \checkmark/$\times$ & \checkmark \\
 Data Re-uploading QNN  & \checkmark & \checkmark  & \checkmark/$\times$ & \checkmark \\
\bottomrule[1.5pt]
\end{tabular}
\caption{Summary of the properties satisfied by the metrics used in each benchmark in the QuSquare suite. The table indicates whether each metric is practical, repeatable, reliable, and consistent. The check mark indicates that the metric satisfies the attributes; otherwise, we use the cross mark.} \label{T1}
\end{table*}

\paragraph{Selection of parameters and strategies}
The strategies and parameter restrictions are as follows:
\begin{itemize}

    \item  Setting $\varepsilon \leq 0.05$ and $\delta \leq 0.1$ yields accurate estimates with limited statistical fluctuations. These parameters keep the effective shot count below $3 \cdot 10^3$, which is practical for current quantum hardware.
    
    \item The datasets referenced in the initialization phase of Protocol~\ref{protocol:4} are publicly available at \cite{aguirre2025datasets}. To construct the training and test sets, users should randomly partition the data such that $30\%$ of the samples are assigned to the test set and the remaining $70\%$ to the training set. Despite the random selection, both subsets should maintain an approximately balanced distribution of samples belonging to classes -1 and 1. While the use of these datasets is mandatory, users may additionally incorporate their own datasets, provided that these are clearly described and properly documented.

    \item Before initiating the training of the single-qubit QNN, the user must first choose the initial values of the trainable parameters. The strategy for selecting these initial values is left to the user. Once the initialization is chosen, the QNN is trained as described in Appendix \ref{app:Apendix_QML} to obtain the optimal parameters for the one-qubit model. The user may repeat this procedure multiple times with different initializations in order to search for better optima, provided that all runs are properly reported.

    \item The number of training epochs for each QNN is defined by the user and must be properly reported. For example, training may stop when the gradient of the cost function falls below a chosen threshold, or after a fixed number of epochs selected by the user.

    \item The learning rate parameter $\eta$ used to update the parameters during the training process as described in Appendix \ref{app:Apendix_QML}, may be freely chosen by the user. Moreover, $\eta$ does not need to remain fixed throughout training: the user may adopt any scheduling or adaptive strategy for selecting $\eta$, provided that the chosen approach is reported.
\end{itemize}
A summary of the parameters and execution strategies for this benchmark is provided in Table \ref{tab:QNN-parameters}.

\definecolor{Alto}{rgb}{0.85,0.85,0.85}
\begin{table}
\centering
\begin{tblr}{
  width = \linewidth,
  colspec = {Q[4.3cm,valign=m]Q[500,valign=m]},
  cell{1}{1} = {c},
  cell{1}{2} = {c},
  cell{2}{2} = {c},
  cell{3}{2} = {c},
  hline{2} = {-}{0.05em},
  hline{1,9} = {-}{0.08em},
  }
 \textbf{Parameters and strategies} & \textbf{Rules}  \\ 
         $\delta$ &  $ \leq 0.1$ \\ 
         $\varepsilon$ & $\leq 0.05$\\ 
         $\mathcal{X}^{(k)}$ and $\mathcal{T}^{(k)}$ & The datasets from \cite{aguirre2025datasets} and any user added datasets. \\
         Strategy for one-qubit parameter initialization & Is left to the user.\\
         Number of initialization and training rounds for the one-qubit QNN & Is left to the user.\\
         The strategy for selecting the training epoch & Is left to the user.\\
         The learning rate parameter $\eta$ & Is left to the user. 
\end{tblr}
\caption{Summary of the parameters and strategies for the Data Re-Uploading QNN Benchmark. The table includes the rules for the confidence level $\delta$, the accuracy $\varepsilon$, and the 
datasets $\mathcal{X}^{(k)}$ and $\mathcal{T}^{(k)}$. The strategies for choosing 
the one-qubit parameter initialization, the number of training rounds, the selection of 
the training epoch, and the learning rate parameter $\eta$ are left to the user.}
\label{tab:QNN-parameters}
\end{table}

\paragraph{Metric definition.}\label{sec:metric_definition_qml}
The metric reported by the Data Re-uploading QNN Benchmark for a given number of qubits $N$ and circuit depth $L$ is the average test accuracy over all dataset partitions, denoted $\overline{\text{Acc}}_{\text{test}}(N,L)$, together with its corresponding standard deviation $\sigma_{\text{Acc}}(N,L)$.\textcolor{blue}{ }This metric is intended to quantify the capability of the quantum computer to perform binary classification tasks.

\paragraph{Reporting requirements.}
To ensure reproducibility, experimental groups or providers must report 
sufficient information about their implementation. In particular, they are required to submit a comprehensive report that includes the elements specified in Section \ref{Benchmark_Suite_design}, as well as the following additional items, for each $N$-qubit QNN architecture with $L$ layers:

\begin{itemize}
    \item The test accuracy of every dataset, $\mathrm{Acc}^{(k)}_{\mathrm{test}}(N,L)$.

    \item The optimized parameters for each dataset and number of qubits $\{\boldsymbol{\theta}^{*},\boldsymbol{\varphi}^{*}\}_k^N$.

\end{itemize}
  
\subsubsection{Benchmark and metric property analysis}
\begin{enumerate}
    \item {\bf Relevance:} This benchmark is relevant because it evaluates one of the most widely used QML architectures in the quantum-computing field. These architectures are applied to binary classification, supervised learning, and regression tasks, and serve as a natural reference point for assessing the capability of quantum processors in machine-learning applications.
    In addition, data re-uploading QNNs have well-established universality and expressivity properties, and have been proposed as natural candidates for demonstrating potential advantages in classification tasks. Consequently, the benchmark captures a practically meaningful and technologically significant dimension of quantum-hardware performance.

    \item {\bf Reproducibility:} The benchmark is reproducible because the entire training–testing workflow can be executed under identical conditions using the same qubits, circuit architecture, datasets, and parameter settings. These well-defined steps ensure that the resulting test accuracies are statistically consistent across repeated runs. Although small fluctuations may arise due to both statistical estimation error and hardware effects such as calibration drift or noise, independent executions of the protocol are expected to yield closely aligned accuracy values. This demonstrates that the method remains stable and repeatable across multiple trials within the expected statistical uncertainty.

    \item {\bf Fairness:} The benchmark does not favor any particular quantum-hardware architecture. Providers are free to choose many parameters and strategies such as, the learning rate, the transpilation strategy, the native-gate decomposition, or the initial parameters before training, enabling each platform to achieve its best possible performance. However, the datasets, training/testing methodology, and protocol rules cannot be modified, preventing device-specific optimizations that would introduce bias. This guarantees that the benchmark enables fair comparison across platforms. The linear qubit connectivity required by the QNN circuit architecture is a deliberate design choice to ensure broad applicability across all hardware platforms, including those with restricted qubit topologies. Devices with richer connectivity are not penalized by this choice — they simply do not gain an additional advantage in this particular benchmark. Connectivity between qubits is probed more directly by other benchmarks in the suite, such as the Partial Clifford Randomized and TFIM Hamiltonian Simulation benchmarks.

    \item {\bf Verifiability:} The benchmark is verifiable because the quantum circuit of each $N$-qubit QNN architecture is well defined. Moreover, the reported test accuracy depends solely on single-qubit Pauli-$Z$ measurements, whose statistical uncertainty is characterized by $\varepsilon$ and $\delta$.

    \item {\bf Usability:} The benchmark is practical to implement on current quantum devices. It requires only multi-qubit parameterized quantum circuits with standard single-qubit rotations and controlled gates, three-feature datasets that avoid high-dimensional encodings, and gradient estimation via the parameter-shift rule, which is widely supported on existing platforms. Because the number of qubits $N$ and the circuit depth $L$ are user-configurable, the benchmark scales to devices of different sizes. This makes the benchmark suitable both for small NISQ processors and for larger systems, as increasing $N$ and $L$ adds layers of the same circuit structure. Moreover, the total number of trainable parameters grows linearly with the number of qubits, $3(2N-1)L$, which makes the training overhead manageable even as the architecture increases in size. Consequently, the benchmark is easy to implement and scalable, allowing evaluation of hardware performance as system size grows.
\end{enumerate}

Using the data obtained from the quantum device, the metric defined in Sec.~\ref{sec:metric_definition_qml} is computed classically. This metric is analyzed with respect to the following properties, as shown in Table~\ref{T1}:

\begin{enumerate}
    \item \textbf{Practical:} The metric is practical because all the steps required for its evaluation, such as computing the fidelity and the test accuracy, can be carried out efficiently.
    
    \item \textbf{Repeatable:} The metric is repeatable because all computations are deterministic, provided that the same measurement data are obtained from the device.
    
    \item \textbf{Reliable:} The metric is reliable for the task of binary classification when a sufficient number of initial points are used to mitigate convergence to local minima. However, it is not reliable for determining which device is less noisy for this type of circuit, since noise can, in some cases, facilitate the training of the QNN \cite{bagaev2025regularizingquantumlosslandscapes, somogyi2024methodnoiseinducedregularizationquantum}.
    
    \item \textbf{Consistent:} No hardware-specific bias is introduced, and the metric is computed identically across all architectures; therefore, the metric is consistent.
\end{enumerate}

  \section{Conclusion \& outlook}
\label{Conclusions}
We have introduced the QuSquare benchmark suite designed to provide a rigorous and well-defined methodology for evaluating the performance of pre-fault-tolerant quantum devices. The suite encompasses both system- and application-level benchmarks, such as Partial Clifford Randomized, Multipartite Entanglement, TFIM Hamiltonian Simulation, and Data Re-uploading Quantum Neural Network. At the system level, the Partial Clifford Randomized benchmark stands out for its scalability and independence from the native gate set, while the Multipartite Entanglement benchmark assesses the ability of quantum processors to generate genuine multipartite entanglement and provides an efficient procedure for its estimation. At the application level, the TFIM Hamiltonian Simulation benchmark employs quantum signal processing, a cutting-edge approach for simulating quantum many-body dynamics. The Data Re-Uploading QNN benchmark offers a scalable and adaptable model that can be efficiently mapped onto different quantum hardware architectures and provides provable performance in classification tasks.

QuSquare has been designed to satisfy key quality attributes such as scalability, fairness, relevance, verifiability, and reproducibility. Subsequently, the quality attributes of the associated metrics have also been analyzed and discussed. 

Scalability is ensured by design across all benchmarks through the use of protocols and metrics whose computational requirements grow at most polynomially with the relevant benchmark parameters. At the system level, this includes benchmarks based on quantum circuits that are efficiently simulatable on classical hardware, as well as fidelity estimation procedures that scale efficiently with system size. At the application level, scalability is supported by an analytical reference solution and an efficiently trainable model, enabling meaningful performance evaluation as quantum processors grow.

Fairness is promoted by controlling the influence of hardware-specific optimizations and by reducing ambiguities in benchmark protocols. While certain parameters and selection strategies are {\it intentionally} left flexible to avoid bias between quantum hardware, clear compilation/transpilation rules and reporting requirements are enforced to prevent artificial shortcuts that could otherwise lead to overestimated performance. Furthermore, QuSquare generally specifies benchmarks at the algorithmic level while allowing flexibility in their implementation at the specific gate-decomposition level, provided that the algorithmic structure is preserved. This balance aims to support fair and meaningful comparisons across quantum computing platforms.

The relevance of QuSquare is reflected in the choice of benchmarks that target both fundamental resources and novel promising applications in quantum computing. At the system level, the benchmarks are designed to evaluate the quantum processor through resources that play a crucial role in quantum computing, such as entanglement generation and the implementation of Clifford gate operations. At the application level, the benchmarks assess the capabilities of quantum devices through applications that are promising for quantum computing, including the simulation of quantum many-body dynamics and the use of quantum processors for classification tasks.

Finally, verifiability and reproducibility are ensured through clearly defined reporting requirements that specify the data, the results of the metrics, and execution details to be provided for each benchmark. By enforcing standardized reporting practices, QuSquare enables benchmark results to be independently reproduced and verified by users and providers. In addition, to facilitate consistent and reproducible execution of the benchmark suite and to reduce the risk of human error in protocol implementation, a reference open-source implementation of all four benchmarks is publicly available~\cite{qusquare_repo}. The implementation is modular and parameterizable within the bounds defined by each benchmark protocol.

Overall, QuSquare establishes a rigorous and hardware-agnostic framework for the evaluation of pre-fault-tolerant quantum processors. It addresses key limitations of existing benchmarking approaches, which are often based on a single test and ambiguously defined protocols. Looking ahead, this framework provides a core to which additional benchmarks may be incorporated in the future, extending the range of hardware capabilities that can be assessed. For instance, the inclusion of a non-Clifford benchmark targeting the $T$ gate is a natural direction for future extension. Furthermore, future work will also focus on the experimental application and validation of QuSquare across real quantum computing platforms and hardware technologies.

  \section{ACKNOWLEDGMENT}
The authors thank R. Ibarrondo, X. Gutierrez de la Cal and C. Cristiano Romero for useful discussions. We acknowledge financial support from OpenSuperQ+100 (Grant No. 101113946) of the EU Flagship on Quantum Technologies, from Project Grant No. PID2024-156808NB-I00 and Spanish Ramón y Cajal Grant No. RYC-2020-030503-I funded by MICIU/AEI/10.13039/501100011033 and by “ERDF A way of making Europe” and “ERDF Invest in your Future”, from the Spanish Ministry for Digital Transformation and of Civil Service of the Spanish Government through the QUANTUM ENIA project call-Quantum Spain, and by the EU through the Recovery, Transformation and Resilience Plan--NextGenerationEU within the framework of the Digital Spain 2026 Agenda, and from Basque Government through Grant No. IT1470-22, and the Elkartek project KUBIBIT - kuantikaren berrikuntzarako ibilbide teknologikoak (ELKARTEK25/79). Grant FPU23/02196 funded by MCIN/AEI/ 10.13039/501100011033 and, as appropriate, by ``ESF Investing in your future'' or by ``European Union NextGenerationEU/PRTR''. This research is supported by the Basque Government through the BERC 2022-2025 program and by the Ministry of Science and Innovation: BCAM Severo Ochoa accreditation CEX2021-001142-S / MICIN / AEI / 10.13039/501100011033.
  \bibliographystyle{iopart-num}

  \bibliography{bibliography}

  \appendix

  \section{Partial Clifford Randomized Benchmark is reliable if the probability of error cancellation
is negligible}
\label{app:Apendix Clifford RB}

We follow the same ideas as in section IV of Ref. \cite{polloreno2023theorydirectrandomizedbenchmarking}. For this analysis, we make two
assumptions. First, we neglect state-preparation and measurement (SPAM) errors,
and focus exclusively on the errors introduced by the circuit implementation.
Second, we assume that stochastic Pauli errors are dominant. In particular, we
assume that after the application of one layer $L$ of a $\mu$-fraction random
Clifford circuit, a Pauli error may occur with some probability. Formally, let
$\{\epsilon_{L,P}\}_{P \in \mathcal{P}_N}$ denote the probability distribution
over all Pauli strings $P \in \mathcal{P}_N$ acting on $N$ qubits, where
$\epsilon_{L,P}$ is the probability that $P$ occurs after the application of
layer $L$. For fixed $L$, this distribution satisfies
\begin{equation}
    \sum_{P \in \mathcal{P}_N} \epsilon_{L,P} = 1.
\end{equation}
The probability that no error occurs after the application of layer $L$ is
given by $\epsilon_{L,I}$. Consequently, the probability of an error is
\begin{equation}
    \sum_{P \in \mathcal{P}_N, \, P \neq I} \epsilon_{L,P} = \epsilon_L.
\end{equation}
It is straightforward to show that $\epsilon_L$ corresponds to the entanglement
infidelity between the ideal and noisy implementations of $U(L)$ under this
error model. Specifically,
\begin{equation}
  \epsilon_L = 1 - F_e(\mathcal{U}(L), \Phi(L)) 
= 1 - \frac{1}{4^N} \mathrm{Tr}\!\left(\mathcal{U}(L)^\dagger \, \Phi(L)\right),
\end{equation}
where $\mathcal{U}(L)$ is the superoperator representing the ideal
implementation of $U(L)$, and $\Phi(L)$ is the noisy superoperator
incorporating the Pauli errors. We define the mean entanglement infidelity as
\begin{equation}
    \epsilon_\mu = \mathbb{E}_{L \in C_\mu}(\epsilon_L),
\end{equation}
where $C_\mu$ denotes the set of all $\mu$-fraction circuits for a fixed $\mu$. 

\vspace{1em}

Consider the $n$-qubit circuit defined in Protocol \ref{protocol:2}. Assuming
perfect state preparation and measurement, and a total of $d$ layers, the outcome
under the stochastic Pauli error model is given by
\begin{align}
  B_d &= \mathrm{Tr}\!\left( U(L_1 \cdots L_d) P U(L_1 \cdots L_d)^\dagger \, 
  \tilde{U}(L_1 \cdots L_d) \ketbra{\psi(P)}{\psi(P)} 
  \tilde{U}(L_1 \cdots L_d)^\dagger \right) \\
      &= \bra{\psi(P)} \tilde{U}(L_1 \cdots L_d)^\dagger 
      U(L_1 \cdots L_d) P U(L_1 \cdots L_d)^\dagger 
      \tilde{U}(L_1 \cdots L_d) \ket{\psi(P)}.
\end{align}
Here,
\begin{equation}
    U(L_1 \cdots L_d) = U(L_d) \cdots U(L_1), 
    \hspace{0.3cm}
    \tilde{U}(L_1 \cdots L_d) = P_d U(L_d) P_{d-1} U(L_{d-1}) \cdots P_1 U(L_1),
\end{equation}
where each $U(L_i)$ is drawn at random and each $P_i$ is a Pauli string
occurring with probability $\epsilon_{L_i,P_i}$, which depends on $L_i$. The operator $P$ is a Pauli string chosen uniformly at random (with a random $\pm$
sign), and $\ket{\psi(P)}$ is a stabilizer state of $P$, modeled as a random
variable independent of the circuit layers.

\vspace{1em}

We now compute the expected value of $B_d$ with respect to the choice of layers
$L_1, \dots, L_d$, the initial Pauli string $P$, and the stabilizer state
$\ket{\psi(P)}$.
\begin{equation}
  F_d = \mathbb{E}_{L_1,\dots, L_d, P, \ket{\psi(P)}}(B_d).
\end{equation}
The quantity $F_d$ is the value estimated by repeating the experiment multiple
times. In what follows, we show that $F_d$ can be used to estimate
$\epsilon_\mu$ under certain conditions.

\vspace{1em}

Since all the $U(L_i)$ are Clifford unitaries, we can express
\begin{equation}
  \tilde{U}(L_1 \cdots L_d) = \tilde{P} U(L_1 \cdots L_d),
\end{equation}
where $\tilde{P}$ is a Pauli-operator-valued random variable. The operator
$\tilde{P}$ is obtained by commuting each $P_i$ through the subsequent Clifford
layers and then multiplying all the resulting operators together. Therefore, we can write
\begin{align}
  B_d &= \bra{\psi(P)}U(L_1\cdots L_d)^\dagger \tilde{P} U(L_1\cdots L_d) P U(L_1\cdots L_d)^\dagger \tilde{P}U(L_1\cdots L_d) \ket{\psi(P)} \\
      &=  \bra{\psi(P)}\tilde{P'} P \tilde{P'} \ket{\psi(P)}
\end{align}
where $\tilde{P'} = U(L_1\cdots L_d)^\dagger \tilde{P} U(L_1\cdots L_d)$. The
quantity $B_d$ equals 1 whenever at least one of the following situations
occurs:
\begin{enumerate}
  \item[$S1$] All the $P_i = I$ which implies that $\tilde{P} = I$ and finally
    $\tilde{P'} = I$.
  \item[$S2$] Not all $P_i$ are equal to $I$, but they cancel each other out,
    so that ultimately $\tilde{P} = I$ and $\tilde{P}' = I$.
  \item[$S3$] $\tilde{P'}$ and $P$ commutes.
\end{enumerate}
If we denote by $S$ the event that $B_d = 1$, then we can see that
\begin{equation}
  F_d = \mathbb{E}_{L_1,\dots L_d, P, \ket{\psi(P)}}(B_d) = \mathrm{Pr}_d(S) -
  (1-\mathrm{Pr}_d(S)) = 2\mathrm{Pr}_d(S) - 1 
  \label{eq:Fd_S}
\end{equation}
where the probability of $S$ can be expressed in terms of the events $S1$,
$S2$, and $S3$ as
\begin{equation}
  \mathrm{Pr}(S) = \mathrm{Pr}_d(S1) + \mathrm{Pr}_d(S2) + \mathrm{Pr}_d(S3)
\end{equation}
Using basic probability theory we have that 
\begin{equation}
  \mathrm{Pr}_d(S) = s_1 + s_2 + (1 - s_1 - s_2) s_3
  \label{eq:ProbS}
\end{equation}
where 
\begin{equation}
  s_1 = \mathrm{Pr}_d(S1), s_2 = \mathrm{Pr}_d(S2), s_3 = \mathrm{Pr}_d(S3 |
  \overline{S1} \cap \overline{S2})
\end{equation}
We now proceed to analyze each term, beginning with $s_3$. By definition of
$s_3$, we have $\tilde{P}' \neq I$. Here, $\tilde{P}'$ represents a Pauli
operator resulting from the random errors $P_i$ and can thus be treated as a
random variable independent of the Pauli operator $P$. Since $P$ is selected
uniformly at random, the probability that two randomly chosen Pauli operators
commute is $\frac{1}{2}$; this follows from the fact that, for any fixed Pauli
operator, exactly half of all Paulis commute with it while the other half
anticommute. Consequently, the contribution of $s_3$ is
\begin{equation}
  s_3 = \frac{1}{2}.
\end{equation}

The analysis of $S_1$ is also straightforward. The probability of no error
occurring in a single layer is $1-\epsilon_\mu$, so the probability of no
errors over $d$ layers is given by
\begin{equation} 
  s_1 = (1 - \epsilon_\mu)^d.
\end{equation}
Substituting this expression into equation~\ref{eq:ProbS}, and then using the
result in equation~\ref{eq:Fd_S}, we obtain
\begin{align}  
  F_d &= \mathbb{E}_{L_1,\dots L_d, P, \ket{\psi(P)}}(B_d) = 2\mathrm{Pr}_d(S) -1 = 2(s_1 + s_2 + (1-s_1 - s_2)s_3) -1 \\
      &= s_1 + s_2 \\
      &= (1-\epsilon_\mu)^d + s_2.
\end{align}
If $s_2 \approx 0$, then $F_d$ will decay exponentially with $d$, allowing us
to estimate $\epsilon_\mu$ from this behavior. We note that this analysis does
not account for state preparation and measurement errors. Since these errors
are independent of the circuit depth $d$, they remain constant as the depth
increases. Therefore, the final expression for the fidelity, including these
errors, is
\begin{equation}
  F = AF_d= A(1-\epsilon_\mu)^d = Ap^d
\end{equation}
This is the curve that we ultimately obtain with our procedure, from which we
can estimate $\epsilon_\mu$.

We now analyze how $s_2$ can be upper-bounded in order to determine the
conditions under which the obtained results are reliable.

\subsection{Deriving a formula for error cancellation}

We begin by analyzing the probability of observing exactly $l$ errors in the
noisy evolution $\tilde{U}(L_1 \cdots L_d)$. Since the probability of an error
occurring after a single layer is $\epsilon_\mu$, the probability of obtaining
$l$ errors and $d-l$ error-free layers is given by
\begin{equation}
  \Pr[\text{$l$ errors in $d$ layers}] = (1 - \epsilon_\mu)^{d-l} \, \epsilon_\mu^{\,l} \binom{d}{l}.
\end{equation}
where the binomial coefficient accounts for the number of possible positions at
which the errors can occur. To study error cancellation, we will reformulate
this expression. First, observe that when a single error occurs, it cannot be
canceled
\begin{equation}
  U(L_d \cdots L_i)P_{i-1} U(L_{i-1} \cdots L_1) = U(L_d \cdots L_1) \implies P_{i-1} = I
\end{equation}
where the equalities hold up to an overall phase factor. 

We now assume that at least two errors occur, i.e., $l > 1$. To analyze this
case, we reorganize the possible error configurations into groups according to
the separation between the first and last errors in $\tilde{U}(L_1 \cdots
L_d)$. Let this separation be $k$ layers, with $k$ ranging from $l-1$ to $d-1$.  

First, consider fixing the positions of the first and last errors such that
they are separated by $k$ layers. It is straightforward to verify that there
are $d-k$ possible choices for these two positions. Next, there are $k-1$
intermediate positions available in which to place the remaining $l-2$ errors,
which can be done in $\binom{k-1}{\,l-2}$ distinct ways. Considering all possible separations $k$, the probability of observing $l > 1$ errors whose first and last occurrences are separated by $k$ layers can be expressed as 
\begin{equation}
  (1- \epsilon_\mu)^{d-l} \epsilon_\mu^l \sum_{k=l-1}^{d-1}(d-k)\binom{k-1}{l-2}.
\end{equation}

\begin{figure}[ht]
 \begin{center}
   \includegraphics[width=0.9\columnwidth]{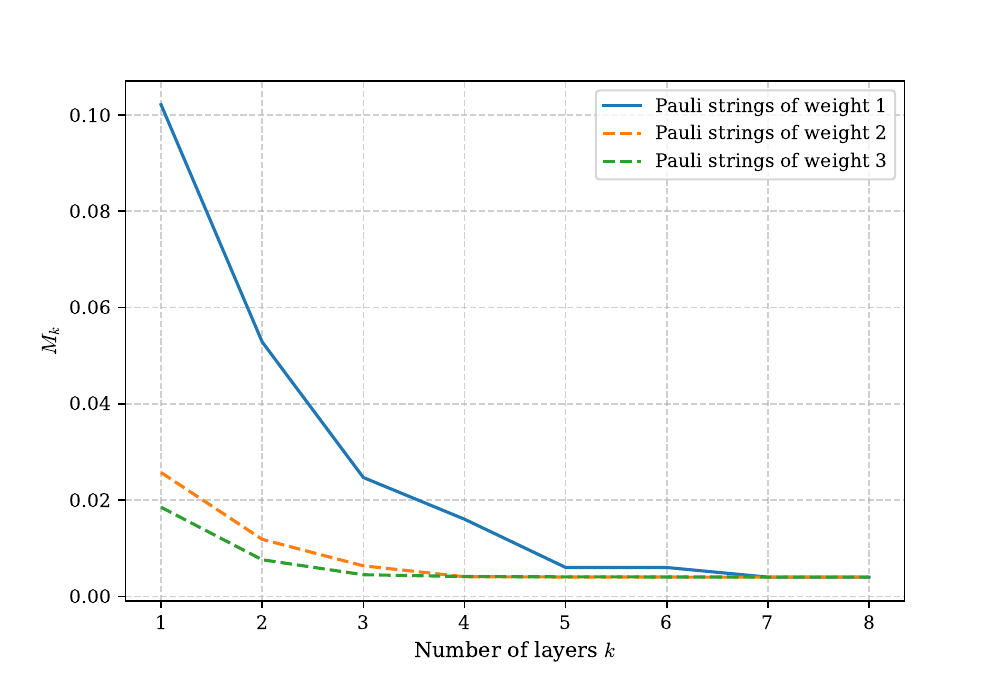}
 \end{center}
   \caption{Estimation of the values $M_k$ defined in Eq.~\ref{eq:M_kDefinition}, which bound the probability that errors whose first and last occurrences are separated by $k$ layers cancel. For each value of $k$, the quantities $M_k$ are estimated using 500 different unitaries $U(L_k \cdots L_1)$, where each layer is constructed by taking $20\%$ of the circuit obtained by compiling, with Qiskit~2.0.0, a fully random Clifford unitary acting on 10 qubits.}
  \label{fig:M_k_values}
\end{figure}

Consider now the case where $l$ errors occur and the two extreme errors, $P_1$
and $P_{k+1}$, which are both different from the identity, are separated by $k$
layers. In this case, the evolution can be written as
\begin{equation}
  U(L_d \cdots L_{i+k+1}) P_{k+1} U(L_{i+k}) P_{k} U(L_{i+k-1}) \cdots U(L_{i+2})P_2 U(L_{i+1}) P_{1} U(L_i \cdots L_1),
\end{equation}
where only $l-2$ of the operators $P_2, \dots, P_{k}$ are non-identity Paulis.
By commuting these $l-2$ Pauli operators through the Clifford unitaries, they
can be shifted to the left, resulting in an effective Pauli operator
$\tilde{P}$ acting at the same position as $P_{k+1}$, so that the expression
becomes
\begin{equation}
  U(L_d \cdots L_{i+k+1}) P_{k+1}\tilde{P} U(L_{i+k} \cdots L_{i+1}) P_{1} U(L_i \cdots L_1).
\end{equation}

Let us denote $P = P_1$ and $Q = P_{k+1}\tilde{P}$. In this notation, all of these gates cancel if and only if
\begin{align}
  U(L_d \cdots L_{i+k+1}) \, Q \, U(L_{i+k} \cdots L_{i+1}) \, P \, U(L_i \cdots L_1) &= U(L_d \cdots L_1) \label{eq:cancellation_condition1} \\
  \iff \quad U(L_{i+k} \cdots L_{i+1})^\dagger \, Q \, U(L_{i+k} \cdots L_{i+1}) &= P, \label{eq:cancellation_condition2}
\end{align}
up to an overall phase.

Since $Q$ depends on the unitary block $U(L_{i+k} \cdots L_{i+1})$, while $P$
depends only on $U(L_i)$, and these unitaries are chosen independently at
random, $U(L_{i+k} \cdots L_{i+1})^\dagger  Q  U(L_{i+k} \cdots L_{i+1})$ and
$P$ can be regarded as independent. Hence, we can bound the probability of
error cancellation between errors separated by $k$ layers by computing
\begin{equation}
M_k = \max_{P,Q \in \mathcal{P}\setminus \{I\}}
\left\{
\mathop{\mathrm{Pr}}\limits_{L_1, \dots, L_k \;\mid\; \mathrm{supp}(Q)\subseteq \mathrm{supp}(U(L_1\cdots L_k))}
\big[ U(L_{k}\cdots L_{1})^\dagger Q U(L_{k} \cdots L_{1}) = \pm P \big]
\right\},
\label{eq:M_kDefinition}
\end{equation}
where $\mathrm{supp}(U) = \{\, i \in \{1, \dots, n\} \mid U \text{ acts
non-trivially on qubit } i \,\}$.  The reasoning for this assumption is that if the sequence $L_1 \cdots L_k$ does not act on a given qubit, then that qubit cannot incur an error. 

The quantity $M_k$ provides an upper bound on the probability that errors
whose first and last occurrences are separated by $k$ layers cancel. Using the
combinatorial decomposition introduced above, the probability that exactly
$l>1$ errors occur and that the two extreme errors (separated by $k$ layers)
cancel is bounded by
\[
(1-\epsilon_\mu)^{d-l}\,\epsilon_\mu^{\,l}\,(d-k)\binom{k-1}{l-2}\,M_k,
\]
since there are $(d-k)$ choices for the positions of the first and last error
and $\binom{k-1}{l-2}$ ways to place the remaining $l-2$ errors between them.
Consequently, the probability that \(l>1\) errors produce cancellation (for
any allowed separation \(k\)) satisfies
\begin{figure}[ht]
 \begin{center}
   \includegraphics[width=0.8\columnwidth]{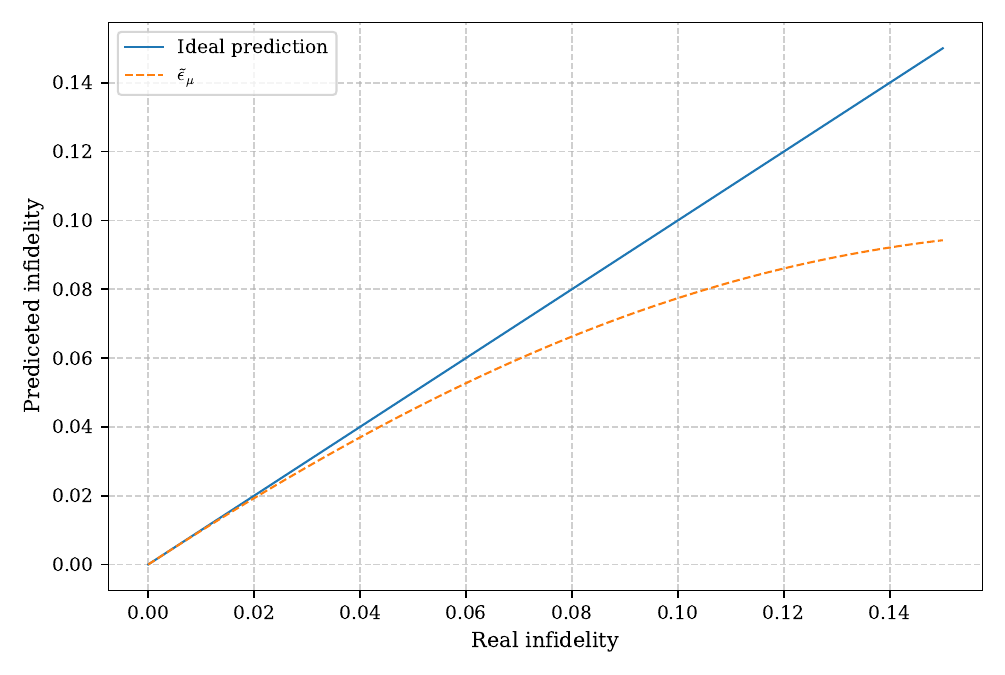}
 \end{center}
   \caption{Prediction of the device infidelity using 10 layers under the
     worst-case cancellation assumption where the probability of error
     cancellation is bounded by Eq. \ref{eq:worstCase}. The blue curve
     represents the prediction assuming no error cancellation, while
     $\tilde{\epsilon}_\mu$ correspond to the predicted values of the real
     infidelities, considering error cancellation. For infidelities below $0.1$,
    the predicted values deviate by less than $0.025$ from the true infidelity,
    consistently underestimating the actual error.}
  \label{fig:worst_case}
\end{figure}
\begin{equation}
  \Pr[\text{cancellation}\,\mid\,l\, \text{errors}] \le (1-\epsilon_\mu)^{d-l}\,\epsilon_\mu^{\,l}
\sum_{k=l-1}^{d-1} (d-k)\binom{k-1}{l-2}\,M_k.
\label{eq:prob_cancel_l}
\end{equation}
Finally, summing over all possible numbers of errors \(l\ge 2\) yields an
upper bound on the total cancellation probability (the contribution that we
called \(s_2\)):
\begin{equation}
s_2 \le \sum_{l=2}^{d} (1-\epsilon_\mu)^{d-l}\,\epsilon_\mu^{\,l}
\sum_{k=l-1}^{d-1} (d-k)\binom{k-1}{l-2}\,M_k .
\label{eq:bound_S2}
\end{equation}
Equations \eqref{eq:prob_cancel_l} and \eqref{eq:bound_S2} are the desired
bounds expressing how the separation-dependent cancellation probabilities
\(M_k\) control the overall cancellation contribution \(s_2\). We can
reorganize the expression so that the summation over $k$ appears first.  In
this way, we obtain
\begin{equation}
  s_2 \leq \sum_{k=1}^{d-1}
  M_k(d-k)\sum_{j=0}^{k-1}\binom{k-1}{j}(1-\epsilon_\mu)^{d-j-2}\epsilon_\mu^{2+j}
  = \epsilon_\mu^2\sum_{k=1}^{d-1}M_k(d-k)(1-\epsilon_\mu)^{d-1-k}.
  \label{eq:upperBounds2}
\end{equation}
Therefore, given an estimate of $M_k$ for every $k$, we obtain a bound on the
probability of error cancellation as a function of $\epsilon_\mu$. This bound
characterizes the regime in which our procedure can provide a reliable estimate
of $s_1$.

\subsection{Verification of Conditions via Simulations}

To evaluate the reliability of our procedure, we need to estimate $M_k$ for
different values of $k$, ranging from 1 up to one less than the maximum number
of layers considered in the protocol. This can be achieved through classical
simulations designed to approximate $M_k$ efficiently. 
The simulation procedure for a given $k$ consists of randomly selecting $L_1,
\dots, L_k$ $\mu$-fraction circuits a total of $R$ times, and computing
\begin{equation}  
  P = U(L_1 \cdots L_k) Q U(L_1 \cdots L_k)^\dagger,
\end{equation}
with global phases removed, for all $Q \in \mathcal{P}$ such that
$\mathrm{supp}(Q) \subseteq \mathrm{supp}(U(L_1 \cdots L_k))$. Each resulting
Pauli operator $P$ is then recorded. Among the $R$ trials, we identify the pair
$(Q,P)$ that occurs most frequently. 

A direct implementation of this procedure is not scalable, since the Pauli
group contains $4^N$ elements. However, scalability can be achieved by
restricting the analysis to Pauli operators of small weight, e.g., weight less
than 4. Intuitively, this restriction corresponds to the worst-case scenario,
as low-weight errors cannot fully spread across all qubits. For higher-weight
Paulis, errors tend to delocalize more efficiently and the probability of
cancellation decreases, as discussed in
\cite{polloreno2023theorydirectrandomizedbenchmarking}. This intuition is
supported by numerical simulations showing that the probability of error cancellation
decreases with increasing Pauli weight, as we can see in Fig.~\ref{fig:M_k_values}.
Restricting the analysis to Pauli operators of weight 1, 2, and 3 yields a
simulation that scales as $O(N^3)$ in the number of qubits, making the
procedure computationally efficient. Examples of these simulations for
different numbers of qubits and values of $\mu$ are presented in Fig.~\ref{fig:M_k_values}
\begin{figure}[t]
  \begin{center}
    \includegraphics[width=0.8\columnwidth]{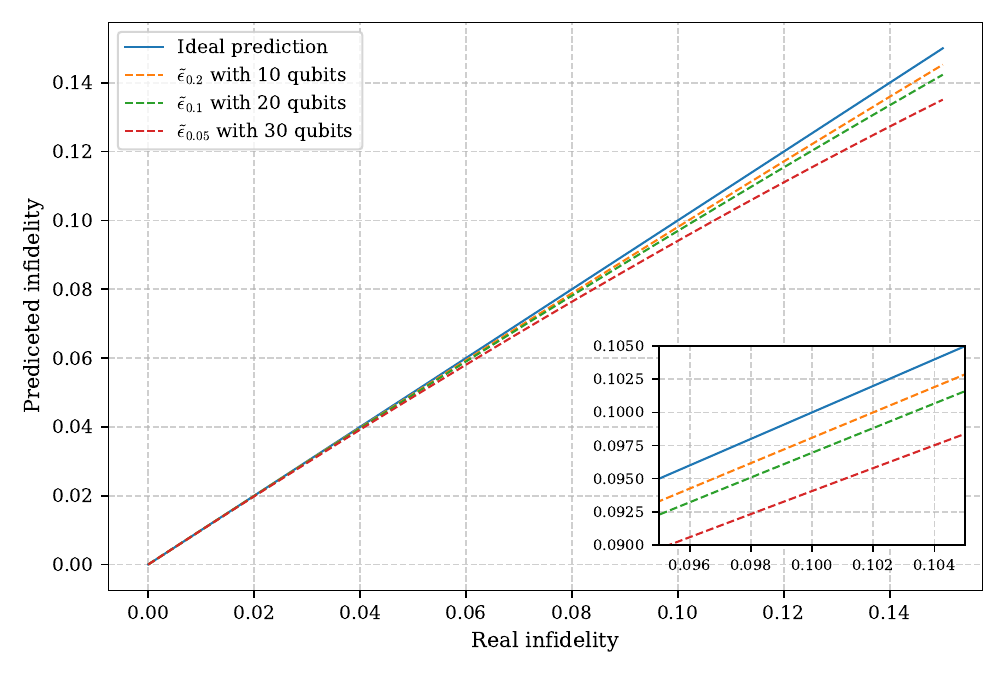}
  \end{center}
    \caption{Prediction of the device infidelity using 12 layers where the
      probability of error cancellation is computed according to
      Eq.~\ref{eq:upperBounds2}. The values of $M_k$ are estimated via
      simulations using pauli strings of weight 1, and the final prediction is
      obtained using linear regression. The blue curve represents the
      prediction assuming no error cancellation, while $\tilde{\epsilon}_\mu$
      corresponds to the predicted values of the real infidelity for different
      numbers of qubits and $\mu$, considering error cancellation.}
      \label{fig:infidelityPrediction}
\end{figure}

\subsection{A Trivial Upper Bound}
Even without resorting to simulations, we can derive an upper bound on the
cancellation error. This bound allows us to predict, for a given value of
$\epsilon_\mu$, the regime in which our protocol is guaranteed to operate
reliably. The worst-case scenario, corresponding to the maximum probability of
error cancellation, occurs when there is only a single layer $U(L)$ between two
errors, and this layer acts on a single qubit where both errors occur. Since
$U(L)$ acts on only one qubit, and every single-qubit Clifford gate can be
decomposed into a small set of basic gates, we can model $U(L)$ as a uniformly
random single-qubit Clifford gate. 

Under this assumption, conjugation by a Clifford maps each non-identity Pauli
operator uniformly to every other non-identity Pauli operator. Consequently, the
probability of error cancellation is simply $\frac{1}{3}$, as all non-identity
Paulis are equally likely. By setting all $M_k = \frac{1}{3}$, we obtain a
conservative upper bound
\begin{equation}
  s_2 \leq 
  \epsilon_\mu^2 \sum_{k=1}^{d-1} \frac{1}{3} (d-k) (1-\epsilon_\mu)^{d-1-k} =
  \frac{1}{3} \left[ 1 - (1-\epsilon_\mu)^d - d \, \epsilon_\mu \, (1-\epsilon_\mu)^{d-1} \right].
  \label{eq:worstCase}
\end{equation}

We now attempt to fit a curve of the form $A f^d$ to the data obtained from
\begin{equation}
  F_d = (1-\epsilon_\mu)^d + \frac{1}{3} \left[ 1 - (1-\epsilon_\mu)^d - d \,
  \epsilon_\mu \, (1-\epsilon_\mu)^{d-1} \right],
\end{equation}
where $s_2$ is evaluated using the worst-case formula \eqref{eq:worstCase}.
This allows us to assess the range of $\epsilon_\mu$ for which this
approximation is reasonable.

To perform the fitting, we apply linear regression to the logarithm of the data:
\begin{equation}
  \log(F_d) \approx \log(A f^d) = \log(A) + d \log(f).
\end{equation}

From this, the slope of the regression line provides an estimate of $\log(f)$,
from which we define an approximate error rate
\begin{equation}
  \tilde{\epsilon}_\mu = 1 - f \approx \epsilon_\mu.
\end{equation}

If we try to estimate $f$ using that $F_d$ is defined as in
\eqref{eq:worstCase}, the formula for the slope of linear regression using $d$
continuous points (corresponding to $d$ layers) is
\begin{equation}
  f = \exp\Biggl[
    \frac{d \sum_{k=1}^{d} k y_k - \sum_{k=1}^{d} k \sum_{k=1}^{d} y_k}
         {d \sum_{k=1}^{d} k^2 - \left(\sum_{k=1}^{d} k\right)^2}
  \Biggr]
  = \exp\Biggl[
    \frac{12 \sum_{k=1}^{d} k y_k}{d(d^2-1)} - \frac{6 \sum_{k=1}^{d} y_k}{d(d-1)}
  \Biggr],
\end{equation}
where $y_k = \log(F_k)$. 

Using $d=10$ total layers for estimating $\epsilon_\mu$ and assuming
$\epsilon_\mu \leq 0.1$, our predictions of $\tilde{\epsilon}_\mu$ are
reasonably close to the true $\epsilon_\mu$ as we can see in Fig.
\ref{fig:worst_case}. For a larger number of layers, however, this
approximation becomes less accurate. This suggests that, at least for
$\epsilon_\mu \leq 0.1$ and a moderate number of layers, we have analytically
shown that our protocol can provide a reliable estimate of $\epsilon_\mu$. In
practice, the actual error is even smaller, as our bound is tighter, a fact
confirmed by numerical simulations and illustrated, for example, in Fig.
\ref{fig:infidelityPrediction}.

  \section{Quantum circuit to implement Quantum Signal processing}
This appendix details the construction of the quantum circuits used in the
TFIM Hamiltonian simulation benchmark protocol described in
Sec.~\ref{protocol:TFIM_protocol}. 

\subsection{Quantum circuit for the block-encoding of the Hamiltonian}

Here, we describe the construction of the operators $\hat{U}_{\text{Prep}}$ and
$\hat{U}_{\text{Select}}$ required for the block-encoding of the Hamiltonian,
defined as
\begin{equation}
\hat{U}_H = \begin{pmatrix}
\hat{H}/\alpha & * \\
* & * 
\end{pmatrix}
=(\hat{U}_{\rm Prep}^\dagger \otimes \mathbb{I}_{s\times s})\,\hat{U}_{\rm Select}\,(\hat{U}_{\rm Prep} \otimes  \mathbb{I}_{s\times s}).
\end{equation}
This decomposition allows the Hamiltonian to be encoded in the upper-left block
of the unitary operator $\hat{U}_H$. In the following subsections, we introduce the protocols used to construct
the quantum circuits.


\subsubsection{Quantum circuit implementing the operator $\hat{U}_{\text{Prep}}$}
\label{UPREP}
\begin{figure}[ht]
    \centering 
    \includegraphics[width=0.95 \columnwidth ]{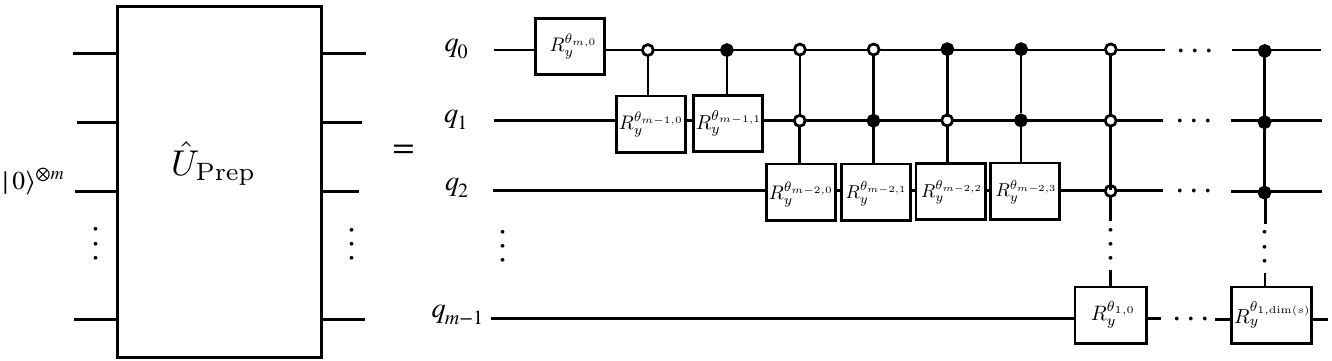}
  \caption{
Construction of the Quantum circuit to implement $\hat{U}_{\mathrm{Prep}}$ following the Protocol \ref{PROTUPREP}.
The circuit consists of a sequence of controlled $R_y$
rotations, where each rotation angle is obtained 
recursively from the vector $\mathbf{P}$. The $R_y$
rotations are denoted by $R_y^{\theta_{k,s}}$, where
$\theta_{k,s}$ are the angles computed at the $k$-th level for index $s$.
}
    \label{fig:PrepState}
\end{figure}

The purpose of $\hat{U}_{\text{Prep}}$ is to load the normalized coefficients
of the Hamiltonian into the amplitudes of a quantum state. 
We employ the quantum state preparation algorithm~\cite{mottonen2004transformationquantumstatesusing,PhysRevA.64.014303,PhysRevA.67.032303}, 
which constructs arbitrary quantum states through a recursive sequence of 
$R_y$ and controlled-$R_y$ rotations. 
Fig.~\ref{fig:PrepState} illustrates the general structure of the circuit.
The protocol is described as follows:

\begin{myprotocol}{Quantum circuit implementing $\hat{U}_{\rm Prep}$.}
\label{PROTUPREP}
  \textbf{Initialization:} Vector $\mathbf{P}$ with dimension $2L$. 
   
  \noindent\textbf{Return:} Quantum circuit implementing $\hat{U}_{\rm Prep}$.

  \noindent\textbf{Procedure:} 
  \begin{enumerate}
    \item \textbf{Dimension normalization:} Let $d = 2L$ and define
    \[
      m = \lceil \log_2 d \rceil, \qquad D = 2^m.
    \]
    If $d < D$, extend vector
    $\mathbf{P}$ by padding with zeros,
    \[
      \mathbf{P}[j] =
      \begin{cases}
        \mathbf{P}[j], & 0 \le j < d,\\[2mm]
        0, & d \le j < D.
      \end{cases}
    \]

    \item \textbf{First level of norms and angles computation:} For each $s
      \in \left\{0, 1, \dots, \frac{\dim(\mathbf{P})}{2} - 1\right\}$, compute the
      norms as follows:
      \[
        \text{norm}_1[s] = \sqrt{\mathbf{P}[2s]^2 + \mathbf{P}[2s+1]^2}.
      \]
      Then, compute the rotation angles:
      \[
        \text{angles}_1[s] = 2 \arccos\left( \frac{\mathbf{P}[2s]}
        {\text{norm}_1[s]} \right).
      \]
      Notice that, if $\text{norm}_1[s] = 0$, the corresponding angle is set to $0$.

    \item \textbf{Recursive norm calculation:} Repeat the same procedure
      recursively using vectors obtained in the previous step. At the $k$-th level, compute the norms as follows:
    \begin{align}
      \text{norm}_{k}[s] &= \sqrt{\text{norm}_{k-1}[2s]^2 +
    \text{norm}_{k-1}[2s+1]^2},\\
    \text{angles}_k[s] &= 2 \arccos\left(
    \frac{\text{norm}_{k-1}[2s]}{\text{norm}_k[s]} \right),  
    \end{align}
    for $k=2,...,m$ and $s \in \left\{0, 1, \dots,
    \frac{\dim(\text{norm}_{k-1})}{2} - 1\right\}$.

     Notice that, if $\text{norm}_k[s] = 0$, the corresponding angle is set to $0$.

\item \textbf{Circuit construction:} Consider a quantum
circuit with $m$ ancillary qubits, labeled $q_0, \ldots,
q_{m-1}$. The circuit is built as a sequence of controlled-$R_y$ rotations. Each rotation is written as $R_y^{\theta_{k,s}}$, where $\theta_{k,s}$ denotes the angle computed at level $k$ for index $s$. The quantum circuit implementing $\hat{U}_{\text{Prep}}$ is described as follows:

    \begin{itemize}
        \item Apply the quantum gate $R_y^{\theta_{m,0}}$ to qubit $q_0$.
        
        \item Then, for each level $k = m - 1, \dots, 1$, apply the controlled-$R_y^{\theta_{k,s}}$ rotations to the ancillary qubit $q_{m-k}$, covering all possible controlled
        conditions of the ancillary qubits $q_0, \dots,q_{m-k-1}$. Fig.~\ref{fig:PrepState} shows the quantum circuits to implement $\hat{U}_{\rm Prep}$.
    \end{itemize}
  \end{enumerate}
\end{myprotocol}

\vspace{3cm}

\subsubsection{Quantum circuit implementing the operator $\hat{U}_{\text{Select}}$}
\label{USELECT}
\begin{figure}[ht]
    \centering 
    \includegraphics[width=0.8 \columnwidth]{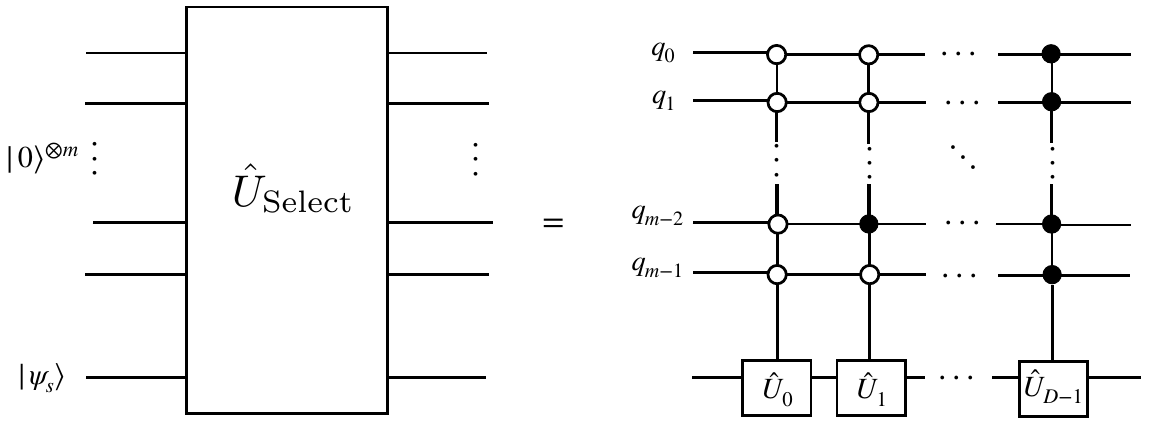}
    \caption{Quantum circuit implementing the operator $\hat{U}_{\mathrm{Select}}$. Each controlled operator $\hat{U}_i$ (for $i=0,\dots,D-1$) acts on the system register and is applied conditionally on the state of the $m$ ancillary qubits. Black and white control circles indicate conditioning on the $\ket{1}$ and $\ket{0}$ states of the ancilla qubits, respectively.
    }
    \label{fig:USELECT}
\end{figure}

The operator $\hat{U}_{\rm Select}$ is implemented using multi-controlled Pauli operators acting on main system, conditioned on the state of the $m$ ancillary qubits. Fig.~\ref{fig:USELECT} shows the quantum circuit used to implement $\hat{U}_{\rm Select}$, where each unitary $\hat{U}_i$ corresponds to a Pauli operator acting on the main system, conditioned on the ancilla register being in the state $|i\rangle$. The black circle denotes a control on the ancillary qubit being in the state $|1\rangle$ while denotes a control on the ancillary qubit being in the state $|0\rangle$. The protocol used to construct the quantum circuit is described as follows:

\begin{myprotocol}{Quantum circuit implementing $\hat{U}_{\text{Select}}$}

\textbf{Initialization:} Vector $\mathbf{U} = [\hat{U}_0, \hat{U}_1, \dots,
  \hat{U}_{2L-1}]$ with dimension $2L$.
  
\noindent\textbf{Return:} Quantum circuit implementing $\hat{U}_{\text{Select}}$.\\

\noindent\textbf{Procedure:}
\begin{enumerate}

\item \textbf{Dimension normalization:} Let $d = 2L$ and define
\[
  m = \lceil \log_2 d \rceil, \qquad D = 2^m.
\]
If $d < D$, extend the vector $\mathbf{U}$ by padding with identity operators,
\[
  \mathbf{U}[j] =
  \begin{cases}
    \mathbf{U}[j], & 0 \le j < d,\\[2mm]
    \mathbb{I}_{2^L}, & d \le j < D.
  \end{cases}
\]

\item \textbf{Circuit construction:} Consider a quantum circuit with $m$
ancillary qubits labeled $q_0, \dots, q_{m-1}$. For each $i
= 0, \dots, D-1$, apply the controlled operator $\hat{U}_i$ to the system register, where $\hat{U}_i$ is conditioned on the state of the ancillary qubits, as illustrated in Fig.~\ref{fig:USELECT}. A black circle indicates conditioning on the ancillary qubit being in the state $\ket{1}$, whereas a white circle indicates conditioning on the state $\ket{0}$.

\end{enumerate}
\end{myprotocol}

\vspace{3cm}

\subsubsection{Protocol to compute phase angles}
\label{Angles}

In the TFIM Hamiltonian simulation benchmark, we employ the Fixed-Point Iteration method to compute the phase angles \cite{dong2023}. The algorithm used to compute the phases is described as follows:

\begin{myprotocol}{Algorithm for Computing Phase angles}
\textbf{Initialization:} Vector $\mathbf{C}$ containing the coefficients of the Chebyshev expansion and the parity $p \in \{0,1\}$ of the target function, where $p = 0$ corresponds to an even function and $p = 1$ to an odd function.

\noindent\textbf{Return:} Sequence of phase angles $\boldsymbol{\phi}$.

\noindent\textbf{Procedure:}
\begin{enumerate}
    \item \textbf{Apply fixed-point iteration:} Initialize the phase vector as
      $\boldsymbol{\phi}^{(0)} = 0$ and set the iteration counter $t = 0$.
      Iteratively update the phase vector according to
      \begin{equation}
        \boldsymbol{\phi}^{(t+1)} = \boldsymbol{\phi}^{(t)} - \frac{1}{2} \left[ F(\boldsymbol{\phi}^{(t)}) - \mathbf{C} \right],
      \end{equation}
      where $F(\boldsymbol{\phi}^{(t)})$ returns the Chebyshev coefficients generated by the
      current phase vector. If $\|F(\boldsymbol{\phi}^{(t)}) - \mathbf{C}\|_1 < \epsilon$
      with $\epsilon = 10^{-10}$, stop the iteration and return the phase vector.
      
      The Chebyshev coefficients, $F(\boldsymbol{\phi}^{(t)})$, generated by the phase vector are computed as follows:
      \begin{enumerate}
        \item \textbf{Evaluate $P_{\mathrm{im}}(x_j)$ at Chebyshev nodes:}  
        Given the phase vector \(\boldsymbol{\phi}^{(t)}\) and the parity \(p\) of
        the target function, evaluate the imaginary part of the $(1,1)$
        element of the QSP unitary at the Chebyshev nodes  
        \[
            x_j = \cos\left(\theta_j \right), \quad j = 0, 1, \dots, \ell.
        \] where $\theta_j =\frac{j\pi}{2\ell}$ and $\ell$ denotes the length of the phase vector $\boldsymbol{\phi}$. 

         Then, compute the imaginary component as follows:
\[
    P_{\mathrm{im}}(x_j) =
    \Pi_\ell \!\left( \prod_{k=1}^{\ell-1} B(\theta_j)\, R_z\!\big(2\phi_k\big) \right) R_0(\theta_j),
\]
where the initial vector is defined as
\[
    R_0(\theta_j) =
    \begin{cases}
        (1,\,0,\,0)^\top, & \text{if } p = 0 \ \text{(even)}, \\[4pt]
        (\cos\theta_j,\,0,\,\sin\theta_j)^\top, & \text{if } p = 1 \ \text{(odd)}.
    \end{cases}
\]
The rotation matrices are defined as
\[
    R_z(2\phi_k) =
    \begin{pmatrix}
        \cos(2\phi_k) & -\sin(2\phi_k) & 0 \\[4pt]
        \sin(2\phi_k) &  \cos(2\phi_k) & 0 \\[4pt]
        0 & 0 & 1
    \end{pmatrix},
\]
and
\[
    B(\theta_j) =
    \begin{pmatrix}
        \cos(2\theta_j) & 0 & -\sin(2\theta_j) \\[4pt]
        0 & 1 & 0 \\[4pt]
        \sin(2\theta_j) & 0 & \cos(2\theta_j)
    \end{pmatrix}.
\]
The projection vector is defined as
      \[
          \Pi_\ell = \big(\sin(2\phi_\ell),\,\cos(2\phi_\ell),\,0\big).
      \]

        \item \textbf{Extend the sequence with parity symmetry:}
        Build the vector $\{y_j\}_{j=0}^{2\ell}$ by reflecting the
        imaginary component with the required parity,
        \[
        y_j =
        \begin{cases}
          S_j, & j=0,1,\dots,\ell,\\[4pt]
          (-1)^p\, S_{\,2\ell-j}, & j=\ell+1,\dots,2\ell,
        \end{cases}
        \]
        where $S_j = P_{\mathrm{im}}(x_j)$.
      This even/odd symmetric extension enforces the desired parity in the
      Chebyshev expansion.

    \item \textbf{Compute Chebyshev coefficients via Discrete Cosine Transform (DCT):}
      The DCT is defined by
      \[
        Y_k \;=\; \sum_{j=0}^{2\ell} y_j \,\cos\!\left(\frac{\pi j k}{2\ell}\right),
        \qquad k=0,1,\dots,2\ell.
      \]
      Normalize the first and last coefficients of the vector $\{Y_k\}_{k=0}^{2\ell}$ by \( \frac{1}{4\ell} \) and all
      intermediate coefficients by \( \frac{1}{2\ell} \).

    \item \textbf{Select coefficients matching the parity:}
      Extract the subsequence depending of the parity
      \[
        F(\boldsymbol{\phi}^{(t)}) \;=\;
        \begin{cases}
            \big(Y_{0},\,Y_{2},\,Y_{4},\dots\big)_{[0:\ell-1]}, & p=0\ \text{(even)},\\[4pt]
            \big(Y_{1},\,Y_{3},\,Y_{5},\dots\big)_{[0:\ell-1]}, & p=1\ \text{(odd)}.
        \end{cases}
      \].

      \end{enumerate}     

      \item \textbf{Construct the full phase sequence.}
After convergence of the fixed-point iteration, the reduced phase vector
$\boldsymbol{\phi} = (\phi_1,\phi_2,\dots,\phi_\ell)$ is obtained. 
The full complete QSP sequence is generated as follows:
\[
\boldsymbol{\phi} =
\big(\phi_\ell+\frac{\pi}{4},\dots,\phi_{1},
\ \phi_{1},\dots,\phi_{\ell}+\frac{\pi}{4}\big)
\ .
\]

    \end{enumerate} 
\end{myprotocol}

\subsubsection{Quantum circuit to implement the projector-controlled phase gate $\Pi_{\phi_k}$}
\label{Projector}
\begin{figure}[ht]
    \centering 
    \includegraphics[width=0.73\columnwidth]{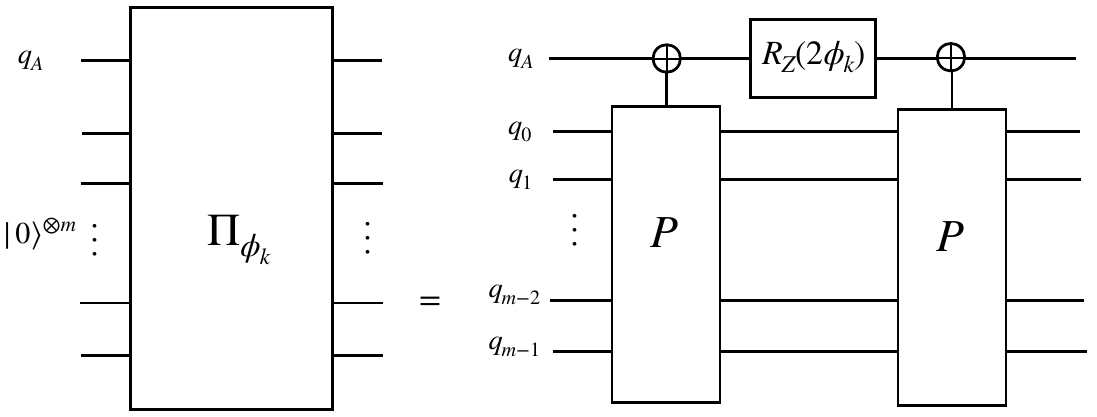}
    \caption{Quantum circuit implementing the projector $\Pi_{\phi_k}$ using multi-controlled phase operations. The circuit applies a controlled projector $P = \ket{0}^{\otimes m}\!\bra{0}^{\otimes m}$ onto the ancillary register, followed by a single-qubit rotation $R_Z(2\phi_k)$ on the ancilla qubit $q_A$, and a second controlled application of $P$. }
    \label{fig:PPhases_Pi}
\end{figure}

The quantum circuit to implement the phase-controlled projector
$\Pi_{\phi_k}$ is described as follows:
\begin{myprotocol}{Quantum circuit to implement the projector $\Pi_{\phi_k}$}
  
  \textbf{Initialization:} Sequence of phase angles $\boldsymbol{\phi}$.

  \noindent\textbf{Return:} Quantum circuit implementing $\Pi_{\phi_k}$.

  \noindent\textbf{Procedure:}
  \begin{enumerate}

    \item \textbf{Circuit construction:} For each $\phi_k$, apply the following quantum gates consecutively:
    \begin{enumerate}

      \item \textbf{Apply the controlled $\mathrm{C}_{P}\mathrm{NOT}$ gate:}  
      Apply a multi-controlled NOT gate defined as
      \[
      \mathrm{C}_{P}\mathrm{NOT} = X \otimes P + \mathbb{I} \otimes (\mathbb{I} - P),
      \]
      where \( P = \ket{0}^{\otimes m}\!\bra{0}^{\otimes m} \) projects onto the all-zero $m$-qubit ancilla state.

      \item \textbf{Apply the phase rotation:}  
      Apply a single-qubit rotation around the $Z$ axis on the ancilla qubit $q_A$:
      \[
      R_Z(2\phi_k).
      \]

      \item \textbf{Reapply the controlled $\mathrm{C}_{P}\mathrm{NOT}$ gate:}  
      Apply again the same multi-controlled NOT gate:
      \[
      \mathrm{C}_{P}\mathrm{NOT} = X \otimes P + \mathbb{I} \otimes (\mathbb{I} - P).
      \]

    \end{enumerate}

  \end{enumerate}
  Fig.~\ref{fig:PPhases_Pi} shows the quantum circuits to implement $\Pi_{\phi_k}$.
\end{myprotocol}

\section{Error Bounds and Success Probability in QSP}
\label{sec:ErrorBound}
\subsection{Polynomial Approximation Error in QSP}
Quantum Signal Processing (QSP) is a powerful technique that enables
the implementation of polynomial approximations of matrix functions using a sequence of
single-qubit rotations controlled by an ancilla qubit~\cite{Low_2017}. 
In the context of Hamiltonian simulation, given a target function 
$e^{-i \hat{H} t/\alpha}$, QSP constructs a polynomial 
$P(t\hat{H}/\alpha)$ that approximates this target operator 
over the normalized spectral domain 
$\lambda_k \in [-1,1]$, where $\lambda_k$ denotes an eigenvalue 
of the rescaled Hamiltonian $\hat{H}/\alpha$. 
The approximation error is bounded by
\begin{equation}
\|\,P(t\hat{H}/\alpha) - e^{-i\hat{H} t/\alpha}\,\|_2 
\le \epsilon_0,
\label{eq:infinity_norm}
\end{equation}
where $\|\cdot\|_2$ denotes the spectral norm. Since any Hamiltonian can be diagonalized in its eigenbasis as
\begin{equation}
\hat{H} = \sum_k \lambda_k |\lambda_k\rangle \langle \lambda_k|,
\end{equation}
the functions $P(t\hat{H}/\alpha)$ and $e^{-i\hat{H}t/\alpha}$ can be expressed in terms of their eigenvalues as
\begin{equation}
P(t\hat{H}/\alpha) = 
\sum_k P(t\lambda_k/\alpha)\,|\lambda_k\rangle \langle \lambda_k|,
\qquad
e^{-i\hat{H}t/\alpha} = 
\sum_k e^{-i\lambda_k t/\alpha}\,|\lambda_k\rangle \langle \lambda_k|.
\end{equation}

\noindent
Therefore, the operator norm in Eq.~\eqref{eq:infinity_norm} is equivalently given by the 
maximum deviation between the two complex functions evaluated on the eigenvalue spectrum of the Hamiltonian:
\begin{equation}
\|\,P(t\hat{H}/\alpha) - e^{-i \hat{H} t/\alpha}\,\|_2
=\max_{\lambda_k \in \mathrm{spec}(\hat{H})}
|\,P(t\lambda_k/\alpha) - e^{-i\lambda_k t/\alpha}\,|
\le \epsilon_0.
\label{eq:eigenvalue_norm}
\end{equation}
This definition ensures that the polynomial approximation holds uniformly for all eigenvalues of the rescaled Hamiltonian
\begin{equation}
|\,P(t\lambda_k/\alpha) - e^{-i\lambda_k t/\alpha}\,| \le \epsilon_0, 
\qquad \forall\,\lambda_k \in [-1,1].
\end{equation}

\subsection{Success Probability in the QSP Circuit}
To implement the polynomial $P(t\hat{H}/\alpha)$, QSP constructs a unitary operator represented as
\begin{equation}
\hat{U}_{\mathrm{QSP}} =
\begin{pmatrix}
P(t\hat{H}/\alpha) & * \\
* & *
\end{pmatrix},
\label{eq:UQSP_block}
\end{equation}
which acts on the main system and ancilla qubits. As we discussed in the main text, to implement the approximation of the  time-evolution operator 
$e^{-i\hat{H}t/\alpha}$ we use two polynomials that approximate 
$\cos(t\hat{H}/\alpha)$ and $\sin(t\hat{H}/\alpha)$. 
Figure~\ref{fig:UTOTAL} shows the quantum circuit used to approximate
the time-evolution operator. 
The Hadamard gates acting on the ancillary qubit $q_B$ enable the coherent 
implementation of both $U_{\rm QSP}^{(\cos)}$ and $U_{\rm QSP}^{(\sin)}$. 
After applying the full quantum circuit, the resulting quantum state is given by

\begin{equation}
\begin{aligned}
|\psi\rangle =&
\frac{1}{2}
\Big[
|0\rangle_B\,|0\rangle_A\,|0\rangle^{\otimes m}\,
\big(P_{\cos}(t\hat{H}/\alpha) + P_{\sin}(t\hat{H}/\alpha)\big)\,|\psi_s\rangle \\
&+
|1\rangle_B\,|0\rangle_A\,|0\rangle^{\otimes m}\,
\big(P_{\cos}(t\hat{H}/\alpha) - iP_{\sin}(t\hat{H}/\alpha)\big)\,|\psi_s\rangle
\Big]
+
|\psi'_{\perp}\rangle,
\end{aligned}
\end{equation}
where $|\psi'_{\perp}\rangle$ is an orthogonal state. Measuring the ancilla qubits and 
post-selecting the measurement outcome $|1\rangle_B|0\rangle_A|0\rangle_m$ projects the 
main system onto the state
\begin{equation}
\frac{P(t\hat{H}/\alpha)\,|\psi_s\rangle}{\|P(t\hat{H}/\alpha)\,|\psi_s\rangle\|},
\end{equation}
where $\|P(t\hat{H}/\alpha)|\psi_s\rangle\|$ is lower bounded by $1-\epsilon_0$. 
Here $P(t\hat{H}/\alpha)=P_{\cos}(t\hat{H}/\alpha) - i\,P_{\sin}(t\hat{H}/\alpha)$ denotes 
the QSP polynomial approximation of $e^{-i\hat{H}t/\alpha}$. 
The success probability $P_{\rm succ}$ to obtain this outcome reads
\begin{equation}
P_{\rm succ} = 
\frac{1}{4}\,\|P(t\hat{H}/\alpha)|\psi_s\rangle\|^2.
\label{eq:success_prob_def}
\end{equation}

Using triangular inequality and the Eq.~\eqref{eq:infinity_norm}, the success probability is lower bounded by
\begin{equation}
P_{\rm succ} \ge \frac{1}{4}(1 - \epsilon_0)^2.
\label{eq:success_prob_bound}
\end{equation}
This result provides a direct connection between the polynomial 
approximation error $\epsilon_0$ and the success probability of the Hamiltonian simulation. 

\subsection{Error in expectation values}
\label{sec:observable_error}

In the TFIM Hamiltonian simulation benchmark, the observable of interest is the
average total magnetization, defined as
\begin{equation}
\hat{M}_z = \frac{1}{L}\sum_{j=1}^{L}\hat{Z}_j,
\end{equation}
where $\hat{Z}_j$ denotes the local Pauli-$Z$ operator acting on qubit $j$.
Since the quantum state prepared on the quantum processor using the Quantum
Signal Processing (QSP) technique is an approximation of the exact
time-evolved state, the estimated magnetization deviates from the exact value.
Using the triangle inequality and the Cauchy–Schwarz inequality, the deviation
in the expectation value of the average magnetization can be upper bounded as follows:
\begin{equation}
\big\|\langle \psi'|\hat{M}_z|\psi'\rangle -
\langle \psi|\hat{M}_z|\psi\rangle\big\|
\le 2\,\|\hat{M}_z\|\;
\big\||\psi'\rangle - |\psi\rangle\big\|,
\end{equation}
where $\|\hat{M}_z\|=1$, the states $|\psi\rangle$ and $|\psi'\rangle$ denote the exact time-evolved state
$e^{-i\hat{H}t/\alpha}|\psi(0)\rangle$ and the
approximate state $P(t\hat{H}/\alpha)|\psi(0)\rangle$ obtained through the QSP implementation, respectively. Using the  bound in Eq. \eqref{eq:infinity_norm} and  the triangle inequality, the norm of the distances between the exact and approximate state can be bounded in
terms of the polynomial approximation error $\epsilon_0$ as follows:
\begin{equation}
\big\||\psi'\rangle - |\psi\rangle\big\|=\big\|P(t\hat{H}/\alpha)|\psi(0)\rangle - e^{-i\hat{H}t/\alpha}|\psi(0)\rangle\big\| 
\le \epsilon_0.
\label{eq:state_error}
\end{equation}
Therefore, the deviation in the expectation value of the average magnetization can be upper bounded by
\begin{equation}
\big\|\langle \psi'|\hat{M}_z|\psi'\rangle -
\langle \psi|\hat{M}_z|\psi\rangle\big\|
\le 2\,\epsilon_0.
\end{equation}

  \section{Iterative training of the multi-qubit QNN architecture}~\label{app:Apendix_QML}
\begin{figure}[ht]
    \centering     
    \includegraphics[width=0.75\columnwidth]{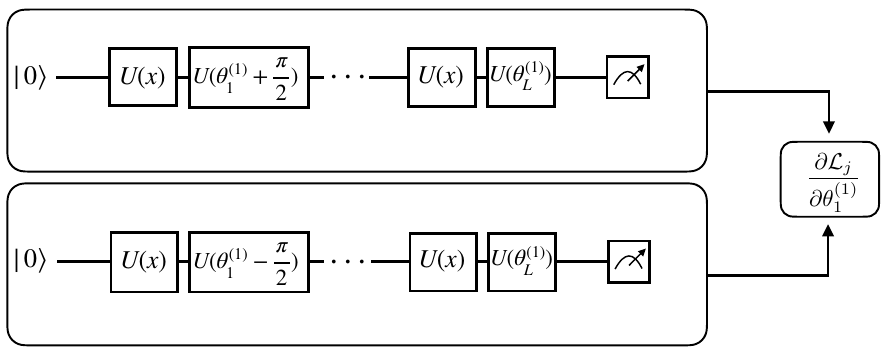}
    \caption{Gradient estimation using the parameter-shift rule.
The QNN architecture is evaluated twice with the parameter 
$\theta^{(1)}_{1}$ shifted by $\pm \frac{\pi}{2}$, while all other 
parameters are kept fixed.}
    \label{fig:GradienEstimation}
\end{figure}

In this appendix, we describe the protocol used to train the one-qubit QNN and the multi-qubit QNN architectures employing data re-uploading. The protocols are described as follow:

\begin{myprotocol}{Train the one-qubit QNN architecture}\label{protocol:QML_training_1Q}
   \textbf{Initialization:} Training dataset $\mathcal{X} =
   \{(\boldsymbol{x}^{(i)},y_x^{(i)})\}_{i=1}^{M_x}$. \\
   \textbf{Return:} The optimized parameters $\{\boldsymbol{\theta}^{*}\}^1$ for one-qubit QNN architecture where $\boldsymbol{\theta}^{*}$ contains the parameters $\{\theta_{l,1}^{(1)},\theta_{l,2}^{(1)},\theta_{l,3}^{(1)}\}$ for ${l \in \{1,\dots,L\}}$.\\

  \noindent\textbf{Procedure:}
    \begin{enumerate}
        \item For $t=1$ to the number of attempts chosen by the user.
    \begin{enumerate}
      \item \textbf{Initialize parameters.} The parameters $\boldsymbol{\theta} = \{\boldsymbol{\theta}_l^{(1)}\}=\{\theta_{l,1}^{(1)},\theta_{l,2}^{(1)},\theta_{l,3}^{(1)}\}$ are initialized within range $[-\pi, \pi]$, for all $l \in \{1,\dots,L\}$, according to a strategy selected by the user.
      
      \item \textbf{Train the QNN architecture.}\label{step:training1q}
      For each input $\boldsymbol{x}^{(i)} \in \mathcal{X}$:
      \begin{enumerate}
        \item \textbf{Construct the circuit.} Implement the circuit given by the Eq. \ref{eq:one-qubit-QNN}, where the data encoding of $\boldsymbol{x}^{(i)}$ is given by Eq.~\ref{eq:encode_data} and the parameters encoding is given by Eq.~\ref{eq:encode_parameters}.

        \item\label{step:Measure}\textbf{Measure.} Measure the qubit in the computational basis and
          estimate the expectation value of the Pauli $Z$ operator. This is achieved by repeatedly measuring in the computational basis, assigning the value $+1$ to outcome $\ket{0}$ and $-1$ to outcome $\ket{1}$, and computing the sample mean $\hat{\mu}_{\boldsymbol{\theta}}^{(i)}$. The total number of shots required depends on the desired estimation precision: to guarantee a maximum error $\varepsilon$ with confidence $1-\delta$, one requires
          \begin{equation}
            \frac{2}{\varepsilon^2}\,\log\left(\frac{2}{\delta}\right)
          \end{equation}
          shots.

        \item~\label{step:costfunction} \textbf{Cost function evaluation.} 
          Compute the  cost function: 
          \[
          \mathcal{L}(\boldsymbol{\theta}) 
          = 1 - \frac{1}{M_x} \sum_{i=1}^{M_x} F^{(i)}(\hat{\mu}_{\boldsymbol{\theta}}^{(i)}),
          \]
          where $F^{(i)}$ denotes the fidelity with respect to the corresponding label
          $y_x^{(i)}$ as follows
          \[ 
            F^{(i)}(\mu) = \frac{1+y_x^{(i)}\mu}{2},
          \quad y_x^{(i)} \in \{-1,1\}. 
          \]

        \item \textbf{Gradient estimation.}~\label{step:gradient}
        For each parameter $\theta_{l,j}^{(1)}\in \boldsymbol{\theta}$ with $j\in \{1,2,3\}$ and $l \in \{1,\dots,L\}$, the gradient
        of the cost function is estimated using the parameter-shift rule
        \cite{Schuld_2019}. This requires estimate the expectation value of the
        Pauli $Z$ operator after shifting only the parameter by
        $\theta_{l,j}^{(1)}\pm \tfrac{\pi}{2}$, while keeping all other parameters
        fixed (see Fig. \ref{fig:GradienEstimation}).
        \[
          \frac{\partial \mathcal{L}}{\partial \theta_{l,j}^{(1)}} =
          -\frac{1}{M_x}\sum_{i=1}^{M_x} \frac{y_x^{(i)}}{4}\Big(\langle Z \rangle_{+}
          - \langle Z \rangle_{-}\Big),
        \]
        where $\langle Z \rangle_{\pm}$ denote the expectation values of Pauli
        $Z$ operator after running the QNN architecture with $\theta_{l,j}^{(1)} \pm
        \tfrac{\pi}{2}$ respectively. These expectation values are estimated according to the procedure described in the previous step.

        \item \textbf{Parameter update.} Update all the parameters as follows
        \[
         \theta_{l,j}^{(1)} = \theta_{l,j}^{(1)} - \eta \, 
         \frac{\partial \mathcal{L}}{\partial \theta_{l,j}^{(1)}},
        \]
        denotes the learning rate, selected by the user, which may vary across training rounds. Go to Step~\ref{step:training1q} and repeat with the new parameters, until the convergence criterion chose by the user is satisfied.
      \end{enumerate}

      \item \textbf{Optimal parameter update.} If $t=1$, or if $t>1$ and $\mathcal{L}(\boldsymbol{\theta}) < \mathcal{L}(\boldsymbol{\theta^*})$ then update $\boldsymbol{\theta}^{*}=\boldsymbol{\theta}$.
    \end{enumerate}
    \end{enumerate}
\end{myprotocol}

\begin{figure}[ht]
    \centering     
    \includegraphics[width=0.75\columnwidth]{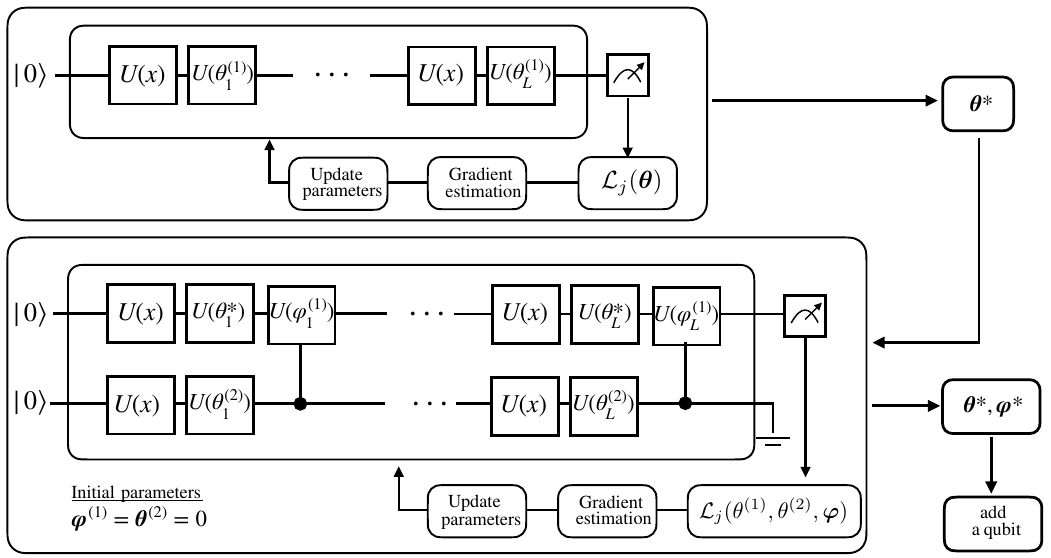}
    \caption{Iterative training of a two-qubit QNN architecture. First, a
    single-qubit QNN is trained to obtain the optimal parameters, denoted by
    $\boldsymbol{\theta}^{*}$. Next, the two-qubit QNN is initialized by
    assigning these optimal parameters to the first qubit, while the additional
    parameters are initialized to zero.}
    \label{fig:TrainingQNN}
\end{figure}

\begin{myprotocol}{Train the N-qubit QNN architecture}\label{protocol:QML_training_Nq}
   \textbf{Initialization:} Training dataset $\mathcal{X} =
   \{(\boldsymbol{x}^{(i)},y_x^{(i)})\}_{i=1}^{M_x}$, and the optimal parameters of the ($N-1$)-QNN $\{\boldsymbol{\theta}^{*},\boldsymbol{\varphi}^*\}^{N-1}$.  \\
   \textbf{Return:} The optimized parameters $\{\boldsymbol{\theta}^{*},\boldsymbol{\varphi}^*\}^N$ for the $N$-QNN architecture where $\boldsymbol{\theta}^{*}$ contains the parameters $\{\theta_{l,1}^{(r)},\theta_{l,2}^{(r)},\theta_{l,3}^{(r)}\}$ for $r\in \{1,\dots N\}$ and $l \in \{1,\dots,L\}$; and $\boldsymbol{\varphi}^*$ contains the parameters $\{\varphi_{l,1}^{(s)},\varphi_{l,2}^{(s)},\varphi_{l,3}^{(s)}\}$ for $s \in \{1,\dots,N-1\}$ and $l \in \{1,\dots,L\}$. \\

  \textbf{Procedure:}
   \begin{enumerate}
      \item \textbf{Initialize parameters.} The parameters of the $N$-QNN will be $\boldsymbol{\theta} = \{\boldsymbol{\theta}_l^{(1)},\dots,\boldsymbol{\theta}_l^{(N)}\}$ where $\boldsymbol{\theta}_l^{(r)} = \{\theta_{l,1}^{(r)},\theta_{l,2}^{(r)},\theta_{l,3}^{(r)}\}$ and $\boldsymbol{\varphi} = \{\boldsymbol{\varphi}_l^{(1)},\dots,\boldsymbol{\varphi}_l^{(N-1)}\}$ where $\boldsymbol{\varphi}_l^{(s)} = \{\varphi_{l,1}^{(s)},\varphi_{l,2}^{(s)},\varphi_{l,3}^{(s)}\}$ for all $l\in\{1,\dots,L\}$. The parameters of the $N$-QNN acting on qubits from 1 to $N-1$ will be equals to the optimal parameters of the $(N-1)$-QNN; and the parameters of the new qubit will be initialized to 0, i.e. 
          \begin{itemize}
            \item $ \theta_{l,j}^{(r)} = \theta_{l,j}^{*(r)},\ \forall j\in \{1,2,3\},\ r \in \{1,\dots,N-1\},\ l\in\{1,\dots,L\}.$
            \item $\varphi_{l,j}^{(r)} = \varphi_{l,j}^{*(r)},\ \forall j\in \{1,2,3\},\ r \in \{1,\dots,N-2\},\ l\in\{1,\dots,L\}.$
            \item $\theta_{l,j}^{(N)} = 0,\ \forall j\in \{1,2,3\},\ l\in\{1,\dots,L\}.$
            \item $\varphi_{l,j}^{(N-1)} = 0,\ \forall j\in \{1,2,3\},\ l\in\{1,\dots,L\}.$     
          \end{itemize}
          where $\theta_{l,j}^{*(r)}$ and $\varphi_{l,j}^{*(r)}$ are the optimal parameters of the $(N-1)$-QNN, $\{\boldsymbol{\theta}^{*},\boldsymbol{\varphi}^*\}^{N-1}$.
      
      \item \textbf{Train the QNN architecture.}\label{step:trainingNq}
      For each input $\boldsymbol{x}^{(i)} \in \mathcal{X}$:
      \begin{enumerate}
        \item \textbf{Construct the circuit.} Implement the circuit given by the Eq. \ref{eq:one-qubit-QNN}, where the data encoding of $\boldsymbol{x}^{(i)}$ is given by Eq.~\ref{eq:encode_data} and the parameters encoding is given by Eq.~\ref{eq:encode_parameters} and Eq.~\ref{eq:encode_parameters_multi}.

        \item\label{step:Measure_multi}\textbf{Measure.} Measure the first qubit in the computational basis and
          estimate the expectation value of the Pauli $Z$ operator, calculating $\hat{\mu}_{\boldsymbol{\theta},\boldsymbol{\varphi}}^{(i)}$ as in Step \ref{step:Measure} of the one-qubit QNN training protocol.

        \item \textbf{Cost function evaluation.} 
          Compute the  cost function: 
          \[
          \mathcal{L}(\boldsymbol{\theta}) 
          = 1 - \frac{1}{M_x} \sum_{i=1}^{M_x} F^{(i)}(\hat{\mu}_{\boldsymbol{\theta},\boldsymbol{\varphi}}^{(i)}),
          \]
          where $F^{(i)}$ is the same as in Step \ref{step:costfunction} of the one-qubit QNN training protocol.
        \item \textbf{Gradient estimation.} 
        For each parameter $\theta_{l,j}^{(r)}\in \boldsymbol{\theta}$ and $\varphi_{l,j}^{(s)}\in \boldsymbol{\varphi}$ with $j\in \{1,2,3\}$, $l \in \{1,\dots,L\}$, $r\in \{1,\dots,N\}$ and $s\in \{1,\dots,N-1\}$, the gradient
        of the cost function is estimated using the parameter-shift rule
        \cite{Schuld_2019} following the procedure described in Step \ref{step:gradient}, but extended to account for the increased number of parameters in the $N$-qubit QNN
        \begin{align}
          \frac{\partial \mathcal{L}}{\partial \theta_{l,j}^{(r)}} &=
          -\frac{1}{M_x}\sum_{i=1}^{M_x} \frac{y_x^{(i)}}{4}\Big(\langle Z \rangle_{+}
          - \langle Z \rangle_{-}\Big), \label{eq:diff_theta}\\
          \frac{\partial \mathcal{L}}{\partial \varphi_{l,j}^{(s)}} &=
          -\frac{1}{M_x}\sum_{i=1}^{M_x} \frac{y_x^{(i)}}{4}\Big(\langle Z \rangle_{+}
          - \langle Z \rangle_{-}\Big),\label{eq:diff_phi}
        \end{align}
        where $\langle Z \rangle_{\pm}$ denote the expectation values of Pauli
        $Z$ operator after running the QNN architecture with $\theta_{l,j}^{(r)} \pm
        \tfrac{\pi}{2}$ respectively, in Eq.~\ref{eq:diff_theta} and with $\varphi_{l,j}^{(r)} \pm
        \tfrac{\pi}{2}$ respectively, in Eq.~\ref{eq:diff_phi}. These expectation values are estimated according to the procedure described in the previous step.

        \item \textbf{Parameter update.} Update all the parameters as follows
        \begin{align}
              \theta_{l,j}^{(r)} &= \theta_{l,j}^{(r)} - \eta \, 
         \frac{\partial \mathcal{L}}{\partial \theta_{l,j}^{(r)}},\\
        \varphi_{l,j}^{(s)} &= \varphi_{l,j}^{(s)} - \eta \, 
         \frac{\partial \mathcal{L}}{\partial \varphi_{l,j}^{(s)}},
        \end{align}
        where $\eta$ denotes the learning rate, selected by the user, which may vary across training rounds. Go to Step~\ref{step:trainingNq} and repeat with the new parameters, until the convergence criterion chose by the user is satisfied.
      \end{enumerate}

      \item \textbf{Return optimal parameters.} Store and return the parameters $\boldsymbol{\theta}, \boldsymbol{\varphi}$ as the optimal parameters $\{\boldsymbol{\theta}^*, \boldsymbol{\varphi}^*\}^N$ of the $N$-QNN.
    \end{enumerate}
 Figure~\ref{fig:TrainingQNN} illustrates the iterative training procedure of a two-qubit QNN architecture.

\end{myprotocol}

\end{document}